\documentclass{SciPost}

\usepackage{hyperref}
\hypersetup{
    bookmarks=true,         
    unicode=false,          
    pdftoolbar=true,        
    pdfmenubar=true,        
    pdffitwindow=false,     
    pdfstartview={FitH},    
    pdftitle={},    
    pdfauthor={},     
    pdfsubject={},   
    pdfcreator={},   
    pdfproducer={}, 
    pdfkeywords={keyword1} {key2} {key3}, 
    pdfnewwindow=true,      
    colorlinks=true,       
    linkcolor=[rgb]{0.55,0,0.5},          
    citecolor=[rgb]{0,0,0.6},        
    filecolor=magenta,      
    urlcolor=blue           
}

\begin{document}

\begin{center}{\Large \textbf{
A microscopically motivated renormalization scheme for the MBL/ETH transition
}}\end{center}

\begin{center}
Thimoth\'ee Thiery\textsuperscript{1}, 
Markus M\"uller\textsuperscript{2}, 
Wojciech De Roeck\textsuperscript{1}
\end{center}

\begin{center}
{\bf 1} Instituut voor Theoretische Fysica, KU Leuven, Belgium
\\
{\bf 2} Paul Scherrer Institute, PSI Villigen, Switzerland
\\
\end{center}

%
%
%
%
%
%
%
%

\begin{center}
\today
\end{center}


\section*{Abstract}
{\bf 
We introduce a multi-scale diagonalization scheme to study the transition  between the many-body localized and the ergodic phase in
disordered quantum chains. 
The scheme assumes a sharp dichotomy between subsystems that behave as localized{,}  and resonant spots that obey the Eigenstate Thermalization Hypothesis (ETH). 
We {define the scheme by} a set of microscopic principles and use them to numerically study the transition in {
} large systems. To a large extent the results are in agreement with an analytically tractable mean-field {approximation}: We find that at the critical point the system is almost surely localized in the thermodynamic limit, hosting a set of thermal inclusions whose sizes are power-law distributed. 
On the localized side the  {\em typical} localization length is bounded from above. The bound saturates upon approach to criticality,  entailing that a finite ergodic inclusion  thermalizes a region of diverging diameter. The dominant thermal inclusions have a fractal structure, implying that {{\em average}} correlators decay as stretched exponentials throughout the localized phase. Slightly on the ergodic side thermalization occurs through an avalanche instability of the nearly localized  bulk, whereby rare, supercritically large ergodic spots  eventually thermalize the entire sample. Their size diverges at the transition, while their density vanishes. The non-local, avalanche-like nature of this instability entails a breakdown of single-parameter scaling and puts the delocalization transition outside the realm of standard critical phenomena. 
}

\vspace{10pt}
\noindent\rule{\textwidth}{1pt}
\tableofcontents\thispagestyle{fancy}
\noindent\rule{\textwidth}{1pt}
\vspace{10pt}

\section{Introduction and main results}

\paragraph{Introduction} 

Many-body-localization (MBL), i.e. the absence of thermalization in isolated interacting quantum systems \cite{anderson_absence_1958,fleishman_interactions_1980,basko2006metal,gornyi2005interacting,znidaric_many-body_2008,oganesyan_localization_2007,pal_many-body_2010,ros2015integrals,kjall2014many,luitz_many-body_2015,nandkishore_many-body_2015,altman_universal_2015,abanin_recent_2017,luitz_ergodic_2017,agarwal_rare-region_2017,imbrie_review:_2016}, is a fascinating topic at the border between statistical mechanics and condensed matter physics. An important step forward in understanding the fully localized phase was the realization that it hosts a robust form of integrability: MBL quantum Hamiltonians possess a complete set of quasi-local integrals of motions (LIOMs) \cite{huse2014phenomenology,serbyn_local_2013}, which naturally prevent thermalization. {In contrast, in the context of interacting quantum systems, the occurrence of thermalization is now widely accepted as being closely related with the applicability of  random matrix theory to the matrix elements of local observables, as described by the eigenstate thermalization hypothesis (ETH) }\cite{deutsch1991quantum,
srednicki1994chaos,rigol2008thermalization,d2015quantum}.  In {lattice models in} $d=1$ it was proven (up to an assumption) that the MBL phase exists \cite{imbrie2016many}, in the sense that a set of LIOMs can be rigorously constructed. The existence of a complete set of LIOMs is now often taken as a definition of many-body localization. As reviewed in \cite{imbrie2016review} {this characterization is useful in that many distinguishing properties of the MBL phase are directly implied by this property.}

\smallskip

In $d=1$, the main outstanding issue is the nature of the {dynamical} phase transition \cite{grover2014certain,serbyn2015criterion,potter2015universal,vosk2015theory,khemani2017critical,PhysRevLett.119.075702,PhysRevB.96.104205,kulshreshtha2017behaviour,parameswaran_eigenstate_2017,zhang2016many,dumitrescu2017scaling,devakul2015early,PhysRevB.94.045111} that separates the MBL from the ETH phase. Since current exact numerical studies are mostly restricted to exact diagonalization and small sample sizes \cite{kjall2014many,luitz_many-body_2015,khemani2017critical,devakul2015early,PhysRevB.94.045111}, it is of great interest to establish effective descriptions that permit to access the thermodynamic limit. Several phenomenological renormalization schemes have  already been introduced in this spirit \cite{potter2015universal,vosk2015theory,zhang2016many,dumitrescu2017scaling}, with, however, partially conflicting predictions. In this paper we develop a {new} multi-scale diagonalization procedure \cite{wegner1994flow} to construct the LIOMs iteratively \cite{imbrie2016many,PhysRevLett.116.010404,PhysRevLett.119.075701,monthus2016flow,rademaker2017many}, {in which we  build in the possibility of quantum avalanches: the growth of ergodic regions around rare spots of higher chaoticity (as resulting from locally weaker disorder, e.g.), a tendency which we find to become explosive at criticality}. For a localized system our procedure provides the complete set of LIOMs. {For ergodic samples instead it elucidates how thermal mixing occurs in space, and permits, e.g., to estimate the local thermalization time at any position along the chain.}

\smallskip

Our diagonalization procedure is based on two microscopic principles. The first principle, used to diagonalize perturbative couplings, is spectral perturbation theory. The second principle is the use of random matrix theory/ETH  for nonperturbative (resonant) couplings \cite{de2017stability,chandran2016many,ponte2017thermal}, i.e. ergodic inclusions. In \cite{de2017stability} it was shown that this second principle remarkably leads to an `avalanche' instability \cite{de2017stability}: an infinite localized system can be thermalized by a finite {ergodic seed} if the typical localization length $\zeta$ of the MBL system exceeds a critical value $\zeta_c$. If $\zeta$ is smaller, the size of the region thermalized by an ergodic inclusion is proportional to the size of the inclusion. {That is, it is sensitive to the bulk of the inclusion, not only its surface, which is} a signature of the non-local features of thermalization. These striking conclusions were recently verified with accurate numerical simulations in \cite{luitz2017}. These unconventional effects were not taken into account in previous RG treatments of the transition \cite{potter2015universal,vosk2015theory,zhang2016many,dumitrescu2017scaling} or in the existing numerical diagonalization schemes \cite{PhysRevLett.116.010404,PhysRevLett.119.075701,monthus2016flow,rademaker2017many}  to construct LIOMs. That distinguishes our work fundamentally from previous approaches. It seems rather natural to take these two principle as the basis of our diagonalization scheme: spectral perturbation theory is (as in  \cite{ros2015integrals,imbrie2016many}) relevant as it entails that the LIOMs must be small deformations of the physical local degrees of freedom, while ETH is the best available description of ergodic systems. Our diagonalization scheme therefore leads to results that are consistent with what is currently understood about the MBL and ETH phases, {while providing new predictions about the nature and the characteristics of the transition.}

\smallskip

A mean-field analysis of our diagonalization scheme was already reported in \cite{ThieryHuveneersMullerDeRoeck2017}. The goal of the present paper is {\em (i)} to give a careful microscopic derivation of the mean-field flow equations presented in \cite{ThieryHuveneersMullerDeRoeck2017}, and {\em (ii)} to compare the results of the mean-field approach with numerical simulations of the unapproximated scheme. We find the results to be in agreement to a large extent. Below we summarize the main conclusions of our approach:

\paragraph{Main results}
\begin{enumerate}
\item
The typical localization length $\zeta$ for the {\it norm} of LIOMs remains bounded throughout the MBL phase, with $\zeta \leq \zeta_c$, where $1/\zeta_c$ is the {entropy density} at infinite temperature.
\item
The critical point is localized with probability $1$ in the thermodynamic limit. The size $S$ of thermal inclusions at the critical point is distributed with a power-law exponent: $p(S) \sim S^{-\tau}$, with $\tau >2$. The typical localization length $\zeta$ saturates the bound $\zeta = \zeta_c$ at criticality. The latter implies an `infinite response' to large thermal inclusions at the critical point: including a thermal region of size $k$ in a critical localized system thermalizes $\ell_k$ spins with $\ell_k/k \to_{k \to \infty} \infty$.
\item
Beyond the critical point delocalization is driven by an `avalanche' instability whereby rare ergodic regions of size $k \geq k_*$ thermalize  the system completely (while the system would be localized without these regions). The threshold size $k_*$ diverges as the critical point is approached from the thermal side.
\item
There is no single parameter scaling at the transition: The transition of {eigenstate properties} appears continuous from the MBL side and discontinuous from the thermal side. This arises as a consequence of the fact that the mechanism for delocalization in the thermal phase (the avalanche instability) is absent in the MBL phase. The divergence of $k_*$ leads to important finite size effects on the thermal side where samples of arbitrarily large size can appear localized.
\item
The average localization length and the average eigenstate correlation length diverge algebraically upon approaching the transition from the MBL side. On the thermal side the average eigenstate correlation length diverges algebraically only when measured from end to end, and remains bounded otherwise. The typical thermalization time diverges super-exponentially.
\item
Thermal inclusions display a fractal structure in the MBL phase, implying that average correlators in localized eigenstates decay subexponentially in the MBL phase.
\end{enumerate}

The conclusions $1)$ to $5)$ were already present in the mean-field analysis of \cite{ThieryHuveneersMullerDeRoeck2017} and are confirmed here. The new prediction $6)$ follows by accounting for (simple) effects neglected in the mean-field analysis. The prediction $5)$ will not be detailed and we refer to \cite{ThieryHuveneersMullerDeRoeck2017} for more details. To avoid confusion, in this paper we only focus on the {\it typical} localization length which, as in \cite{ThieryHuveneersMullerDeRoeck2017}, remains bounded at the transition.

\medskip

Let us discuss here the relation between our results and the literature. Predictions 1)-2) about the bound on $\zeta$ in the MBL phase, that is  saturated only at the transition (and drives the transition) is specific to our work, which thus differs qualitatively from previous RG approaches \cite{vosk2015theory,potter2015universal,dumitrescu2017scaling,zhang2016many}. For example, \cite{dumitrescu2017scaling} predicts a divergence of $\zeta$ at the transition and while \cite{vosk2015theory} also set a bound on $\zeta$, the bound is weaker and not linked to the transition. The prediction 2) that the critical point is  localized with probability $1$ was also to our knowledge never put on the foreground and is in clear contradiction with \cite{dumitrescu2017scaling,zhang2016many}, but in agreement with exact diagonalization studies \cite{kjall2014many,luitz_many-body_2015,khemani2017critical} where a subvolume scaling of the average half-chain entanglement entropy at the transition was reported. The property of having a power-law distribution of thermal spots at criticality is common to other approaches. The prediction 3) is specific to our approach as this is the first study that takes the avalanche instability into account. The prediction 4) is also new (we note however that the importance of finite size effects on the thermal side was already discussed in \cite{devakul2015early}). The prediction 5) that only the end-to-end average correlation length can diverge at the transition on the thermal side is original to our work. The divergence of the average localization and correlation length on the MBL side was already suggested before, but the mechanism for this divergence that is present in our work is also original. Finally the prediction 6) coincides with the one of \cite{zhang2016many}, with the same explanation.

\paragraph{Outline} The outline of the paper is as follows. Sec.~\ref{sec:thescheme} introduces our RG scheme in a general setting. Sec.~\ref{sec:estimates} discusses the main estimates that permits in Sec.~\ref{sec:AnalysisGen} to analyze in more depth the different (perturbative and non-perturbative) aspects of the scheme. Sec.~\ref{sec:MF} discusses the mean-field approximation to our scheme (already analyzed in \cite{ThieryHuveneersMullerDeRoeck2017}). Sec.~\ref{sec:numerics} finally gives the results of the simulations of the scheme. Sec.~\ref{sec:estimates}-\ref{sec:AnalysisGen}-\ref{sec:MF} are the most technical and can be skipped in a first reading, and we repeat that  \cite{ThieryHuveneersMullerDeRoeck2017} is a good introduction to this work.

\section{The diagonalization procedure in principle} \label{sec:thescheme}

In this section we describe our diagonalization procedure. The procedure can be applied in principle to any system potentially displaying an MBL/ETH transition (otherwise it is not informative as it focuses on this dichotomy) and here we focus on a one-dimensional model with short-ranged interactions and no specific microscopic structure: the model is built on random matrices.

\subsection{Model, roadmap and notations} \label{subsec:scheme:roadmap}

\paragraph{Model} We consider a spin-$S$ chain of length $L$ with local couplings. At each site $i \in [1,L]$, we have hence a copy of the space $\mathbb{C}^{{d}}$ where ${d}=2S+1$. In what follows, the spin does not play any role (it is not conserved) and only the dimension ${d}$ remains meaningful. We choose a distinghuished basis at each site (for example corresponding to the operator $S^z$) and we write for the tensor products of such base vectors 
$$
\ket{\eta} =  \ket{\eta_{1}} \otimes \ket{\eta_{2}} \otimes \ldots  \ket{\eta_{L}},\qquad \ket{\eta_j} \in \mathbb{C}^{{d}} 
$$
We also refer to this basis as the `localized basis'. 
We consider an Hamiltonian of the form
\bea \label{Eq:defH}
H && = \sum_{i} (D_i + D_{\{ i , i+1 \}} + V_{\{i,i+1 \}}) \\ 
&& = \sum_I ( D_I +V_I  )\nonumber \, ,
\eea
where:
\begin{enumerate}
\item the operators $D_i$ act on site $i$ and are diagonal in the basis $\eta_i$. They have independent, identically distributed (iid) random eigenvalues drawn from a distribution of width $W>0$;
\item  the operators $D_{\{ i , i+1 \}}$ act on  two neighboring sites $i,i+1$ and they are diagonal in the $\eta$-basis, with random eigenvalues also iid with a distribution of width $W$; 
\item   the $V_{\{ i ,i+1\}}$ are coupling terms acting non-trivially on the spins at site $i,i+1$, taken as iid GUE random matrices with the variance of off-diagonal matrix elements taken as $g >0$ and zero diagonal elements.
\end{enumerate}
In the second line of \eqref{Eq:defH} we have introduced the notation $I$ to label stretches of sites, i.e. sets of consecutive sites.  We always write $D_I$ for diagonal operators and $V_I$ for operators with vanishing diagonal, and we refer to the $V_I$ as couplings.  We will also often write $d_I = d^{|I|}$ for the total dimension of the local Hilbert space on the stretch $I$.
None of these specifics is essential for our treatment, but for the above model, our analysis applies in the most direct way.  For example, we certainly do not require that the disorder is both on-site and in the couplings $V_I$, but after a single step of our scheme, this will be the case anyhow.

\paragraph{Roadmap} The goal of our procedure is to diagonalize $H$ in the localized basis: we aim at constructing an unitary operator $U$ such that $H' = U H U^{\dagger}$ is diagonal in the basis of the eigenvectors of $D_i$. This is always possible and we say that the system is in the MBL phase if $U$ can be constructed in such a way that it is quasi-local (see def in e.g. \cite{imbrie2016review}). In that case $\hat{D}_i = U^{\dagger} D_i U$ defines a complete set of local integrals of motions (LIOMs) for $H$, the defining feature of the `fully' MBL phase. If $U$ is not quasi-local, then we say that the system is in the thermal phase.

We construct $U$ recursively in a similar way as done in the work of Imbrie \cite{imbrie2016many}: $U = \prod_{k}U_k $, where at each elementary step the Hamiltonian is `locally' diagonalized, i.e. non-diagonal operators in the running Hamiltonian $H_n = (\prod_{k=1}^{n} U_k)^\dagger H (\prod_{k=1}^{n} U_k)$ are recursively eliminated. Taking the example of the bare Hamiltonian \eqref{Eq:defH} $H_0=H$ the coupling operators are the $V_{\{i,i+1\}}$. While small (perturbative, see definition in Sec.~\ref{subsec:perturbativeVsnonpert}) couplings will be eliminated using perturbation theory (see Sec.~\ref{subsec:scheme:elimination}), with a small rotation $U_i$ that is close to the identity, large (nonperturbative or resonant) couplings will be eliminated with a unitary rotation $U_i$ that is far from the identity and that will be taken as a random matrix (see Sec.~\ref{subsec:scheme:fusion}). While perturbative rotations preserve the quasi-locality of $U$, non-perturbative rotations scramble the degrees of freedom locally, eventually leading the system to the thermal phase. We first define precisely these elementary steps and discuss in Sec.~\ref{subsec:scheme:order} the order in which perturbative and non-perturbative rotations should be applied, thereby fully defining our diagonalization scheme.


\subsection{Perturbative vs. Nonperturbative couplings} \label{subsec:perturbativeVsnonpert}

The distinction between perturbative and nonperturbative (resonant) couplings is at the heart of our procedure. 
Let us fix a stretch $I$. We say that the coupling $V_I$ is `perturbative' if the eigenstates of 
$$
H_I= D_I+V_I
$$
are `small' perturbations of the eigenstates of ${D}_I$, i.e. of the localized base-states $\ket{\eta}$. Writing $E_I(\eta)=\braket{\eta|D_I|\eta}$ for the eigenvalues of these eigenstates, spectral perturbation theory applies if, for any $\eta,\eta'$
\bea \label{Eq:TooStrongCriterion}
\cag_{I}(\eta,\eta'):= \left|\frac{\braket{\eta |V_I | \eta'}}{E_I(\eta)-E_I(\eta')}\right| <1.
\eea
If this condition is violated for a pair $(\eta,\eta')$, then these two states are hybridized by the interaction. However, hybridization of a few pairs can not cause real delocalization.
Indeed, if a single pair of eigenstates $(\ket \eta,\ket{\eta'})$ (resp. $(\ket{\eta''} , \ket{\eta'''})$) are hybridized by $V_{\{ i,i+1 \}}$ (resp. $V_{\{ i+1,i+2 \}}$), then most likely no pairs of eigenstates other than those are hybridized by $V_{\{ i,i+1 \}}+V_{\{ i+1,i+2 \}}$.

 Instead, we declare a coupling `resonant' if for a typical choice of $\eta$, there exist $\eta'$ such that $\cag_{I}(\eta,\eta') \geq 1$.  This condition is sufficient to ensure that (all)\footnote{We are thus looking in this work at the MBL/ETH transition at infinite temperature. It was however argued in \cite{de2016absence} that a transition as a function of the energy density is impossible, thus implying that our work is not restrictive.} the eigenstates of $D_I+V_I$ are indeed delocalized in the basis of the eigenstates of $D_I$. Conversely a coupling is thus said to be perturbative iff.\  $
 \max_{\eta'} {\cal G}_I(\eta,\eta') < 1 $ for \emph{typical} $\eta$. By \emph{typical} we mean here, for example,  logarithmic averaging:
\bea \label{Eq:DefCalGI}
{\cal G}_I  = \exp \left( {\rm Av}_{\eta}  {\rm max}_{\eta'} \log |\frac{\braket{\eta |V_I | \eta'}}{E_I(\eta)-E_I(\eta')}| \right),
\eea
with ${\rm Av}_{\eta} =   \frac{1}{{d}^{|I|}}\sum_{\eta}$ the average with respect to the $\eta$-states.  Hence, our final condition for a coupling to be perturbative reads:
$$
\cag_I <1 \, .
$$

{\it Remark } Thus in the strict sense we should not be able to apply perturbation theory for some perturbative couplings. We will nevertheless do it but our procedure should in principle be supplemented by local non-perturbative rotations dealing with possible isolated local resonances to pave the way to the use of perturbation theory.

\subsection{Elimination and creation of perturbative couplings} \label{subsec:scheme:elimination}

If a coupling is perturbative, it can be eliminated by a `small' base change, which is simply one way of implementing spectral perturbation theory. 
Let us make this more precise. Let $\eta,\eta'$ label eigenstates  of ${ D}_I$. One can hope to diagonalize $H_I :={ D}_I+V_I$ in the $\eta$ basis by conjugating it with a rotation $U$: the transformed Hamiltonian is
$$
H'_I= U H_I U^\dagger \, .
$$
Making the Ansatz $U=\e^{ A_I}$ with $A_I$ `small' and anti-Hermitian we get 
\beq \label{eq: expansion of hprime}
H'_I= {D}_I +[ A_I, { D}_I] + V_I +\text{higher orders}  \, .\nonumber
\eeq
Before commenting on the higher order, let us choose $A_I$ 
such that it eliminates the first-order term in $V_I$, i.e.\ 
\beq \label{eq: first order elimination}
V_I +[ A_I, { D}_I] =0 \, . \nnn
\eeq
This equation is solved as
\be \label{eq:defA}
\braket{\eta|A_I |\eta' }= \frac{\braket{\eta|V_I |\eta' }}{E_I(\eta)-E_I(\eta')} (1 - \delta_{\eta,\eta'}) \, ,
\ee
and we have thus indeed eliminated $V_I$ at first order in perturbation theory. Let us now inspect the higher order terms in \eqref{eq: expansion of hprime}. They are given by (upon taking into account \eqref{eq: first order elimination})
$
\sum_{k \geq 1} \frac{1}{k!} \ad^k_{A_I} V_I ,
$
where $\ad_A B = [A,B]$. To go further this terms should be separated into `true' coupling terms and diagonal terms (renormalizing the eigenvalues). The coupling terms are then smaller than the original coupling and iterating the procedure the Hamiltonian is indeed locally diagonalized.

\begin{figure}
\centerline{\includegraphics[width=8cm]{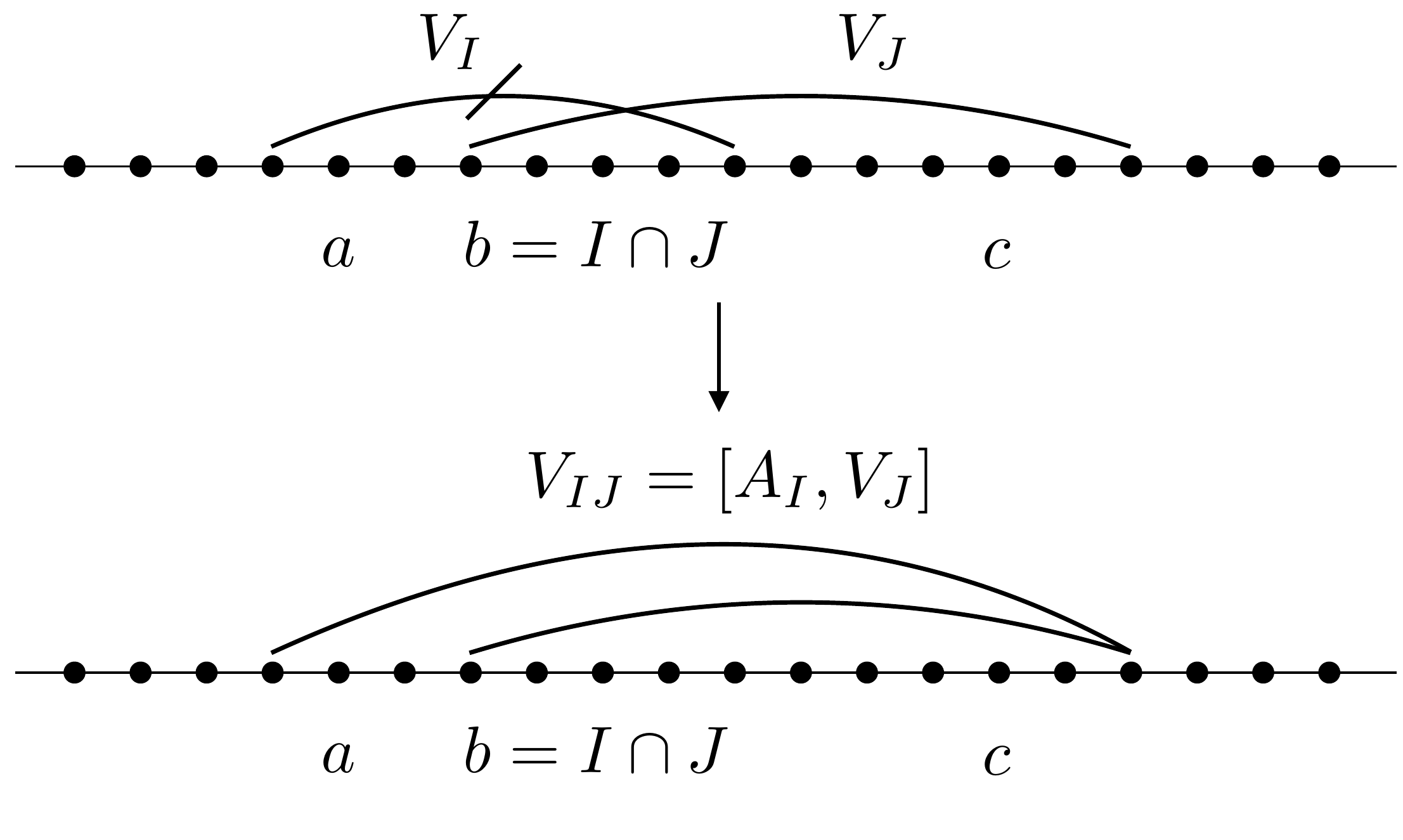}} 
\caption{Couplings $V_I$ and $V_J$ are indicated by an arc bridging a stretch of the chain. The couplings act on all sites of the stretch.  The coupling $V_{I}$ is eliminated perturbatively, which is indicated by the slash through the arc. This generates a new coupling $V_{IJ} = [A_I,V_J]$.}
\label{Fig:VIJ}
\end{figure}

\begin{figure}
\centerline{\includegraphics[width=10cm]{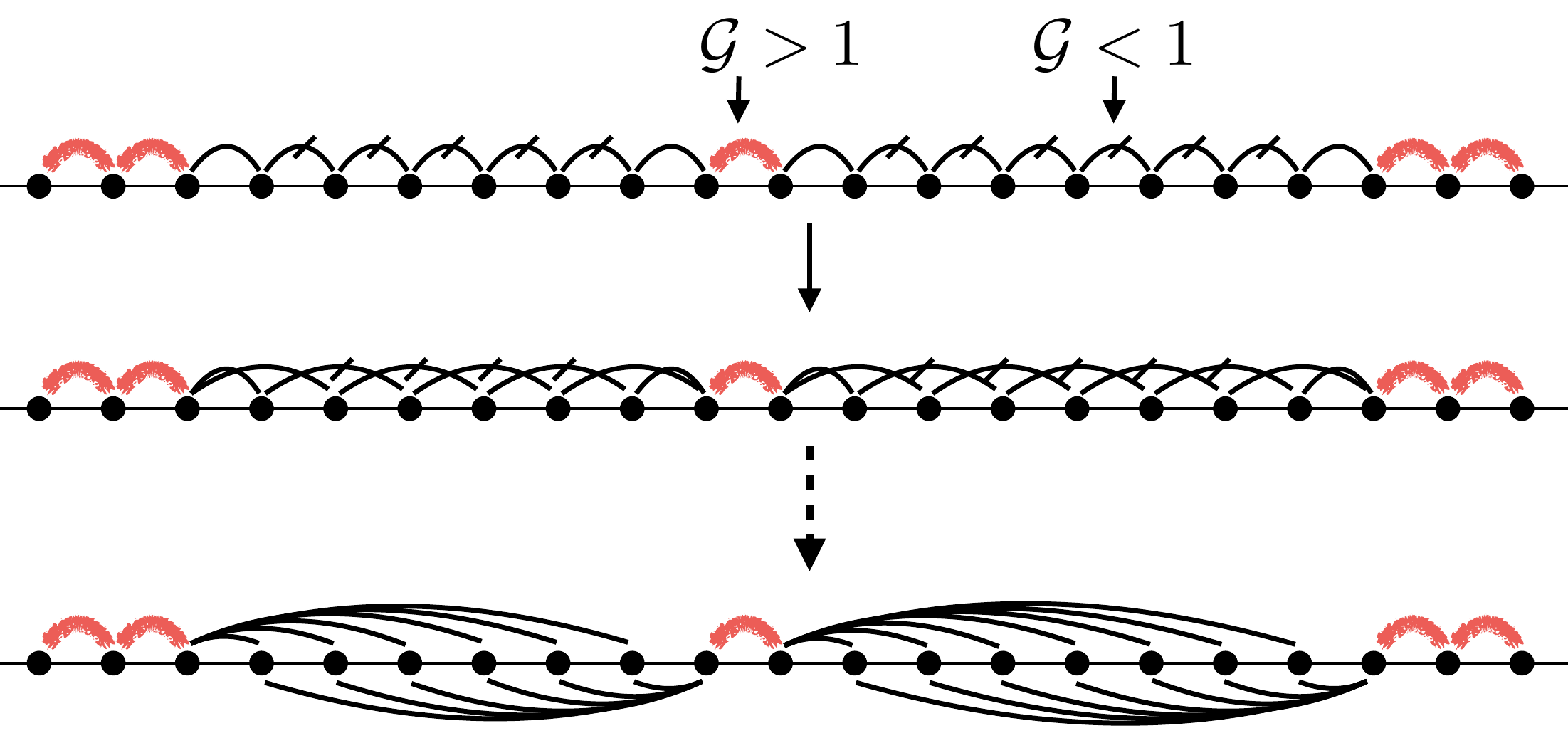}} 
\caption{Elimination of perturbative couplings in between resonant regions at the beginning of the procedure. The red arcs are resonant couplings that are not touched here, nor are perturbative couplings (black arc) that connect to these resonant regions. The upper picture is the spin chain at the beginning. Successively eliminating the perturbative coupling not touching the resonant regions we arrive at the bottom picture where we have drawn for more clarity the couplings linking the resonant regions to the localized material on their left (resp. right) above (resp. below) the chain. Note that the coupling linking a resonant region to the spin at the distance $\ell$ also acts on all the spins in between.}
\label{Fig:FirstBlob}
\end{figure}

In the above discussion we have restricted ourselves to the stretch $I$. If we consider also the environment, there will be couplings $V_J$ that overlap with $I$, i.e.\ $I \cap J \neq \emptyset$ and the rotation $\e^{A_I}$ affects them (see Fig.~\ref{Fig:VIJ}): 
$$
\e^{A_I} V_J \e^{- A_I}= V_J+ [A_I,V_J]+\text{higher orders} \, .
$$
Hence, at first order, new couplings and diagonal terms are created and we write 
$$
D_{IJ}+V_{I J} =  [A_I,V_J] \, .
$$
Here, we wrote schematically $IJ$ for $I\cup J$, the stretch that is the union of $I,J$ and we \emph{defined} $D_{IJ}$ as the diagonal part of the right-hand side. Note that the roles of $V_I$ and $V_J$ in $V_{IJ}$ are not symmetric: $V_{IJ} = [A_I,V_J]$, and energy denominators only appear on the stretch $I$.  The procedure is illustrated on Fig.~\ref{Fig:VIJ}. 
We will iterate this procedure and Fig.~\ref{Fig:FirstBlob} shows how the result of such iterations looks locally in between two resonant couplings (it will be clear later on that perturbative couplings acting on resonant regions should not be eliminated). 

\subsection{Fusion of resonant couplings} \label{subsec:scheme:fusion}

\subsubsection{The fusion operation}
Above we described how to eliminate perturbative couplings. Necessarily, there will also be resonant, ie non-perturbative couplings. These can be eliminated by a non-perturbative unitary $U$, the properties of which are a priori unknown. In this work we will assume that the region spanned by the resonant couplings is fully thermalized and the unitary $U$ is one that diagonalizes a fully ergodic Hamiltonian satisfying ETH.
This assumption and its  implications have been very well confirmed in recent numerics \cite{luitz2017}.  We explain below (see Sec.~\ref{subsec:scheme:order}) when the assumption is justified in the context of our scheme, but first we state it explicitly. 

\medskip

\begin{figure}
\centerline{\includegraphics[width=12cm]{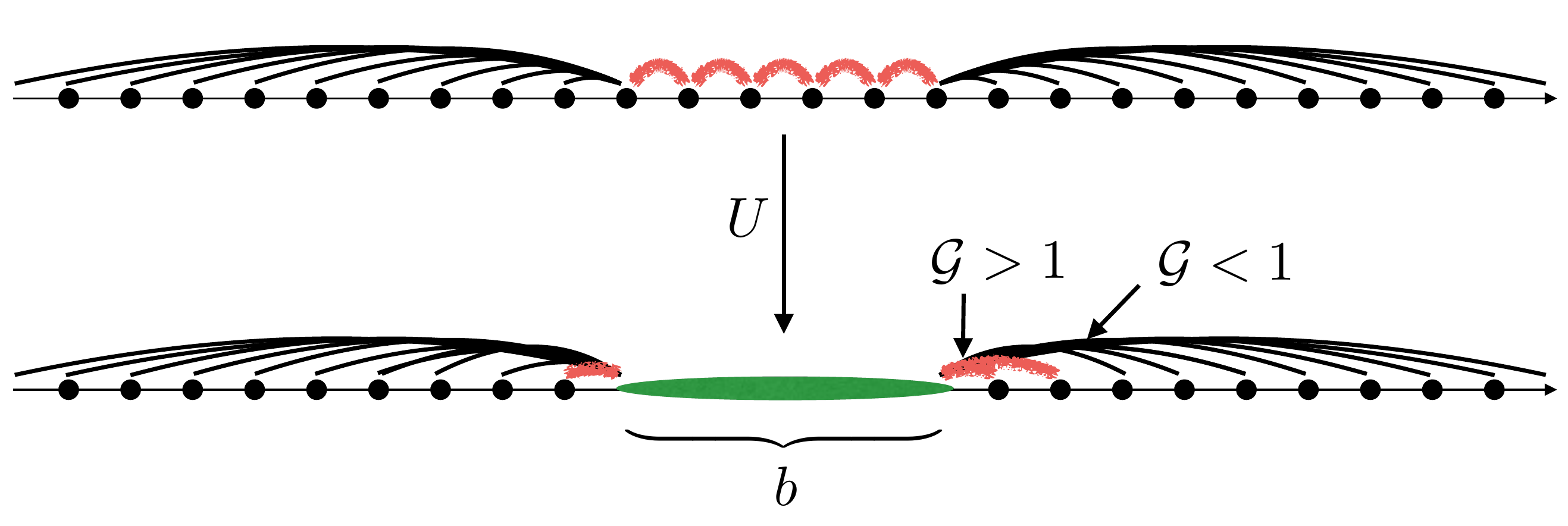}} 
\caption{A resonant region is diagonalized by a random unitary $U$. The system is locally thermal and we say that the spins inside the resonant regions were `fused' into a `block' $b$, symbolized by a green area. The perturbative couplings linking the resonant spot to the surrounding spins are modified by the rotation. These notably now see a smaller level spacing and might become resonant.}
\label{Fig:firstnonpert}
\end{figure}

An {\it active resonant spot} is defined as a stretch of blocks $I=I_1 \cup I_2 \cup \ldots \cup I_r$ that is the union of a connected cluster of resonant couplings $V_{I_1},\ldots, V_{I_r}$, i.e.\  $V_{I_j} \cap V_{I_{j+1}} \neq \emptyset$ and all $\cag_j >1$.
We say that we `fuse' these stretches, meaning that we abandon all information contained in $I$ and we assume that the Hamiltonian inside $I$, e.g.\ 
$$
{ D}_I + \sum_{j=1}^r V_{I_j} \, ,
$$
is a perfectly ergodic system, satisfying in particular the ETH. This implies that the unitary $U$ that diagonalizes this local Hamiltonian in the localized basis is essentially a random unitary. We set
\bea
D_b := U \big(  D_I + \sum_{j=1}^r V_{I_j} \big) U^\dagger  \, ,
\eea
with $D_b$ diagonal in the $\eta$-basis. This diagonal operator has generically no locality on smaller scales than $I$ and for this reason we think of the sites as being fused into a block $b=I$ (see Fig.~\ref{Fig:firstnonpert}).  In the following $I$ might refer to a stretch of indices and blocks, as it would be too tedious to keep an explicit distinction.

\subsubsection{Influence on the surrounding couplings}

The unitary rotation $U$ also acts non trivially on all perturbative couplings that were connected to a block inside the resonant region. If the fused blocks consists in $b=I_1 \cup I_2$ and a perturbative coupling $V_{I_2 I_3} $ acts on $I_2 \cup I_3$ with $I_3 \cap b = \emptyset$, then the coupling is modified as
\bea
V'_{bI_3} = U V_{I_2I_3} U^\dagger \, .
\eea
This coupling now acts on $b \cup I_3$, and since $U$ is unitary, the operator norm of the coupling is conserved
\bea
|| V'_{bI_3}  || = || V_{I_2I_3}  || \, ,
\eea
where the operator norm is defined as $||A|| = \sup_{||\ket{\psi}||=1} ||A \psi ||$. Since the unitary operation $U$ is approximated by a random matrix that acts non-trivially only on the block $b$, the created operator $V'_{bI_3}$ are now structureless on $b$, but the possible precedent structure (see Sec.~\ref{sec:estimates}) that was present on the remaining sites is retained in the transformation. 
\emph{It is exactly the fact that the couplings become structureless on fused regions, that gives our theory a universal and predictive character. } 
For example, if the length of $b$ is much larger than $I_3$ (which could simply be a stretch consiting of two sites), then the matrix elements of  $V'_{bI_3}$ can be estimated as 
\begin{equation}\label{eq: rmt surroundings}
\langle \eta| V'_{bI_3} | \eta' \rangle \sim  d^{-1/2}_{bI_3}  || V'_{bI_3} ||,
\end{equation}
corresponding to the assumption that $V'_{bI_3}$ is a random matrix, i.e.\ a matrix that for all practical purposes is a GUE matrix. The new coupling $V'_{bI_3}$ now acts on a space of larger dimension (it now acts on the full block and not only on one of the spin at the boundary) and thus sees a smaller level spacing: it might become resonant (see Fig.~\ref{Fig:firstnonpert}). If that occur the fusing operation must be repeated (eventually later in the scheme, see Sec.~\ref{subsec:scheme:order}).

\subsection{The order of fusion/elimination} \label{subsec:scheme:order}

In general, the system will contain both resonant and perturbative couplings. The question then arises whether we should first fuse (nonperturbative rotations) or first eliminate (perturbative rotations), and which couplings to treat first. The schematic answer is that we should first eliminate perturbative couplings, because perturbative eliminations are exact operations whereas fusion incurs a loss of information.  Moreover, the simple RMT treatment of ergodic spots remains valid only if the surrounding (localized) Hamiltonian has been rewritten in a LIOMs form: the long-range couplings between the ergodic spots and the a priori localized spins are exactly the one that are relevant to decide whether or not the spins are thermalized by the spot.
This is illustrated by the example of Sec.~\ref{subsubsec:allpertfirst}. Thereafter (see Sec.~\ref{subsubsec:notallpertfirst}), we also give an example that shows that doing \emph{all} perturbative eliminations first is inconsistent. Based on these two examples we formulate in Sec.~\ref{subsubsec:summaryoftthescheme} the order in which we alternate perturbative and non-perturbative rotations, thereby fully defining our diagonalization scheme.

\subsubsection{The importance of eliminating perturbative couplings first} \label{subsubsec:allpertfirst}
Assume that we have a large number of consecutive resonant nearest-neighbour couplings, i.e.\ a large ergodic spot. Let us fuse these couplings before eliminating any of the surrounding perturbative couplings. The result will be that the couplings linking the nearest neighbours of the spot to the spot will now be resonant with a high probability because of the small level spacing inside the spot. Let us continue and fuse this new, larger resonant spot (the initial spot + the nearest neighbours spins). When doing so, the couplings to the new nearest neighbours has even more chance to be resonantly coupled to the (enlarged) spot.  Iterating one obtains that the spot will with high probability invade the full system. Indeed, the only way that it would fail to do so is if it would encounter along the way an extremely small two-body coupling, the strength of which has to be exponentially smaller the further this coupling occurs from the original spot.  

It is instructive to reflect on where precisely the 'error' has occurred in this procedure. This question was analyzed in details in \cite{de2017stability}. To understand the answer we need more details on the ETH ansatz. ETH declares that the off-diagonal matrix elements of a local operator $V$ are given by 
$$
\langle \eta |V| \eta' \rangle =  e^{-S/2} f(\omega) R_{\eta,\eta'} 
$$
with $R$ random numbers with zero mean and unit variance, $S$ the  entropy at energy $\tfrac{1}{2}(E_\eta+E_{\eta'})$ and $f(\omega)$ a function of $\omega= E_\eta-E_{\eta'} $ that is related to the time dependent correlation function $\langle V(t)V\rangle$, see \cite{d2015quantum,de2017stability}.  In well-coupled systems, $f$ decays rapidly once $\omega$ exceeds the energy per site, and it has some (interesting) structure for smaller $\omega$, with features that scale polynomially with the volume.  Since $e^{-S/2}$ is exponentially small in the volume, it is usually justified to drop $f$, leading, at maximum entropy, to the random matrix prediction (cf.\ \eqref{eq: rmt surroundings})
$$
\langle \eta |V| \eta' \rangle = \frac{|| V||}{\sqrt{d_{\text{tot}}}}. 
$$
We can now return to the question at hand. When the spot grows by `eating its neighbours', one can exhibit that the resulting couplings from the spot to the (new) neighbours have an ever more singular function $f$, whose effect becomes now comparable to the term $e^{-S/2}$, invalidating the use of the RMT ansatz.

Eventually, this analysis reproduces the conclusions of a slicker procedure; namely to first eliminate the perturbative couplings outside the ergodic spot. In the latter case, the couplings that will play an important role are those that connect to spins of the original ergodic spot. These couplings have better behaved $f$ and the RMT approximation remains justified. This is the `strategy' used in this paper: we first use perturbation theory outside resonant spots so that the RMT approach to resonant spots leads to a simple derivation of the extension of the space thermalized by the spots.

\subsubsection{The danger of eliminating all perturbative couplings first} \label{subsubsec:notallpertfirst}

Some perturbative couplings should however not be eliminated immediately. Consider for example the case of three spins at sites $1,2,3$, with the spins $1,2$ being coupled by an unusually large (resonant) coupling $V_{12}$, whereas the coupling $V_{23}$ from $2$ to $3$ is much weaker. 

If we eliminate $V_{23}$ through perturbation theory first, then the created coupling 
$$
V_{123}=[A_{23},V_{12}]
$$
depends linearly on $V_{12}$ and it will hence be unusually large as well. A first clear issue here is that we are creating though perturbation theory couplings that cannot be treated through perturbation theory, invalidating the property that perturbation theory can be used to exactly diagonalized in the localized basis the portions of the system that are devoted of resonant spots. 

In contrast, if we had first diagonalized the spot $b=\{12\}$, then the only remnant of the large coupling $V_{12}$ would now be in the fact that the diagonal operator $D'_{12}=U(D_{12}+V_{12})U^\dagger$ is large.
For the purpose of coupling site $3$ to the spot $b$, this would actually tend to decrease the effect of the coupling $V_{23}$ because of a large denominator, and hence this setup could in fact favor localization of site $3$. 
For conceptual completeness, note that we neglected here a opposite effect, namely the fact that after `fusing' the spot $b$, the relevant level spacing for the calculation of $\mathcal G$ for $V_{23}$ is smaller (by a factor $2$) than it would be if the spot had not been fused. However, if $V_{12}$ was sufficiently large, this opposite effect is irrelevant.  The scenario described here lies of course at the heart of the traditional strong-disorder renormalization scheme where the largest terms are treated first.

\subsubsection{Conclusion} \label{subsubsec:summaryoftthescheme}

We implement the knowledge gained from the above examples to define the sequence of perturbative and nonperturbative rotations that is implemented in our scheme.

\paragraph{{\bf Step I}}  We eliminate all perturbative couplings except those that touch an active resonant spot. 

\paragraph{{\bf Step II}}
We fuse the smallest active resonant spot in the system. 

\begin{figure}
\centerline{\includegraphics[width=10cm]{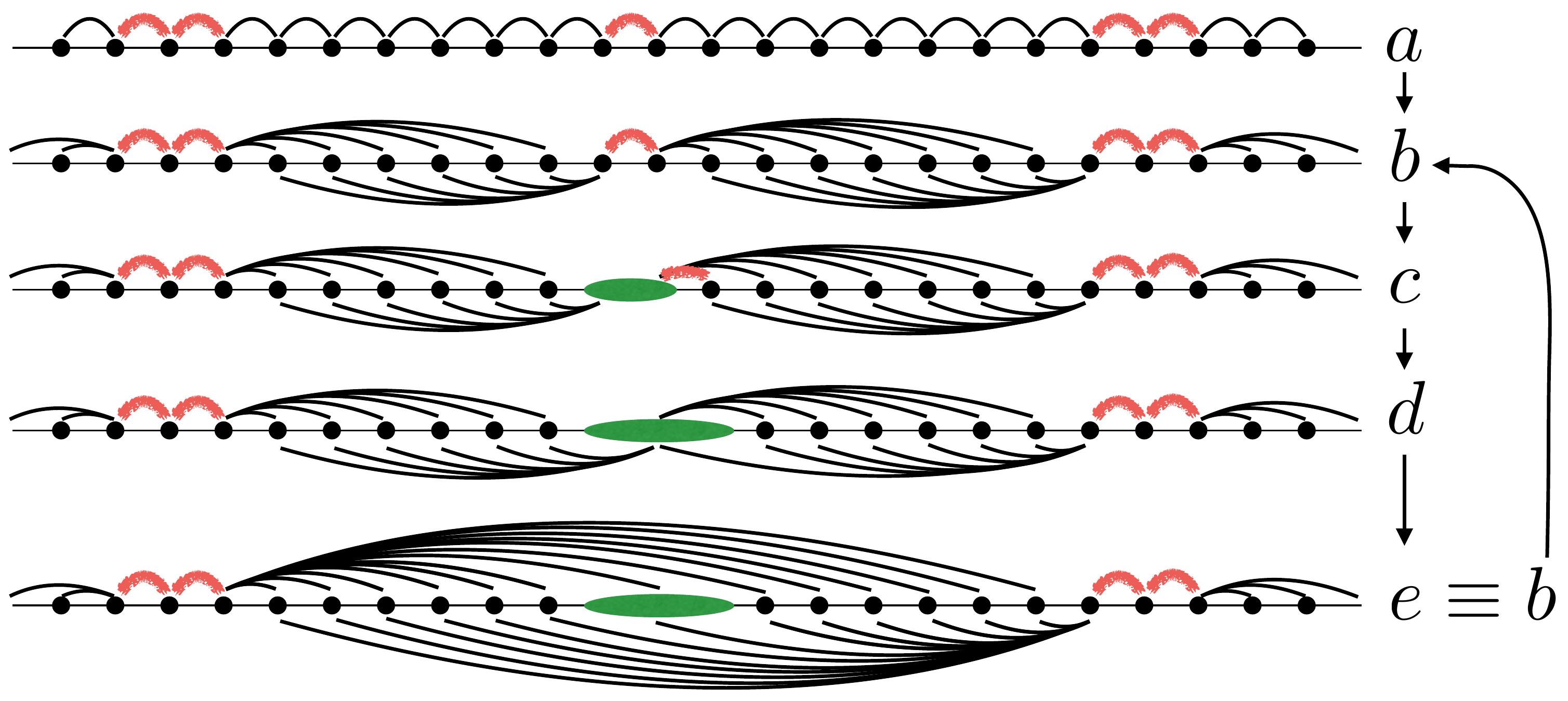}} 
\caption{The diagonalization procedure starts with the identification of the perturbative (black bounds) and resonant (red bounds) couplings between successive spins (black dots) ($a$). At step $a\to b$ all perturbative couplings are eliminated through perturbative rotations. In ($b$), the only coupling remaining are either resonants, or linking a resonant region to the localized material. At step $b \to c$ a (the smallest) resonant region is diagonalized (fused) using a rotation far from the identity. In $(c)$ a coupling that was previously perturbative became resonant and we repeat the fusion operation ($c \to d$). At $d$ only perturbative couplings connect the block (bare resonant region + collar) to the surrounding material. At step $d \to e$ these are perturbatively eliminated. In ($e$) we are back at ($b$), with a few spins (three here) now regrouped in a block. Figure taken from \cite{ThieryHuveneersMullerDeRoeck2017}.}
\label{Fig:Scheme}
\end{figure}

\paragraph{{\bf Recursion}}
If no perturbative coupling surrounding the resonant spot became resonant after step II, we can eliminate new perturbative couplings (step I). Otherwise we go back to step II and fuse the smallest resonant spot in the system. The complete procedure is summarized in Fig.~\ref{Fig:Scheme}

\medskip

The rule of step I is a direct consequence of the examples of Sec.~\ref{subsubsec:allpertfirst}-\ref{subsubsec:notallpertfirst}. The rule of step II comes from the fact that the smallest resonant spots are those that will be completely diagonalized by the smallest number of non-perturbative rotations. That also means that first treating the small resonant spots will eventually allow us to use perturbation theory again. This is an interesting property for the same reason as discussed in Sec.~\ref{subsubsec:allpertfirst}.

\smallskip

{\it Remark} As already mentioned, the sequence of operation that we implement aims at using perturbation theory to a maximal extent in order for the RMT treatment of ergodic spots to remain valid (Sec.~\ref{subsubsec:allpertfirst}). One can however build examples that show that (using the estimates of Sec.~\ref{sec:estimates}-\ref{sec:AnalysisGen}) the strategy defined here is not optimal. Obtaining the true optimal order of fusion/elimination is a difficult combinatorial problem and we take as a working assumption that the predictions of our strategy do not differ qualitatively from the predictions of the optimal procedure.

\medskip

\paragraph{Outcome of the scheme} The scheme as two possible outcomes (after a finite number of operations on a finite system). Either all the spins have been regrouped in a single resonant blocks $b$, i.e. $H'=U^\dagger H U = D_b$, and we conclude that the system is in the thermal phase, with the properties of the Hamiltonian described by ETH. Or at some step only perturbative couplings remained and they can be diagonalized. In that case the Hamiltonian is said to be in the MBL phase and we have constructed the LIOMs as $\hat{D}_i = U^{\dagger} D_i U$. The partition of the system into blocks then contains information on the locality properties of the LIOMs.

\paragraph{Summary of terminology}
In the following we will adopt the following terminology to refer to thermal spots, some of which has already been introduced. Thermal spots initially present in the system are called {\it bare spots}. Completely diagonalized resonant regions are called {\it inactive} spots. These thus consist in the union of the bare spots + the surrounding region they have thermalize through fusion operations, henceforth referred to as {\it collar} (green area at the bottom of Fig.~\eqref{Fig:Scheme}). Inactive spots are region that are thermal but cannot further thermalize their environment. Thermal regions not completely diagonalized are usually referred to as {\it active} spots. Finally in Sec.~\ref{sec:AnalysisGen} we will also discuss the notion of {\it effective bare spots}.

\section{Estimates} \label{sec:estimates}


Implementing the scheme defined in Sec.~\ref{sec:thescheme} requires the knowledge of the norm $||V_I||$ and of the typical matrix elements $m_I$ of couplings $V_I$, in particular of newly-formed couplings. Indeed, as discussed in Sec.~\ref{subsec:perturbativeVsnonpert}, a coupling is declared resonant or perturbative as a function of its typical matrix elements. On the other hand implementing a non-perturbative rotation on a stretch (see Sec.~\ref{subsec:fusionofresonantregions}) generally changes the typical matrix elements of couplings attached to the stretch, but without affecting their norm. We thus need a relation between the norm of perturbative couplings and their typical matrix elements that remains valid at any point in the scheme. If one could assume that the operators $V_I$
 where random matrices then the simple relation $||V_I||\sim\sqrt{d_I}m_I$ would hold.
As it will become clear however the operators that are created during the procedure develop some structure due to the appearance of small energy denominators. In this section we thus discuss the structure of operators that are created during the RG and obtain estimates for their typical matrix elements (Sec.~\ref{subsec:typicalmat}) and their norm (Sec.~\ref{subsec:normvstyp}).


\subsection{Typical matrix elements of new couplings} \label{subsec:typicalmat}

\subsubsection{Recursion for the typical matrix elements}

To describe our estimates, we return to the situation of Section \ref{subsec:scheme:elimination} in particular the situation as depicted in Figure \ref{Fig:VIJ}. 

Let us  evaluate the size of the `typical' matrix elements of $V_{I J}= [A_I, V_J]$.  Let us set  
$$I = a\cup b, \qquad J = b \cup c, \quad \text{so that} \, \, b = I \cap J. $$
Then the matrix elements of $V_{IJ}= [A_I,V_J]$ are evaluated as, using the definition of $A_I$ \eqref{eq:defA}
\bea \label{Eq:VIJmatrixElements}
\braket{\eta'_a \eta'_b \eta'_c |  V_{IJ}| \eta_a \eta_b \eta_c}  \sim  \sum_{\eta''_b} \frac{\braket{\eta'_a \eta'_b \eta'_c |  V_{I}| \eta_a \eta_b'' \eta'_c}  \braket{\eta_a \eta''_b \eta'_c |  V_{J}| \eta_a \eta_b \eta_c} }{E_{b}(\eta_{b}')-E_{b}(\eta_{b}'') + E_a(\eta'_a)-E_a(\eta_a)} . 
\eea
\smallskip
To go further we now discuss the distribution of the denominator
\be
\Delta(\eta_a,\eta'_a,\eta'_b , \eta''_b) = E_{b}(\eta_{b}')-E_{b}(\eta_{b}'') + E_a(\eta'_a)-E_a(\eta_a) \nonumber \ .
\ee
In order not to get lost in notations, we first consider the problem of a general energy difference
$$
\Delta_I= E_I(\tilde\eta)-E_I(\tilde\eta'),
$$
with both $\tilde\eta,\tilde\eta'$ eigenstates of $D_I$.  The typical value of $\Delta_I$ is of the order 
of the bandwidth, $\Delta \sim W_I$ that we assume to be of order $W_I \sim W \sqrt{\ell_I}$ with $\ell_I$ the length of the stetch $I$.
Assuming Poisson statistics, which turns out to be always justified for this purpose\footnote{For example in the case considered in this section, even if eigenvalues of $D_a$ and $D_b$ are drawn from some random matrix ensembles and exhibit level repulsion, that is not the case for the eigenvalues of $D_a+D_b$.}, the values of $\Delta_I$ are uniformly drawn from the interval $[0,W_I]$. In particular, this implies that typically
$$
\min_{\tilde{\eta}} | \Delta_I|  \sim  \frac{W_I}{d_I},\qquad \min _{\tilde{\eta},\tilde{\eta'}} | \Delta_I|  \sim \frac{W_I}{d^2_I},
$$
Analogously, if we minimize over a subset $a \subset I$ gives then 
 $$
\min _{\tilde{\eta}_a} | \Delta_I|  \sim \frac{W_I}{d_a} ,\qquad \min_{\tilde{\eta}_a,\tilde{\eta'}_a} | \Delta_I|  \sim \frac{W_I}{d_a^2},
$$
Now we need to understand how to perform sums like 
$
\sum_{\tilde{\eta}_a} \frac{R(\tilde\eta_a)} {\Delta_I}
$
where $R(\eta_a)$ are viewed as i.i.d.\ random variables with mean zero and variance $\sigma$, and independent from $\Delta_I$. 
The mean of this expression vanishes and higher orders are infinite, but the typical value is of order
$$ \sum_{\tilde{\eta}_a} \frac{R(\tilde\eta_a)} {\Delta_I} \sim   \frac{ K d_a \sqrt{\sigma}}{W_I}
$$
with $K$ a constant that depends on the distribution of the energy denominators.
It is important to note that, perhaps up to the factor $K$, the same result is obtained if one just keep the smallest denominator in the sum. This principle will be used repeatedly in what follows; sums of this type (as relevant in \eqref{Eq:VIJmatrixElements}) are dominated by a single term.

Let us now return to the computation at hand, see Eq.~\eqref{Eq:VIJmatrixElements}.  We assume first that the original couplings $V_I,V_J$ were GUE-like, so that all matrix elements are roughly of the same order, namely $m_I,m_J$. Then we see that the sum over $\eta''_b$ in \eqref{Eq:VIJmatrixElements} is typically dominated by one element and we obtain an estimation of the typical matrix elements of $V_{IJ}$ as
\bea \label{Eq:mIJ1}
m_{IJ} = K \frac{d_{I \cap J} m_I m_J}{W_I} ,
\eea
 In the following we will neglect the variation of the bandwith term, that is we take in \eqref{Eq:mIJ1} $W_I \to W$. Replacing it by the more realistic estimate $W_I = W \sqrt{\ell_I}$ would only make the calculations more tedious, change some non-universal constants (the localization length) without changing the conclusions of our scheme. The formula that we will use in the following thus takes the simple form
\bea \label{Eq:mIJ}
m_{IJ} = K d_{I \cap J} \frac{m_I m_J}{W} ,
\eea
Recall however that we assume that $V_I,V_J$ were GUE like. Below we investigate this assumption further and we will find that the above calculation remains consistent under minimal assumptions. 

But first we describe a relevant application of the rule \eqref{Eq:mIJ}: If $V_I$ {\it and} $V_J$ are perturbative, then the created coupling $V_{IJ}$ is perturbative as well. Indeed, from the previous considerations one can evaluate the dimensionless coupling ${\cal G}_I$ as 
\bea \label{Eq:calGform}
{\cal G}_I   = K d_I \frac{m_I}{W} \, .
\eea
And 
$$
\cag_{IJ}=  K d_{IJ} \frac{m_{IJ}}{W} = \frac{d_{I} d_J }{d_{I \cap J}} \, \times K^2 \, {d_{I \cap J}} m_{I}m_J
$$
which yields 
\beq \label{eq: new coupling is perturbative}
\cag_{IJ}=  \cag_I \cag_J  
\eeq
so not only is the generated coupling perturbative, it is also \emph{more} perturbative than the original couplings. This is obviously a simplification of our scheme which renders fully perturbative region completely harmless to localization, and delocalization will only come from the resonant region initially present in the system.

\subsubsection{Atypical matrix elements of new couplings: a first example} \label{subsubsec:atypV}

As a first example of the kind of structure (the general case is considered in Sec.~\ref{subsec:normvstyp}) that is created through perturbative rotations we consider the precedent case, still assuming that $V_I$ and $V_J$ are GUE random matrices. Even with that assumption, it should be clear from  the expression \eqref{Eq:VIJmatrixElements} and the above discussion that $V_{IJ}$ is not a random matrix: it acquires some structure coming from atypical values of the energy denominators in \eqref{Eq:VIJmatrixElements}.  More precisely we get the scaling laws
\bea \label{Eq:AtypicalElementsVIJ}
&& |\braket{\eta'_a \eta'_b \eta'_c | V_{IJ} | \eta_a \eta_b \eta_c}| \sim m_{IJ} \nn
&& {\rm sup}_{\eta'_b} |\braket{\eta'_a \eta'_b \eta'_c | V_{IJ} | \eta_a \eta_b \eta_c}|  \sim d_b m_{IJ} \nn 
&& {\rm sup}_{\eta'_b , \eta_b } |\braket{\eta'_a \eta'_b \eta'_c | V_{IJ} | \eta_a \eta_b \eta_c}|  \sim d_b m_{IJ} \nn 
&& {\rm sup}_{\eta'_a} |\braket{\eta'_a \eta'_b \eta'_c | V_{IJ} | \eta_a \eta_b \eta_c}|  \sim d_a m_{IJ} \nn 
&&  {\rm sup}_{\eta'_a, \eta_a} |\braket{\eta'_a \eta'_b \eta'_c | V_{IJ} | \eta_a \eta_b \eta_c}|  \sim d_a^2 m_{IJ}
\eea
all these being understood as equalities (up to sub-dominant factors in $d_b,d_a$) holding for typical choice of the states that are not optimized over. Note also that optimizing over the states $\eta_c,\eta'_c$ does not bring any additional terms since the energy denominator in \eqref{Eq:VIJmatrixElements} is independent of them, and one can also combine the above properties as e.g. $ {\rm sup}_{\eta , \eta'} |\braket{\eta'| V_{IJ} | \eta}|  \sim d_a^2 d_b m_{IJ} $.

\subsubsection{Consistency of the recursion}

The operators created by perturbative rotations during the procedure thus acquire some structure and a question is whether the estimate of the typical matrix elements $m_{IJ}$ in \eqref{Eq:mIJ} remains valid when $V_I$ and $V_J$ are also operators that were created during the RG.
\smallskip

This can be answered recursively by making the following assumption for  $V_I$ (and similarly for $V_J$).  
Choose a partition $I = a_1 \cup a_2 \cup \ldots \cup a_r$ and a sequence
$(\tilde\eta_{b_j})_{j=1\ldots k}$ where $\tilde \eta_{b_j}$ stands for $\eta_{b_j}$ or $\eta'_{b_j}$ and $b_j$ is one of the partitioning sets $a_i$, and no $\eta_{a_i}$ or $\eta'_{a_i}$ appears more than once. The assumption is that  (for all partitions and all sequences)
\bea  \label{eq: sup sequence}      {\rm sup}_{ \tilde \eta_{b_k} }   \ldots       {\rm sup}_{ \tilde \eta_{b_2} }  {\rm sup}_{ \tilde \eta_{b_1} } | \braket{ \eta' | V_I |\eta} |   \leq    d_{b_k}  \ldots  d_{b_2}  d_{b_1}  m_I,
\eea
where $m_I$ is again the typical matrix element of $V_I$. Given that it is clear that the estimate \eqref{Eq:mIJ} still holds (i.e. the scaling is correct). This is because optimizing over $\eta''_b$ in \eqref{Eq:VIJmatrixElements} so as to use the atypical matrix elements of $V_I$ does not give a larger contribution to the typical matrix elements of $V_{IJ}$ than the one coming from the `new' denominator. For the same reason it is clear that the assumption \eqref{eq: sup sequence} is preserved and that our computations remain consistent at every order.

\smallskip

We will thus proceed by computing the typical matrix elements of the operators created by perturbative rotations using \eqref{Eq:mIJ}. The precise structure that is present in these operators will be examined below.

\subsection{Relation between norm and typical matrix elements} \label{subsec:normvstyp}

While the rule to apply perturbative rotations were easily formulated at the level of typical matrix elements of couplings (see Eq.~\eqref{Eq:mIJ}), the implementation of non-perturbative rotations requires the knowledge of the norm of the coupling operators. Due to the atypical matrix elements (see the example of Sec.~\ref{subsubsec:atypV}), the relation between norm and typical matrix elements is \emph{not} the one valid for GUE matrices (namely $||V_I|| \sim \sqrt{d_I} m_I$), and, moreover, it depends on the whole history of a created coupling.
 We note that for any operator having atypical matrix elements the norm can be computed as 
$$
\norm V_I \norm :=  \sup_{\psi,\psi'} \frac{1}{|| \psi ||  || \psi' ||}  |  \braket{\psi' | V_I | \psi}   |   \sim   \sup_{\eta,\eta'}   |  \braket{\eta' | V_I | \eta}   |
$$
($\psi,\psi'$ are general states, whereas $\eta,\eta'$ are chosen from the localized basis).  This approximation is justified because the atypical matrix elements dominate the supremum and there is hence no gain in considering superpositions of $\eta,\eta'$.  For a GUE matrix, this approximation is not justified, and then the relation $||V_I|| =\sqrt{d_I} m_I$ holds. 

\medskip

Within our scheme we find that the following estimate holds. Given a perturbative coupling $V_{\ell}$ acting on $\ell$ spins with typical matrix elements $m_{\ell}$ we have
\bea \label{Eq:relation-norm-typicalmatrixelements}
|| V_{\ell} || = {d}^{\ell_{\bullet}/2} {d}^{\ell_{\oneop}} {d}^{b \ell_{\twoop}} m_{\ell} \, ,
\eea
where\footnote{In the following we often keep $b$ arbitrary to easily follow the influence of atypical matrix elements in our estimates.} $b=2$. Here we have partitioned the $\ell$ spins on which $V_\ell$ acts according to three `types': $\ell =\ell_\bullet+\ell_\oneop+\ell_\twoop$. The type of a spin $i$ determine the dimensional factor that is gained (compared to $m_\ell$) when optimizing $\braket{\eta'_i | V_\ell | \eta_i}$ (the other spins on which $V_\ell$ act being taken randomly, i.e. being typical). The three possible types are:
\begin{itemize}
\item[$1)$]
Spins of type $\bullet$ are spins on which $V_{\ell}$ acts as a random matrix, i.e. is structureless, and for which the optimization over $\eta_i, \eta'_i$ does not bring any factor: $\sup_{\eta_i , \eta'_i} \braket{\eta'_i | V_\ell | \eta_i} \sim m_\ell$ .
\item[$2)$]
Spins of type $\oneop$ are spins for which optimization over $\eta_i, \eta'_i$ yields a single factor $d$: $\sup_{\eta_i , \eta'_i} \braket{\eta'_i | V_\ell | \eta_i} \sim  d m_\ell$.
\item[$3)$]
Spins of type $\twoop$ are spins for which optimization over $\eta_i, \eta'_i$ yields two factors $d$: : $\sup_{\eta_i , \eta'_i} \braket{\eta'_i | V_\ell | \eta_i} \sim  d^2 m_\ell$.
\end{itemize}
If one admits (see below) that only those three `types' of spins are possible, it is then immediate to show that the relation \eqref{Eq:relation-norm-typicalmatrixelements} holds. 
Determining the type of each spin requires in general to know the full sequence of perturbative and non-perturbative rotations that led to the coupling $V_\ell$. Given that sequence, following the evolution of the type of each spin is however simple as we now explain.

\begin{table}[!htb] 
\vspace{0.5cm}
\hspace{1cm}
 \begin{minipage}{.3\linewidth}
\centering
\label{my-label}
\begin{tabular}{|c|c|c|}
\hline
$ V_I$  & $V_J$ &  $V_{IJ}$  \\ \hline
 \hspace{0.3cm}  $\zeroop$ \hspace{0.4cm}  & \hspace{0.3cm}  $\zeroop$  \hspace{0.3cm}  & \hspace{0.3cm}  $\oneop$  \hspace{0.3cm}   \\ \hline
$\zeroop$  & $\oneop$ &  $\twoop$  \\ \hline
 $\zeroop$ & $\twoop$  &  $\twoop$  \\ \hline
$\oneop$ &  $\zeroop$ & $\oneop$   \\ \hline
$\oneop$  &  $\oneop$ &  $\twoop$  \\ \hline
$\oneop$  & $\twoop$  & $\twoop$  \\ \hline
$\twoop$  & $\zeroop$ &  $\twoop$ \\ \hline
$\twoop$  & $\oneop$  & $\twoop$  \\ \hline
$\twoop$ & $\twoop$  & $\twoop$  \\ \hline
\end{tabular}
\caption*{sites in $I \cap J$}
 \end{minipage}
  \begin{minipage}{.3\linewidth}
\centering
\begin{tabular}{|c|c|}
\hline
$ V_I$  &   $V_{IJ}$  \\ \hline
 \hspace{0.3cm}  $\zeroop$ \hspace{0.3cm}  & \hspace{0.3cm}  $\twoop$  \hspace{0.3cm}   \\ \hline
$\oneop$  &  $\twoop$  \\ \hline
 $\twoop$ &  $\twoop$  \\ \hline
\end{tabular}
\caption*{sites in $I \setminus J$}
 \end{minipage}
 \begin{minipage}{.2\linewidth}
 \centering
\begin{tabular}{|c|c|}
\hline
$ V_J$  &   $V_{IJ}$  \\ \hline
 \hspace{0.3cm}  $\zeroop$ \hspace{0.3cm}  & \hspace{0.3cm}  $\zeroop$  \hspace{0.3cm}   \\ \hline
$\oneop$  &  $\oneop$  \\ \hline
 $\twoop$ &  $\twoop$  \\ \hline
\end{tabular}
\caption*{sites in $J \setminus I$}
 \end{minipage}
 \caption{Evolution of the `type' of spins due to perturbative rotations. The coupling $V_I$ is eliminated. As it overlaps with $V_J$, a new coupling $V_{IJ}$ is created. We determine for each site in $IJ=I \cup J$, its type, see \eqref{Eq:relation-norm-typicalmatrixelements}. }\label{table rules}
\end{table}

\begin{figure}
\centerline{\includegraphics[width=7cm]{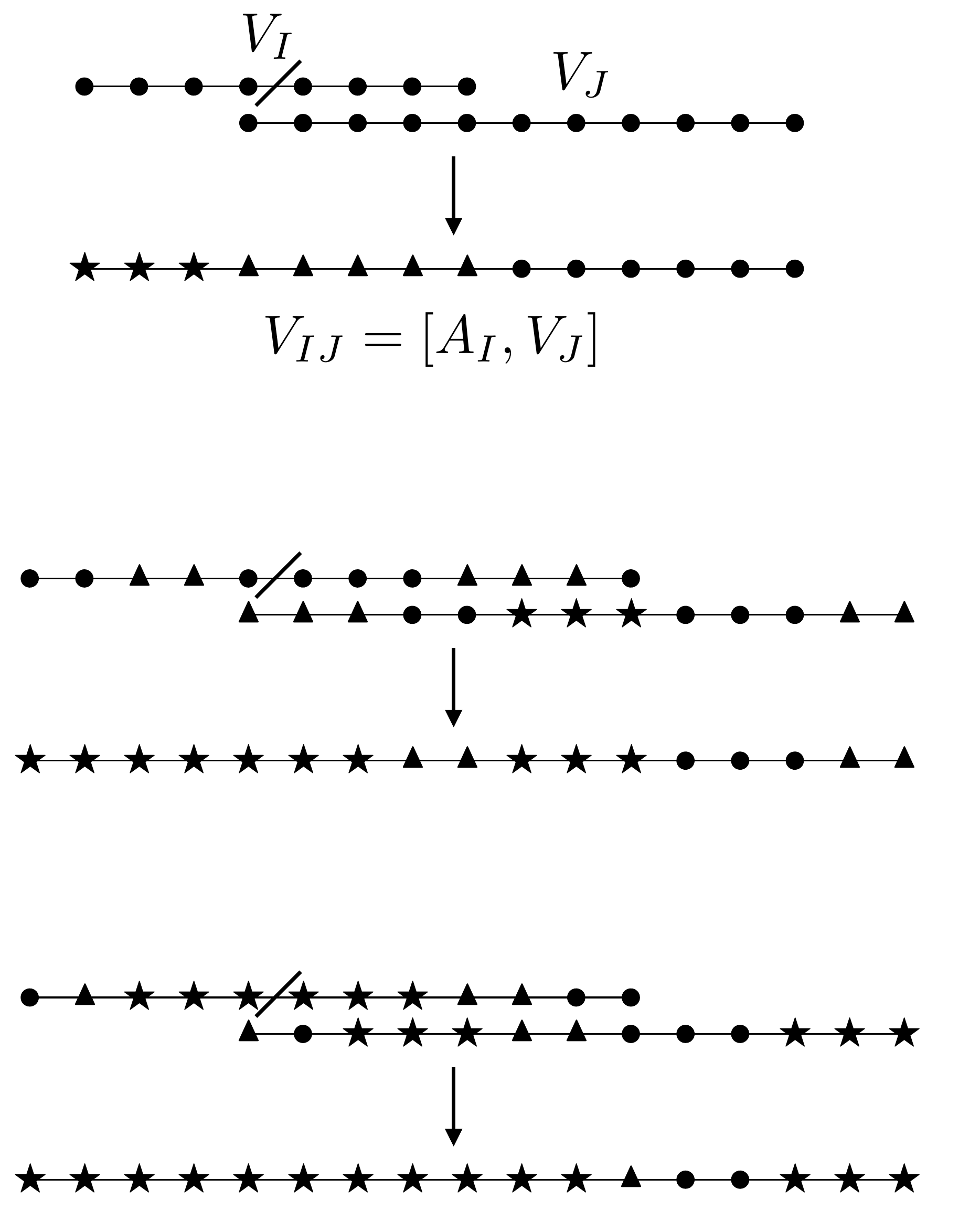}} 
\caption{Illustration of the rules from table 1. The slashed coupling (the one on the left) is eliminated.}
\label{Fig:Ex1}
\end{figure}

\paragraph{For perturbative rotations:}
 We give in Tab.~\ref{table rules} the law that gives the type of spins for $V_{IJ} = [A_I,V_J]$ given the type of spins originally present in $V_I$ and $V_J$. These rules are also illustrated with some examples in Fig.~\ref{Fig:Ex1}. We already saw in Sec.~\ref{subsubsec:atypV} a first example of these rules: there $V_I$ and $V_J$ were random matrices (i.e. acting only on spins of the $\bullet$ type), and we saw that spins in $a$ were of the $\twoop$ types, spins in $b$ were of the $\oneop$ type, and spins in $c$ remained of the $\bullet$ type (see \eqref{Eq:AtypicalElementsVIJ}). The top picture of Fig.~\ref{Fig:Ex1} symbolizes that case. This case thus shows that these three types of spins are already present after the first perturbative rotations. Showing that only those three types are possible and that the rules of Tab.~\ref{table rules} are valid can be done recursively by inspecting \eqref{Eq:VIJmatrixElements}. Note that although it seems intricate, the distinction between $\oneop$ spins and $\twoop$ simply avoids a double counting of energy denominators. It will play only a minor role in the following as it should be clear from Tab.~\ref{table rules} that the perturbative rotations lead after a few iterations to couplings with spins that are mostly of the $\twoop$ type, a simplification we will often use.

\paragraph{For non-perturbative rotations} The case of non-perturbative rotations is simpler. If a non-perturbative rotation $U$ acting on a stretch of sites $I_U$ acts on a coupling $V_I$, the new coupling $V'_I = U^\dagger V_I U$ has the same norm as $V'_I$, with spins of the same type as $V'_I$ for those in $I \backslash I_U$, and structureless spins of the $\bullet$ type for spins in $I \cup I_U$.

\section{Analysis of the scheme: general considerations} \label{sec:AnalysisGen}

In this section we now analyze the general scheme presented in Sec.~\ref{sec:thescheme} (for a particular choice of the distribution of initial couplings, see Sec.~\ref{subsec:simplifiedsetting}) using the estimates \eqref{Eq:mIJ} and \eqref{Eq:relation-norm-typicalmatrixelements} of Sec.~\ref{sec:estimates} giving recursion equations for the typical matrix elements and norm of couplings created during the RG. The goal is to obtain an algorithm that can be efficiently simulated (see results in Sec.~\ref{sec:numerics}) or analyzed using a mean-field approximation (see Sec.~\ref{sec:MF} and \cite{ThieryHuveneersMullerDeRoeck2017}). Sec.~\ref{subsec:simplifiedsetting} describe the initial distribution of couplings we analyze. In Sec.~\ref{subsec:bareLL} we analyze the outcome of the first perturbative rotations which provides an estimate for a `bare localization length' (localization length of a system in the absence of any resonant spots). In Sec.~\ref{subsec:fusionofresonantregions} we analyze in details the fusion of resonant regions through non perturbative rotations. In Sec.~\ref{subsec:haltingrule} we analyze the subsequent perturbative rotations that lead to a `renormalized' localization length (now taking into account the effect of resonances). In Sec.~\ref{sec:effective bare spots} we discuss the notion of effective bare spots. Finally we summarize in Sec.~\ref{subsec:MainAlgo} our RG algorithm.

\subsection{Setting}\label{subsec:simplifiedsetting}

Introducing $\epsilon \in [0,1]$ the probability that an initial two body coupling is resonant, we think of starting with a chain where two-body couplings are either resonant (with proba $\epsilon$) or perturbative, in which case their typical matrix elements are taken uniformly as $g >0$ (with dimensionless coupling ${\cal G} = K {d}^2 \frac{g}{W} <1$). The width $W$ of the distribution of energies is taken as $1$ by a choice of unit. Our scheme thus has two parameters $\epsilon$ and $g$, and the initial distribution of couplings is bimodal. We suppose for simplicity the relation
\bea
\epsilon = C \times g \ , 
\eea
that mimics what would be obtained using a more realistic distribution of couplings where modifying $g$ should also modify $\epsilon$. Here $C>0$ is a constant and we take $\epsilon$ as our control parameter. All the randomness is thus taken into account in the initial distribution of resonant couplings. In general other sources of randomness (in particular in the size of perturbative couplings) should be irrelevant at the transition (the effective disorder in the distribution of couplings that is generated by the scheme is much more pronounced) and we completely neglect it here.

\subsection{First step of the scheme: bare localization lengths} \label{subsec:bareLL}

\begin{figure}
\centerline{\includegraphics[width=9cm]{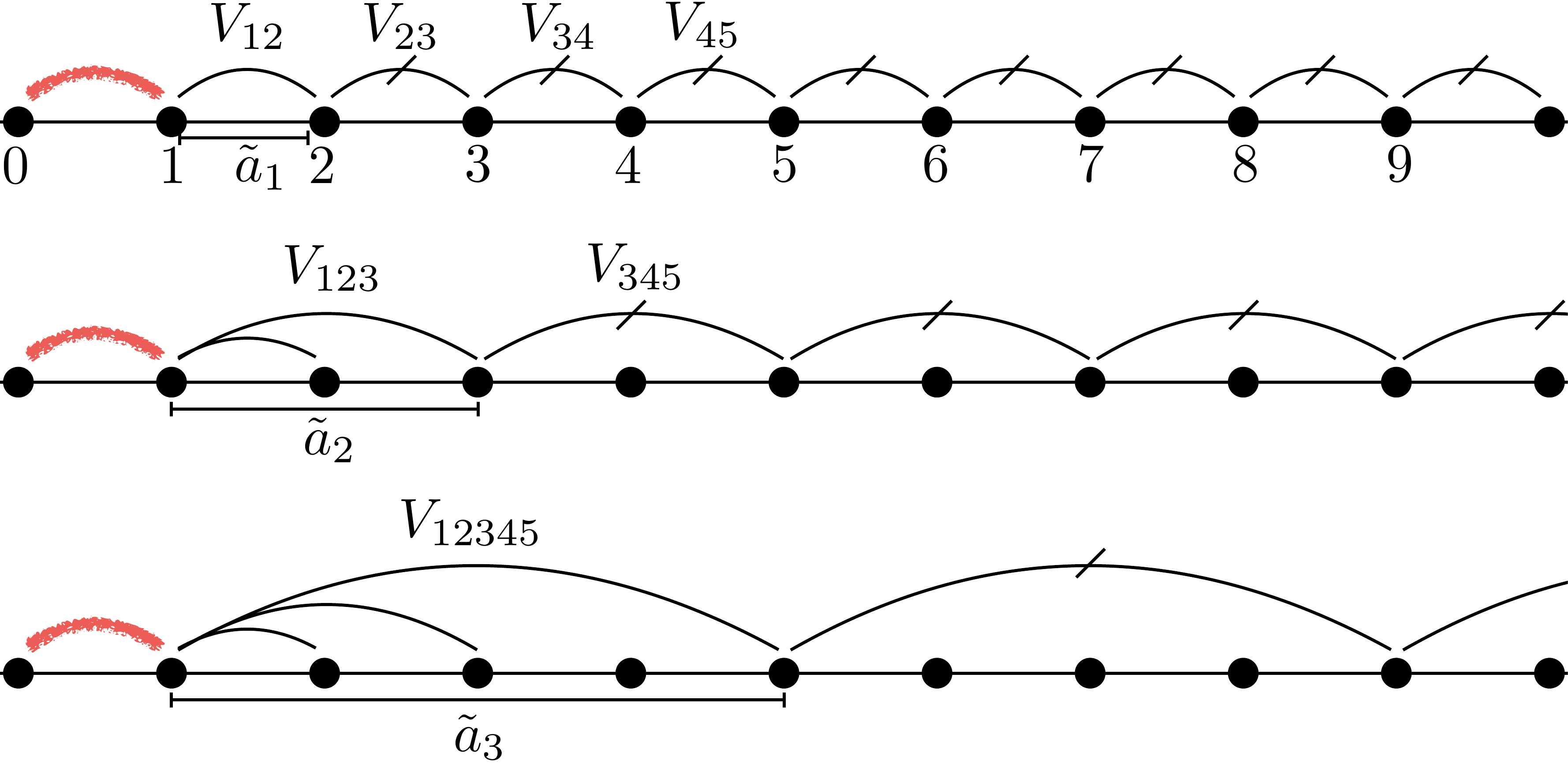}} 
\caption{MNOT construction. The coupling $V_{123}$ of size $m_2$ originates from the action of the perturbative rotation of $V_{23}$ on $V_{12}$. A coupling $V_{345}$ is similarly created by rotating $V_{34}$ (and/or $V_{45}$). Rotating $V_{345}$ generates (by action on $V_{123}$) the coupling $V_{12345}$ of size $m_3$. The construction is iterated and creates the couplings between the resonance and spins at distance $\tilde{a}_n$ that are the smallest in the perturbative expansion in $g/W$. Here only the couplings important to the construction were represented.}
\label{Fig:MNOT}
\end{figure}

In the first step of the scheme, all perturbative couplings are eliminated. That generates perturbative couplings $V_{1+\ell}$ linking spins at the boundary of (but included in) resonant regions to the $\ell$ successive spins in the perturbative regions (on both sides of the resonance). These couplings thus act on $1+\ell$ spins. When successively rotating all perturbative couplings, many couplings acting on the same region are obtained. Because of the simple rule $\eqref{eq: new coupling is perturbative}$, it should be clear that the dominant coupling acting on a given region is the one that is obtained using the minimum number of perturbative rotations (this gives the dominant order in $g$ and can be thought of as an analogue of the forward approximation \cite{ros2015integrals}). Obtaining this dominant sequence of perturbative rotations is immediate for spins at a distance $\ell= \tilde{a}_n$ given by (see Fig.~\ref{Fig:MNOT})
\bea
\tilde{a}_n = 2^{n-1} \, , \nonumber
\eea
which acts on $a_n = 2^{n-1}+1$ successive spins. Writing $m_n$ for the size of their typical matrix elements, we have at the leading order in $g$ that, using \eqref{Eq:mIJ},
\bea
m_{n+1} = K {d} m_n^2 \, , \nonumber
\eea
since the leading order in $m_{n+1}$ comes from the action on a coupling of size $m_n$ of the perturbative rotation of another coupling of size $m_n$ (see Fig.~\ref{Fig:MNOT}). These two couplings are only overlapping on a single spin and we refer to this construction as the MNOT (`maximally non-overlapping terms') construction. Using by definition $m_1=g$ we get
\bea
m_n && = g^{2^{n-1}}  (K {d})^{-1+2^{n-1}} \ . \nonumber
\eea
We thus obtain the decay rate of these couplings as
\bea
\hat{\alpha}_1 := \lim_{n \to \infty} - \frac{\log(m_n)}{a_n} = - \log (K g {d} ) \, . \nonumber
\eea
 For simplicity we now suppose that the decay of the typical matrix elements of perturbative couplings obtained in the end of the first step is homogeneous and given by $\hat{\alpha}_1$. We thus suppose that couplings $V_{1+\ell}$ acting on $1$ spin at the bounday of a resonant region and $\ell$ successive spins outside the region have typical matrix elements\footnote{A factor $g$ is missing in front of the exponential in \eqref{Eq:forfootnote1}. We drop it here for simplicity as our scheme is only accurate to logarithmic accuracy, and adding any sub-exponential term to \eqref{Eq:forfootnote1} would not change the predictions.}
\bea \label{Eq:forfootnote1}
m_{1 + \ell } = e^{-\ell \hat{\alpha}_1} \ ,
\eea
and $\hat{\zeta}_{1} := 1/\hat{\alpha}_1$ is a bare localization length for typical matrix elements. The norm of these couplings also decay exponentially and
\bea \label{Eq:BareLL}
&& ||V_{1+\ell}|| \sim d^{b \ell}m_{1 + \ell }  = e^{-\alpha_1 \ell}  \nn
&& \alpha_1 = \hat{\alpha}_1 - b \alpha_c \, ,
\eea
with
\bea
\alpha_c = \log(d) \, ,
\eea
a constant that will play a crucial role in the following. The above equality is only valid for large $\ell$ with logarithmic accuracy but we promote it to the full space for simplicity. This follows from \eqref{Eq:relation-norm-typicalmatrixelements}, noting that the fraction of spins with at least one energy denominator that was not optimized over approaches one in the large $\ell$ limit. We also define $\zeta_{1} := 1/\alpha_1$ the 
bare localization length for the norm of couplings.

{\it Remark} A first bound on the bare localization length can be obtained by asking that the created perturbative couplings are perturbative. Using \eqref{Eq:calGform} this imposes $K {d}^{1+\ell} e^{-\ell/\hat{\zeta}_1} <1$ $\forall \ell$, which is consistent only if $\hat{\zeta}_1^{-1} > \log({d})$. An equivalent bound was already obtained in \cite{vosk2015theory}. By taking into account the effect of resonant regions, we will in the following obtain a stronger bound (already discussed in \cite{de2017stability,ThieryHuveneersMullerDeRoeck2017})

\subsection{Fusion of resonant regions}  \label{subsec:fusionofresonantregions}

\subsubsection{First fusion}

\begin{figure}
\centerline{\includegraphics[width=9cm]{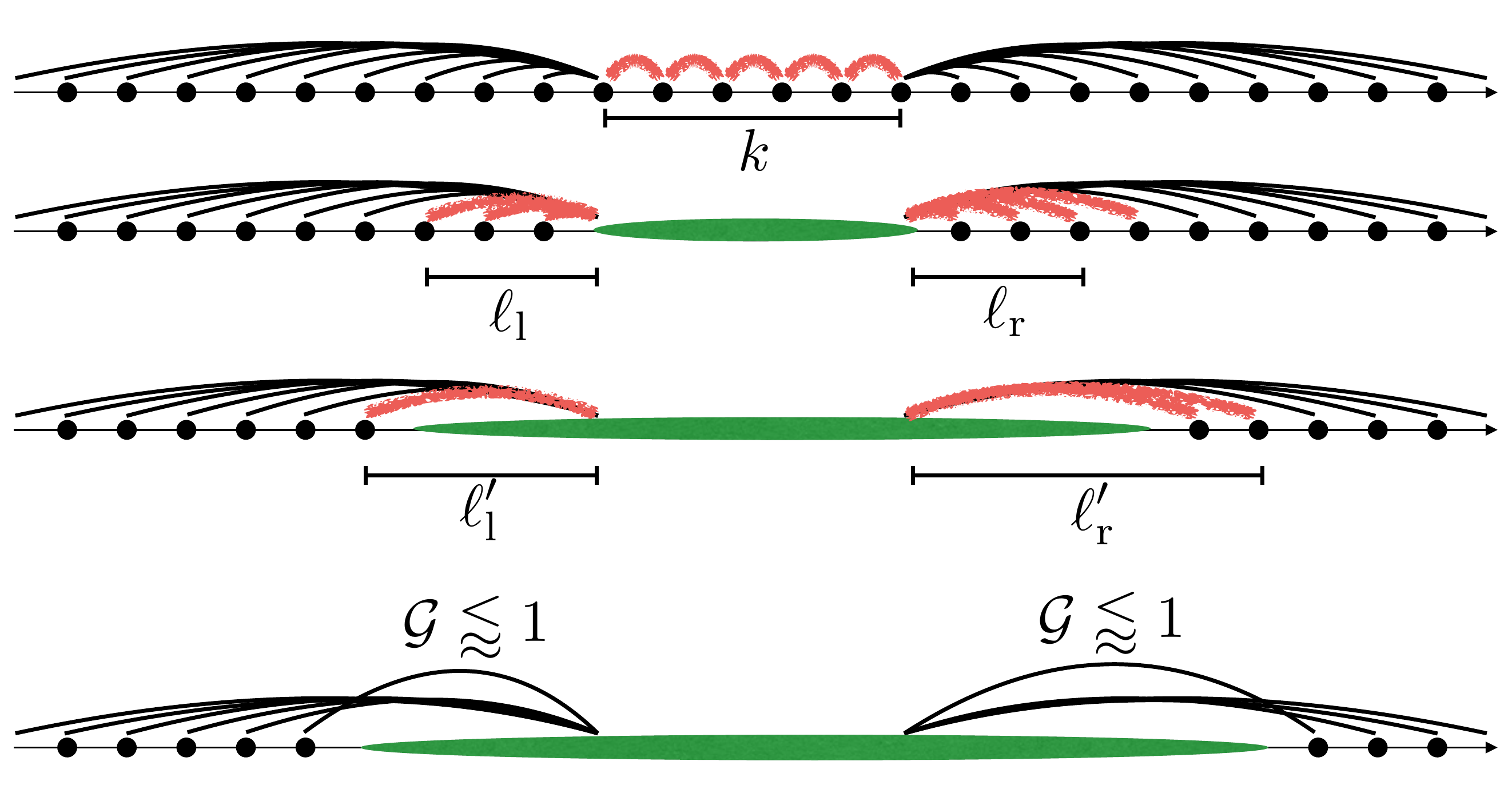}} 
\caption{Successive non-perturbative diagonalization (`fusion') of a growing resonant spot.}
\label{Fig:Fusion}
\end{figure}

We now analyze the process of fusing a resonant region of size $k$ (containing $k$ spins). We suppose that couplings to the right and to the left of the spot have an operator norm $||V_{1+\ell}^{\rr/\rl}|| = e^{-\sum_{i=1}^{\ell} \alpha_i^{\rr/\rl}}$. Here we have written the operator norm introducing individual decay rates $\alpha_i^{\rr/\rl}$ for each spin on each side of the spot. At the beginning of the procedure these are uniformly equal to the bare decay rate, $\alpha_i^{\rr/\rl}= \alpha_1$, but are at later stages renormalized (see Sec.~\ref{subsec:haltingrule}). Upon fusing the $k$ spins inside the resonant spot, the perturbative couplings are transformed as $V_{1+\ell}^{\rr/\rl} \to \tilde{V}_{1+\ell}^{\rr/\rl}$ as they now act on $k + \ell$ spins (see Fig.~\ref{Fig:Fusion}). Since these now act as a random matrix on the $k$ spins absorbed in the resonance we can obtain their typical matrix elements $\tilde{m}_{1+\ell}^{\rr / \rl}$ as (using that their norm is conserved by the rotation)
\bea \nonumber
e^{-\sum_{i=1}^{\ell} \alpha_i^{\rr/\rl}} = ||V_{1+\ell}^{\rr/\rl} || = || \tilde{V}_{1+\ell}^{\rr/\rl}||  = d^{\frac{1}{2} k} d^{b \ell} \tilde{m}_{1+\ell}^{\rr, \rl}  \, .
\eea 
Here we used the rule \eqref{Eq:relation-norm-typicalmatrixelements} assuming that all spins on which $V_{1+\ell}^{\rr/\rl}$ act are of the $\twoop$ type, which should be true for large $\ell$ (see Tab.~\ref{table rules}). 
We obtain
\bea \nonumber
\tilde{m}_{1+\ell}^{\rr/\rl}  = \frac{1}{ d^{\frac{1}{2} k} d^{b \ell}} e^{-\sum_{i=1}^{\ell} \alpha_i^{\rr/\rl}} \ .
\eea
The dimensionless coupling constant are thus obtained as, using \eqref{Eq:calGform}
\bea \nonumber
\tilde{\cag}_{1+\ell}^{\rr/\rl}  =  K \frac{d^{k + \ell}}{ d^{\frac{1}{2} k} d^{b \ell}} e^{-\sum_{i=1}^{\ell} \alpha_i^{\rr/\rl}} \ .
\eea
Neglecting terms that are sub-exponential in $\ell,k$ the spins that become resonant to the right of the couplings are those for which $\tilde{\cag}_{1+\ell}^{\rr} >1$, i.e.,
\bea
\left(\frac{k}{2}-(b-1)\ell \right)\alpha_c - \sum_{i=1}^{\ell} \alpha_i^{\rr} >0 \, .
\eea
In the following we will see that $\alpha_i^{\rr} > (1-b) \log(d)$ is always ensured. The above expression is thus decaying with $\ell$ and the distance upon which couplings become resonant is given by
\be
\ell_{\rr/\rl}= {\rm min}\left\{  \ell \in \mathbb{N} | \left[ \frac{k}{2}- (b-1)\ell \right]\alpha_c - \sum_{i=1}^{\ell} \alpha_i^{\rr/\rl}   \ \leq 0 \right\}- 1 \, .
\ee

\subsubsection{Subsequent fusions} \label{subsubsec:subsequentfusions}

If ${\rm min}(\ell_{\rr} , \ell_{\rl})\geq 1$, at least one new coupling has become resonant and the spot remains active. This means that the resonant region has grown and the fusion operation must be repeated (though maybe after having fused other smaller resonances in the chain first). Upon fusing the new couplings, the perturbative couplings are once again transformed and we now have $\tilde{V}_{1+\ell}^{\rr/\rl} \to \hat{V}_{1+\ell}^{\rr/\rl}$ with new typical matrix elements $\hat{m}_{1+\ell}^{\rr/\rl}$ given by, using the conservation of the norm (see Fig.~\ref{Fig:Fusion}):
\bea \nonumber
e^{-\sum_{i=1}^{\ell} \alpha_i^{\rr/\rl}} = d^{\frac{1}{2} (k + \ell_{\rr} + \ell_{\rl})} d^{b (\ell-\ell_{\rr/\rl})} \hat{m}_{1+\ell}^{\rr/\rl} .
\eea
Again here we used the rule \eqref{Eq:relation-norm-typicalmatrixelements} assuming that all spins on which ${V}_{1+\ell}^{\rr/\rl}$ originally acted were of the $\twoop$ type, and therefore spins on which $ \hat{V}_{1+\ell}^{\rr/\rl}$ are either of the $\bullet$ type (for spins that were non-perturbatively rotated) or of the $\twoop$ type.
The dimensionless couplings are thus given by, using \eqref{Eq:calGform},
\bea \nonumber
\hat{\cag}_\ell^{\rr/\rl} = \frac{e^{-\sum_{i=1}^{\ell} \alpha_i^{\rr/\rl}} }{ d^{\frac{1}{2} (k + \ell_{\rr} + \ell_{\rl})} d^{b (\ell-\ell_{\rr/\rl})} } d^{k + \ell + \ell_{\rl/\rr}} 
\eea
From that we obtain that the new distances $\ell_\rr' \geq \ell_\rr$ and $\ell_\rl' \geq \ell_\rl$ on which perturbative couplings become resonant as:
\bea \label{Eq:propagiterated}
&& \ell_{\rr}' = -1+  {\rm min}\left \{ \ell \geq \ell_\rr \, | \left[ \frac{k + \ell_\rr + \ell_\rl}{2} - (b-1)(\ell  - \ell_\rr) \right] \alpha_c - \sum_{i=1}^{\ell} \alpha_i^\rr \leq 0  \right\} \nn 
&& \ell_{\rl}' = - 1+  {\rm min}\left \{ \ell \geq \ell_\rl \, | \left[ \frac{k + \ell_\rr + \ell_\rl}{2}  - (b-1)(\ell  - \ell_\rl)\right]\alpha_c - \sum_{i=1}^{\ell} \alpha_i^\rl \leq 0  \right\}   \, .
\eea
This `propagation' of the resonance must be repeated until no new resonant coupling is created: we iterate changing $\ell_{\rr/\rl} \to \ell_{\rr/\rl}'$, compute $\ell_{\rr/\rl}''$ etc. The procedure can either converge in  a finite number of spins or diverge in some cases (see below). When it converges, in the end of the procedure, the last spins absorbed to the right and to the left must both be `just' resonant, with the first non-absorbed being `just' perturbative. Denoting $\ell_{\rr}$ and $\ell_{\rl}$ the total number of spins absorbed to the left and to the right we must have $\cag_{\ell_\rr /\ell_\rl}^{\rr/\rl} \gtrapprox 1$ and $\cag_{\ell_\rr /\ell_\rl}^{\rr/\rl} \lessapprox 1$. We therefore obtain the condition, using \eqref{Eq:propagiterated} with $\ell \to \ell_{\rr/\rl}$
\bea \label{Eq:propagend}
&& \left( \frac{k + \ell_\rl + \ell_{\rr}}{2} \right)\alpha_c - \sum_{i=1}^{\ell_\rr} \alpha_i^\rr  \simeq   \left(\frac{k + \ell_\rl + \ell_{\rr}}{2} \right)\alpha_c - \sum_{i=1}^{\ell} \alpha_i^\rl \simeq 0    \, .
\eea 
Once the procedure has converged we say that the thermal spot is inactive.

\subsubsection{Case of a spot in an homoegenous environment}

 If the environment is homogenous with $\alpha_i^{\rr,\rl} = \alpha$ (e.g. in the beginning of the procedure where $\alpha=\alpha_1$), then after the $n-th$ iteration we obtain $\ell_{\rr} = \ell_{\rl} = \ell(n)$ with (discarding now the discrete aspect of the problem)
\bea \nonumber 
\left[\frac{k}{2} + \ell(n) - (b-1)(\ell(n+1)-\ell(n) ) \right] \alpha_c - \alpha \ell(n+1) = 0 \, 
\eea
with $\ell_0=0$ by definition. This is easily solved as 
\bea \nonumber 
\ell(n) = \frac{k \alpha_c}{2(\alpha - \alpha_c)} \left[  1 - \left(  \frac{b \alpha_c}{b \alpha_c + (\alpha-\alpha_c)} \right)^n \right] \, .
\eea
Hence for $\alpha > \alpha_c$ the procedure converges (in a finite number of iteration on a lattice) with
\bea \label{Eq:reselln}
\lim_{n \to \infty} \ell(n) = \frac{k \alpha_c}{2(\alpha - \alpha_c)} \, .
\eea
For $\alpha < \alpha_c$ on the other hand the length thermalized by the spot diverges exponentially with the number of iterations. Hence a finite spot can thermalize an infinite localized system if its localization length $\zeta = 1/\alpha$ exceeds the critical value $\zeta_c = 1/\alpha_c=1/\log(d)$. We call that process an avalanche as the mechanism responsible for that instability is that the spot becomes a better and better bath as it becomes larger.

\subsubsection{General case and link with a first passage time problem} \label{Sec:FirstPassageTimeInterp}

Assuming in general that $\ell_{\rl} \sim \ell_{\rr}$, we obtain that, replacing $\alpha_i^{\rr} \to \overline{\alpha} + \delta \alpha_i$ in \eqref{Eq:propagend},
\bea \label{Eq:FirstPassageTimePRoblem}
\frac{k}{2} \alpha_c  + \ell_{\rr}(\alpha_c- \overline{\alpha}) - \sum_{i=1}^{\ell_\rr} \delta \alpha_i= 0 \, .
\eea
That shows that the problem of determining $\ell_\rr$ is similar to the problem of finding the position at which a discrete time random walker on $\mathbb{R}$ starting at $k/2$ at time $0$ and performing at each time step a jump $\delta \alpha_i$ with a linear bias $\alpha_c- \overline{\alpha}$ hits $0$. If $\overline{\alpha}> \alpha_c$ this time exists with probability $1$ ($\overline{\delta \alpha_i} = 0$). The distribution of this time exhibits in general a power-law (whose precise exponents depend on the details of the distribution of the $\delta \alpha_i$) that is cutoff at a scale diverging with $\frac{1}{\overline{\alpha}-\alpha_c}$. In the other case there is a finite probability that the random walker never return to the origin, a process we refer to as an avalanche. Here we have for the moment not determined what is the distribution of the $\alpha_i$. Treating the resonances recursively in order of increasing size, the environment seen by resonant spots of size $k$ is determined by the distribution of treated spots of size $k' < k$ (see below). It will become clear later on that in the MBL phase the $\alpha_i$ can asymptotically (for large $k$) assumed to be homogeneous with $\alpha_i = \alpha  \in  ] \alpha_c , \alpha_1[$. In the thermal phase the environment seen by spots of size $k>k_*$ with $k_*$ finite is such that avalanches occur with a finite probability, thermalizing the system. Finally at the transition point the environment is such that at large scales $\overline{\alpha_i}$ approaches $\alpha_c$, implying a broad (power-law) distribution of thermal inactive spots at criticality.

\subsection{Generation of new perturbative couplings -- renormalization of the localization length}\label{subsec:haltingrule}

\begin{figure}
\centerline{\includegraphics[width=9cm]{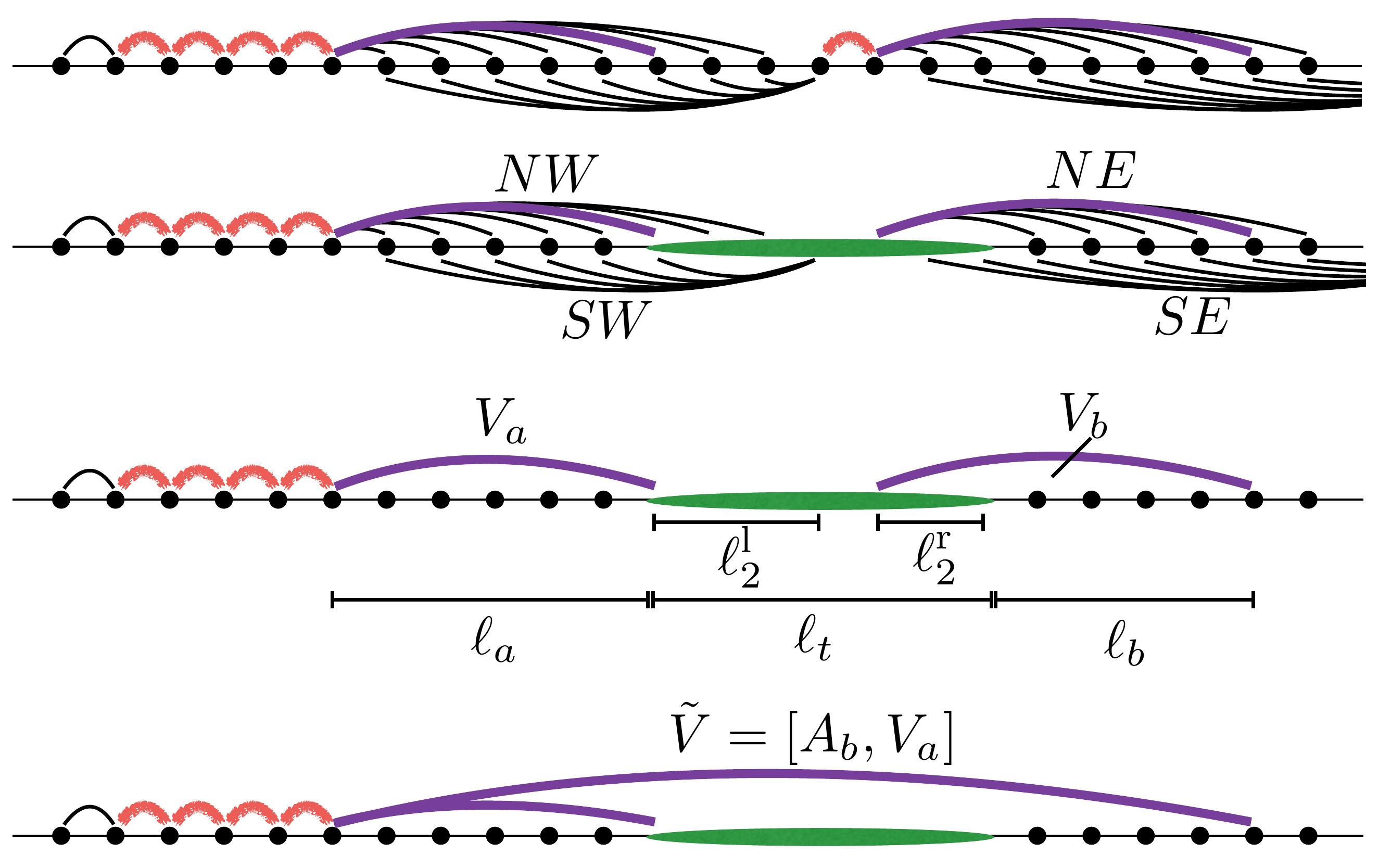}} 
\caption{Derivation of the halting rule}
\label{Fig:Halting}
\end{figure}

We now study the distribution of decay rates $\alpha_i$ precedently introduced. To that aim, we study the new perturbative couplings that are created once a resonant spot was completely digaonalized and became inactive. We suppose that preexistant perturbative couplings decay with the bare localization length $\alpha_1$ (i.e. relevant in the beginning of the procedure), but the generalization to other environments is straightforward. We study the case of having one spot $s_2$ of size $k_2$ to the right of a spot $s_1$ of size $k_1$. We suppose that the spot $s_2$ was fully treated and now covers a distance $\ell_t = k_2+\ell_2^\rr+\ell_2^\rl$, with $\ell_2^\rr$ and $\ell_2^\rl$ the number of spins thermalized to the right and left of $s_2$. Denoting $\ell_a$ the distance between the spot $s_1$ and the leftmost spins absorbed by the spot $s_2$, we are interested in the perturbative coupling that is created in between the spot $s_1$ and a spin on the right of $s_2$, at a distance $\ell_b$ of the rightmost spin absorbed by $s_2$ (see Fig.~\ref{Fig:Halting}). This spin is thus at a distance $\ell_a + \ell_t + \ell_b$ of the spot $s_1$. After having fused $s_2$, $4$ groups of couplings now interact with the treated resonant spot. The couplings that are up for elimination are in groups $NE$ and $SW$ (see Fig.~\ref{Fig:Halting}): those in groups $NW$ are in contact with the spot $s_1$ to the left, and those in the group SE are in contact with another resonance $s_3$ (not depicted in Fig.~\ref{Fig:Halting}). Eliminating the couplings in the $NE$ (resp. $SW$) group creates, upon acting on the couplings in the $NW$ (resp. $NE$) group, perturbative couplings linking the resonant spot $s_1$ (resp. $s_3$) to spins to the right (resp. left) of the treated spot. The two couplings $V_a$ and $V_b$ that are of interest in our case are higlighted in purple in Fig.~\ref{Fig:Halting}: upon eliminating $V_b$ we create a perturbative coupling $\tilde{V}  = [A_b , V_a]$ linking the spot $s_1$ to a spin at a distance $\ell_a + \ell_t + \ell_b$. Using \eqref{Eq:mIJ} we evaluate its typical matrix elements as
\bea \nonumber
\tilde{m} = K  d^{\ell_t} m_a m_b  \, .
\eea
To evaluate $m_a$ and $m_b$, we use that the norm of the couplings $V_a$ and $V_b$ were preserved during the non-perturbative rotations used to diagonalize the resonant spot $s_2$. Hence we have $|| V_a || =e^{-\alpha_1 \ell_a}$ and $||V_b|| = e^{-\alpha_1( \ell_2^{\rr} + \ell_b)}$. Furthermore, using the rule \eqref{Eq:relation-norm-typicalmatrixelements} we get
\bea \nonumber
|| V_a || = m_a d^{b \ell_a} d^{\frac{\ell_t}{2}} \quad , \quad || V_b|| = m_b d^{b \ell_b} d^{\frac{\ell_t}{2}} \, ,
\eea
which again neglects the possibilities of having a few spins of the $\oneop$ type. Hence we obtain
\bea \nonumber
\tilde{m} = K d^{\ell_t} \frac{e^{-\alpha_1 \ell_a} e^{-\alpha_1( \ell_2^\rr + \ell_b)}}{d^{b \ell_a} d^{b \ell_b} d^{\ell_t}} = K \frac{e^{-\alpha_1 \ell_a} e^{-\alpha_1 \ell_2^\rr}  e^{-\alpha_1 \ell_b}}{d^{b \ell_a} d^{b \ell_b}} \, .
\eea
The norm of the coupling $|| \tilde{V} ||$ is now obtained as, using \eqref{Eq:relation-norm-typicalmatrixelements} (note now that the spins in the treated resonant region are precisely spins of the $\oneop$ type here),
\bea \nonumber
|| \tilde{V} || = d^{b(\ell_a + \ell_b)} d^{\ell_t} \tilde{m} = K e^{-\alpha_1(\ell_a+ \ell_b)} d^{\ell_t}e^{-\alpha_1 \ell_2^\rr} \, .
\eea
That can be further simplified by noting that $e^{-\alpha_1 \ell_2^\rr}$ is the norm of the last coupling absorbed by the resonant spot $s_2$. Hence its dimensionless coupling in the end of the non-perturbative rotations was of order $1$, implying $d^{\ell_t} \frac{e^{-\alpha_1 \ell_2^\rr}}{d^{\ell_t/2}} \sim 1$. From this we obtain our final result
\bea \label{eq:haltingrule}
|| \tilde{V} || \sim e^{- \alpha_1 (\ell_a + \ell_b) - \alpha_t \ell_t} \, ,
\eea
with
\bea
\alpha_t = \frac{\alpha_c}{2} = - \frac{\log(d)}{2} \  .
\eea
Hence the decay through the treated resonant spot is replaced from $\alpha_1$ to $\alpha_t$, while the decay around remains unaffected. It should be clear that this rule remains valid throughout the RG. Note that $\alpha_t < \alpha_c$ (here we actually obtain that it is negative), showing that treated resonant spots act as `shortcut' that have the potential to delocalize the system. The decay rates $\alpha_i$ introduced in the precedent section are thus taken as $\alpha_i = \alpha_t$ for spins inside inactive spots, and $\alpha_i = \alpha_1$ otherwise.

\subsection{Effective bare spots}\label{sec:effective bare spots} 

\begin{figure}
\centerline{\includegraphics[width=9cm]{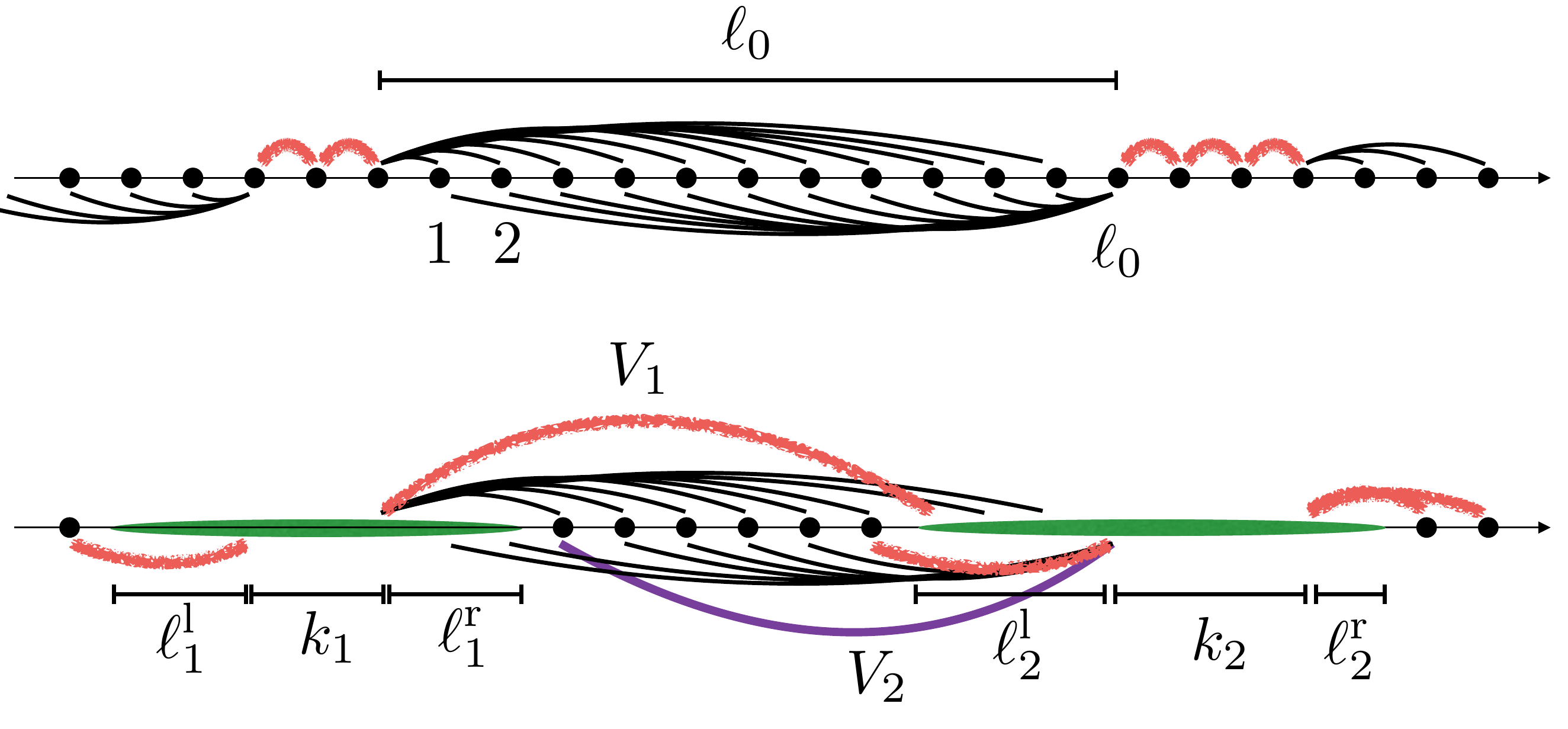}} 
\caption{Creation of renormalized untreated spots. Here the coupling $V_1$, with dimensionless coupling constant given in \eqref{eq:RUS}, becomes resonant.}
\label{Fig:RUS}
\end{figure}

In Sec.~\ref{subsec:fusionofresonantregions} we have described the mechanism upon which the fusion of a resonant region might thermalize perturbative couplings attached to it. A possibility that was discarded up to know is that a perturbative coupling linking two adjacent resonant regions become resonant when fusing one of the region (see Fig.~\ref{Fig:RUS}). While these couplings are not generated after the first set of perturbative rotations, some exist as soon as a resonance has propagated. Consider two resonance of size $k_1$ and $k_2$ that have eventually both been fused and propagated to their left and to their right on a distance $\ell_{1/2}^{\rr/\rl}$ (regions already fused, some other spins might be resonant), initially separated by a distance $\ell_0$. At that stage the two possible couplings $V_1$ and $V_2$ linking the two resonant regions that are the most likely to be resonant are those depicted in Fig.~\ref{Fig:RUS}. Following the usual rules their dimensionless coupling are evaluated as 
\bea \label{eq:RUS}
&& \cag_1=d^{\ell_1^\rl + k_1 + \ell_0 + k_2 + \ell_2^\rr}  \frac{e^{-\sum_{i=1}^{\ell_0 - \ell_2^\rl} \alpha_i}}{\sqrt{d^{k_1 +\ell_1^\rl + \ell_2^\rr} d^{k_1 +\ell_1^\rl + \ell_2^\rr} } d^{b(\ell_0 - \ell_1^\rr - \ell_2^\rl)}} \nn 
&& \cag_2=d^{\ell_1^\rl + k_1 + \ell_0 + k_2 + \ell_2^\rr}  \frac{e^{-\sum_{i=\ell_1^\rr}^{\ell_0} \alpha_i}}{\sqrt{d^{k_1 +\ell_1^\rl + \ell_2^\rr} d^{k_1 +\ell_1^\rl + \ell_2^\rr} } d^{b(\ell_0 - \ell_1^\rr - \ell_2^\rl)}}   \, ,
\eea
where we have used the labelling of the spins as in Fig.~\ref{Fig:RUS}. If one of the two becomes resonant, that creates a large resonant spot of size $k_1+k_2+\ell_0 + \ell_1^\rl + \ell_2^\rr$. Remarkably, fusing this resonant spot has an effect on the perturbative coupling among it that is {\it equivalent}\footnote{In our approach the only relevant feature of a thermal region is its level spacing. Note however that, from a dynamical point of view, an effective bare spot is different from a bare spot, as it takes some time for the effective bare spot to act as a good bath for its surrounding. In this paper we are only interested in eigenstate properties, or equivalently infinite-time static properties, for which this effect is not present.} to having started with a bare resonant spot of size $k_1+k_2+\ell_0$. In the following we will refer to thermal spots that are constructed in this way as `effective bare spots'. The boundaries of an effective bare spots are the outer boundaries of the original bare spots that created it (e.g. in the example of Fig.~\ref{Fig:RUS} the left boundary is located at spin $-2$ and the right at spin $\ell_0+3$).

\subsection{Summary -- Main algorithm} \label{subsec:MainAlgo}

We now formulate our algorithm for a chain of $L$ spins indexed by $i \in [1, \dots , L]$ with periodic boundary condition. Each initial two-body coupling is taken resonant with probability $\epsilon >0$ and we determine the position of all bare resonant spots.

\medskip

Denoting at any point in the scheme $n_{{\rm res}} \geq 0$ the number of active resonant spots, we label resonant spots (according to their position in the chain) by $k \in \{1,\dots ,n_{{\rm res}} \} $. For each resonance $k$ we keep track of: 
\begin{enumerate}
	\item
 $i_{{\rm min}}^k  $ and $i_{{\rm max}}^k $, the position of the leftmost and rightmost spins included in the resonance;
  \item
  $j_{{\rm min}}^k $ and $j_{{\rm max}}^k$, the minimum and maximum spins on which a non-perturbative rotation was already applied (with by convention $j_{{\rm min}}^k = i_{{\rm max}}^k +1$ and $j_{{\rm max}}^k = i_{{\rm min}}^k -1$ at the beginning of the scheme);
  \item $n_k^{\rm l}$ and $n_k^{\rm r}$, the norm of the perturbative coupling linking the resonance to the first non-resonant spin on its left and right (initially $n_k^{\rm r/l}=e^{-\alpha_1}$);
  \item 
  the set ${\cal T}$ of sites covered by treated (inactive) spots (initially ${\cal T}= \emptyset$). 
\end{enumerate}
  Other perturbative couplings can be determined using the rule \eqref{eq:haltingrule}. For example the norm of the perturbative coupling connecting the  $k-$th resonant spot to a spin at a distance $\ell \geq 2$ is $n_k^{\rm r}(\ell) = n_k^{\rm r} \e^{-\sum_{i=i_{{\rm max}}^k+2}^{i=i_{{\rm max}}^k+\ell} \alpha_i}$ with $\alpha_i = \alpha_t$ if $i \in {\cal T}$, and $\alpha_i = \alpha_1$ otherwise. Typical matrix elements are evaluated using their estimate \eqref{Eq:relation-norm-typicalmatrixelements} (always supposing $\ell=\ell_\twoop$ for simplicity). The dimension of the space on which the couplings acts can be deduced from the knowledge of the $i_{{\rm min}}^k, \, i_{{\rm max}}^k \, j_{{\rm min}}^k$ and $j_{{\rm max}}^k$, and from this we determine the dimensionless couplings constants ${\cal G}$ (see \eqref{Eq:calGform}).

  \medskip

This set of information is enough to run the scheme but to analyze the effective bare spots distribution and collar distribution we also keep track of: $\hat{i}_{{\rm min}}^k $ and $\hat{i}_{{\rm max}}^k $, the leftmost and righmost spins that consitute the effective bare spot the resonant region originate from (with at the beginning $\hat{i}_{{\rm min}}^k  =i_{{\rm min}}^k $ and $\hat{i}_{{\rm max}}^k  =i_{{\rm max}}^k $).

\medskip

We alway treat the resonance $k$ whose size $i_{{\rm max}}^k-i_{{\rm min}}^k$ is the smallest (if several have the same size, one is chosen at random). Diagonalizing this resonance means applying a non-perturbative rotation to the sites in between $i_{{\rm min}}^k  $ and $i_{{\rm max}}^k $. We then determine what are the new couplings that become resonant. These can be either couplings linking the resonance to the localized material to its left and to its right (following Eq.~\eqref{Eq:propagiterated}), or couplings linking the resonance $k$ to the resonance $k+1$ or $k-1$ (to its right or to its left, following Eq.~\eqref{eq:RUS}).  Several outcomes are thus possible.
\begin{enumerate}
	\item
If no coupling becomes resonant, the resonant region is declared treated. We update ${\cal T}\to{\cal T}\cup [i_{{\rm min}}^k, \, i_{{\rm max}}^k]$ and $n_{{\rm res}}\to n_{{\rm res}}-1$ and relabel the remaining resonances. We store the size $\hat{i}_{{\rm max}}^k-\hat{i}_{{\rm min}}^k$ of the effective bare spot that led to this thermal region. The size of the collar region now thermalized around the spot is evaluated as $i_{{\rm max}}^k-i_{{\rm min}}^k-(\hat{i}_{{\rm max}}^k-\hat{i}_{{\rm min}}^k)$.
\item
If some new couplings become resonant (without creating a renormalized bare spot, see next case), the resonance propagates on some distance to its left and to its right $\ell^{{\rm l / r}}$ that is determined as in Sec.~\ref{subsubsec:subsequentfusions}. We update $j_{{\rm min}}^k  \to i_{{\rm min}}^k$,  $j_{{\rm max}}^k  \to i_{{\rm max}}^k$,  $ i_{{\rm min}}^k \to i_{{\rm min}}^k - \ell^{{\rm l}}$, $ i_{{\rm max}}^k \to i_{{\rm max}}^k + \ell^{{\rm r}}$ and $n_k^{\rm l/r} \to  n_k^{\rm l/r}(\ell^{\rm l/r}+1)$.
\item
If a coupling linking one resonant region to another one becomes resonant, that creates a bigger resonant spots. For example if a coupling linking the $k-$th resonance to the $k+1$-th resonance becomes resonant, then the two resonant regions are merged and we update (taking also into account the number $\ell^{{\rm l}}$ of spins eventually thermalized by the resonant region to the left) $j_{{\rm min}}^k  \to i_{{\rm min}}^k$, $j_{{\rm max}}^k  \to j_{{\rm max}}^{k+1}$, $ i_{{\rm min}}^k \to i_{{\rm min}}^k - \ell^{{\rm l}}$, $ i_{{\rm max}}^{k} \to i_{{\rm max}}^{k+1}$, $n_{k}^{{\rm r}} \to n_{k+1}^{{\rm r}} $ and $n_k^{\rm l} \to n_k^{\rm l}(\ell^{\rm l}+1)$. We also update $\hat{i}_{{\rm min}}^k \to \hat{i}_{{\rm min}}^k$ and $\hat{i}_{{\rm max}}^k \to \hat{i}_{{\rm max}}^{k+1}$ to keep track of the effective bare spot. Finally the resonance $k+1$ (now taken into account in $k$) is destroyed, resonances are relabeled and $n_{{\rm res}}\to n_{{\rm res}}-1$. We proceed similarly if the resonance must be merged with the resonance $k-1$ to its left, or with both (in which case $n_{{\rm res}}\to n_{{\rm res}}-2$).
\end{enumerate}
The algorithm is iterated until a single resonance remains. In that case it either invade the remaining of the system (thermal phase), or not (MBL phase).

\section{Mean-field flow equations: derivation and critique}\label{sec:MF}

In this section we derive mean-field flow equations `solving' the scheme. These flow equations were already analyzed in \cite{ThieryHuveneersMullerDeRoeck2017} and we recall here the results, that will be compared with the numerical analysis of the scheme in the next section. We also discuss in the end of the section some aspects missed by the analysis.

\subsection{Mean-field hypothesis and flow equations}

To perform a mean-field analysis of our scheme, we suppose that resonances of size $k$: H1) can always be treated entirely, without ever encountering other active resonant regions; H2) see an homogeneous localized environment with inverse localization length (for the norm of perturbative couplings) $\alpha_k$ (computed below). With this homogeneity assumption, the number of spins $\ell_k$ that is included in the region thermalized by resonant spots (this region includes spins already thermalized by other small resonances) of size $k$ is constant and can be obtained by replacing $\alpha \to \alpha_k$ in \eqref{Eq:reselln}. We get 
\bea \label{Eq:flowellk}
\ell_k = \frac{k\alpha_c}{2(\alpha_k-\alpha_c)} \, ,
\eea
if $\alpha_k >\alpha_c$ and $\ell_k = +\infty$ otherwise (avalanche instability and first possibility (a) for thermalization). Since a single coupling is resonant with probability $\epsilon$, the density of bare spots of size $k$ is $\epsilon^{k}$. If $k+2\ell_k > \epsilon^{-k}$ resonances of size $k$ percolates when they propagate (second possibility (b) for thermalization). Otherwise the fraction of space that is covered by treated (inactive) resonant spot originating from bare spots of size $k$ is given by, at this stage of the procedure
\bea \label{Eq:rhok}
\rho_k = \epsilon^k(k+2\ell_k) \, .
\eea
We can then write the inverse localization length felt by spots of size $k+1$ as, using \eqref{eq:haltingrule},
\bea\label{Eq:flowalphak}
\alpha_{k+1}=(1-\rho_k)\alpha_k + \rho_k \alpha_t \, ,
\eea
with $\alpha_t= - \alpha_c/2$ the inverse localization length through treated (inactive) thermal region. We recall that $\alpha_c= \log(d)$. Combining \eqref{Eq:flowellk} and \eqref{Eq:flowalphak} with the initial condition \eqref{Eq:BareLL} for the bare inverse localization length $\alpha_1(\epsilon)=- \log (K g {d} )-b \alpha_c = - \log(\frac{K{d}^{1+b}}{C}\epsilon)$ closes the scheme. 

\smallskip

{\it Remark} There are thus, from the algorithmic point of view, two possibilities for thermalization (defined as (a) and (b) above). Distinguishing the two is however formal (it only makes sense within this mean-field approximation) and in the following we will discard the (b) possibility, always defining $\rho_k$ by \eqref{Eq:rhok} even when $k+2\ell_k >\epsilon^{-k}$. In those cases we get $\alpha_{k+1}<\alpha_c$ and delocalization due to the possibility (a). 

\subsection{The mean-field transition}
The analysis \cite{ThieryHuveneersMullerDeRoeck2017}  of the flow equations Eq.~\eqref{Eq:flowellk}-\eqref{Eq:rhok}-\eqref{Eq:flowalphak} with initial condition $\alpha_1(\epsilon)= - \log(\frac{K{d}^{1+b}}{C}\epsilon)$ reveals the existence of two regimes separated by a critical point at $\epsilon_c \in ]0,1[$:
\begin{enumerate}
\item
For $\epsilon<\epsilon_c$, $\alpha_k$ converges with $\lim_{k \to \infty} \alpha_k =: \alpha(\epsilon) > \alpha_c$. This is the MBL phase.
\item
For $\epsilon>\epsilon_c$, there is a finite $k_*(\epsilon) \in \mathbb{N}$ such that $\alpha_{k^*(\epsilon)} \leq \alpha_c$ with $\alpha_k>\alpha_c$ for $k<k_*(\epsilon)$, implying that spots of size $k_*$ thermalize the system. This is the thermal phase.
\end{enumerate} 
Exactly at the critical point we get
\bea \label{eq:predictionalphaatcriticality}
\alpha(\epsilon_c) := \lim_{k \to \infty} \alpha_k(\epsilon) = \alpha_c \, ,
\eea
showing that (i) the critical point is included in the MBL phase; (ii) the transition is governed by the avalanche instability. The density of (inactive) thermal spots at the critical point $\rho_c$ is found by inverting $\alpha_c = \alpha_1(\epsilon) (1-\rho_c) +\rho_c \alpha_t$: we get $\rho_c = \frac{\alpha_1(\epsilon_c)-\alpha_c}{\alpha_1(\epsilon_c)-\alpha_t} <1$. The critical point displays an infinite response property as $\lim_{k \to \infty}\frac{\ell_k}{k}=+\infty$ (with proba $1$).

\medskip

The critical behavior of the transition is at the mean-field level controlled by several (a priori) independent exponents. On the MBL side we define $\nu_->0$ as giving the approach of the inverse localization length $\alpha(\epsilon) = \lim_{k \to \infty} \alpha_k $ to its critical value:
\bea \label{eq:defMFnumoins}
\alpha(\epsilon)-\alpha_c \sim (\epsilon_c-\epsilon)^{\nu_-} \, .
\eea
Numerical simulations reveal that $\nu_-$ is non-universal already at the mean-field level (here all non-universality is included in $\alpha_1(\epsilon)$), with always $1/2<\nu_- <1$. From the thermal side we find that $k_*(\epsilon)$ diverges logarithmically and we define an exponent $\nu_+$ as
\bea \label{eq:defMFnuplus}
k_*(\epsilon) \sim  \frac{\nu_+}{\log(\epsilon_c)} \log(\epsilon-\epsilon_c) \, ,
\eea
also found to be non-universal at mean-field with $\nu_+ \in [1,2]$. Finally the critical nature of the critical point manifests itself in the thermal (inactive) spot distribution at criticality: denoting $S_k=k+2\ell_k$ the size of thermal spots we find that the $S_k$ are distributed as
\bea
p(S) \sim S^{-\tau} \quad , \quad  \tau =3 \, .
\eea

Away from the critical point the thermal spot distribution is cut off at an $\epsilon-$dependent value $S^*_-(\epsilon) \sim (\epsilon_c-\epsilon)^{-\nu_-}$. Finally denoting $p_{{\rm therm}}(L,\epsilon)$ the probability to observe a chain of length $L$ in the thermal phase, we find at the mean-field level the asympotic behavior (always understood as valid up to sub-leading corrections)
$$
p_{{\rm therm}}(L,\epsilon) \sim_{L \to \infty} \begin{cases} e^{-L (\epsilon_c-\epsilon)^{\nu_-}}   & \epsilon < \epsilon_c \\
 L^{-\beta}     & \epsilon = \epsilon_c    \\
  1 - e^{-L (\epsilon-\epsilon_c)^{\nu_+}}  & \epsilon>\epsilon_c  \, \, \, , 
  \end{cases}
$$
with $\beta=\tau-2=1$. Note that this makes impossible the use of single parameter scaling ansatz of the form $p_{{\rm therm}}(L,\epsilon) \sim \frac{1}{L^\beta} {\cal G}(L(\epsilon-\epsilon_c)^{\nu})$ with ${\cal G}$ a scaling function. We compare in Sec.~\ref{sec:numerics} these results with numerical simulations of the true scheme, and in the remaining of this section we discuss the (non-)validity of the approximation performed at the mean-field level.

\medskip

{\it Remark} It can be  noted that the mean-field results for $\nu_\pm$ break the rigorous Chayes-Harris bound of \cite{chandran2015finite}. We will see in Sec.~\ref{sec:numerics} that that is no longer the case for the full scheme.

\subsection{Critic and possible improvements of the mean-field analysis}

\subsubsection{A non-trivial first passage time problem}

An obvious simplification made in the mean-field analysis is the homogeneity assumption (H2). In practice, the distribution $p(\ell_k)$ of the number of spins thermalized by spots of size $k$ is  non-trivial. Away from the critical point, at large scales, it is clear that the self-averaging assumption is correct and we expect \eqref{Eq:flowellk} to be satisfied with probability $1$. Close to criticality however, fluctuations remain important and one expects $\ell_k$ to display a broad (power-law) distribution. Let us first note that if the random variables $\alpha_i$ were bounded and independently distributed one could use a Brownian approximation for the solution of the first passage time problem \eqref{Eq:FirstPassageTimePRoblem}. That would lead to, noting $\delta \alpha_k= \overline{\alpha_i} - \alpha_c$ and $\sigma_k = \sqrt{\overline{\alpha_i^2}^c}$
\bea \nonumber
p(\ell_k) = \frac{  k \alpha_c}{2\sqrt{2\pi \sigma_k} \ell_k^{3/2}} e^{-  \frac{( k\alpha_c/2 -  \delta \alpha_k \ell_k)^2}{2 \sigma_k  \ell_k}} \, .
\eea
Taking the average value, one retrieves that $\overline{\ell_k}$ is given by the mean-field prediction \eqref{Eq:flowellk}. Note however that close to criticality $\delta \alpha_k \to 0$ while fluctuations $\sigma_k$ grows and the average $\overline{\ell_k}$ is dominated by rare events and is much larger than the typical value $\ell_k^{{\rm typ}}$. The use of the Brownian approximation is however clearly inconsistent since, although the $\alpha_i$ are bounded, they exhibit long-range correlations at criticality (since $\alpha_i=\alpha_t$ for all spins inside an inactive spot). Our problem is thus closer to a first passage time problem for a random walker with power-law distributed jumps, a highly non trivial problem (see \cite{MajumdarLectureNotes2010} for review and \cite{LeDoussalWiese2009} for a rare solvable case). Despite this, it would still seem that taking the mean-field prediction \eqref{Eq:flowellk} for the average $\overline{\ell_k}$ is not an obviously wrong approximation. The mean-field analysis is however plagued by a second issue that we now discuss.

\subsubsection{Effective bare spots} \label{subsec:renormalizedbarespot2}

In the mean-field analysis we suppose that spots of size $k$ can be entirely diagonalized without ever merging with another active spots (H1). However, such processes occur and, in those cases, our procedure requires to treat these new, larger spots, later on in the diagonalization procedure. In fact, we noted in Sec.~\ref{sec:effective bare spots} that merging two active spots creates a larger active spot that affects the rest of the system as if one had started with an effective bare spot with boundaries the outer boundaries of the bare spots we started with. These events occur at any scale, dramatically modifying the probability to observe a bare spot of size $k$ from $\epsilon^k$ to $p_b(k)$, the probability to observe an effective bare spot of size $k$. Given an exact formula for $p_b(k)$, it seems reasonable to modify the mean-field analysis by changing $\epsilon^k \to p_b(k)$. However, as we now explain, the determination of  $p_b(k)$ close to the critical point is an extremely difficult problem.

\medskip

 As a simple example consider first two initial resonant spots of size $1$ separated by $\ell$ spins. From the general rules of our scheme, these resonances should be treated simulatenously. If they are far engouh, these resonances both thermalize a region of size $\ell_1$ given by \eqref{Eq:flowellk}. However, if $\ell \leq 2 \ell_1$ the two resonant regions merge, creating a large resonant region of size $2 + 4 \ell_1$. This region has an effect that is equivalent to having started with and effective bare spot (EBP) of size $2+\ell$ (the two initial bare spots and the thermalized space in between). If $\ell=\ell_1$, we create in such a way an EBP of size $2+2\ell_1$, that is the maximum size of a EBP that can be created from two bare spots of size $1$. The appearance of such an EBP is much more likely to occur than having at the beginning $2+2\ell_1$ consecutive resonant bounds ($\epsilon^2 \gg \epsilon^{2+2\ell_1} $). Two such EBPs can now be used to create a larger EBP. An EBP of size $2+2\ell_1$ can thermalize a region that is devoted of any resonance on a distance $\ell_c = (2+2\ell_1)\times \ell_1$. Using two EBPs of size $2+2\ell_1$ at a distance $\ell = 2 \times (2+2\ell_1) \ell_1$ from one another one obtains a EBP of size $2\times(2+2\ell_1) + 2 \times (2+2\ell_1) \ell_1$. Repeating this construction recursively and arranging bare spots of size $1$ in a fractale way we obtain that an EBP of size $(2(1+\ell_1))^n$ with $n\geq 1$ can be obtained using $2^n$ bare spots of size $1$. This is the smallest number of bare spots that needs to be used to create such a large EBP. We thus obtain that, to lowest order in $\epsilon$, $p_b((2(1+\ell_1))^n) = \epsilon^{2^n} $. Generalizing to any $k$ we obtain, to lowest order in $\epsilon$,
\bea \label{Eq:barespotproba1}
p_b(k) \simeq \epsilon^{k^{\gamma}} \quad , \quad \gamma = \frac{1}{1+\frac{\log(1+\ell_1)}{\log(2)}} <1 \, .
\eea
For larger $\epsilon$, more bare spots can be used to create an EBP more efficiently (having a bare spot in region of size $1/\epsilon$ does not 'cost' any probability). In the MBL phase $\epsilon < \epsilon_c$, one expects the inverse localization length to converge at large $k$ to a value $\alpha(\epsilon) > \alpha_c$. One can then think of modifying the estimate \eqref{Eq:barespotproba1} by renormalizing the localization length: we replace $\alpha_1$ (hidden in $\ell_1$) to $\alpha(\epsilon)$ and obtain
\bea \label{Eq:barespotproba2}
p_b(k) \simeq \epsilon^{k^{\gamma(\epsilon)}} \quad , \quad \gamma(\epsilon)^{-1} = 1+\frac{\log(1+ \frac{\alpha_c}{2(\alpha(\epsilon)-\alpha_c)})}{\log(2)} \, ,
\eea
i.e. a stretched exponential decay with an exponent $\gamma(\epsilon)$ converging to $0$ as the critical point is approached. This should however only be valid asymptotically for $\epsilon< \epsilon_c$ fixed. One expects that for smaller $k$ (and for any $k$ at the critical point), $p_b(k)$ takes a power-law form, that is related to the distribution of $p(\ell_{k'})$ for $k' \leq k$. For example one always expect that $p_b(k +2 \ell_k^{{\rm typ}}) \geq p_b(k)^2$, but $\ell_k^{{\rm typ}}$ appears difficult to evaluate since the first passage time problem \eqref{Eq:FirstPassageTimePRoblem} is basically unsolvable. We do not go further in this direction here.

\medskip

{\it Remark} Note that combining this stretched exponential decay of the effective bare spots probability throughout the MBL phase with the considerations of \cite{ThieryHuveneersMullerDeRoeck2017} leads to the prediction that average correlators in eigenstates decay as stretched exponentials in the MBL phase. Such a  behavior was already predicted from the phenomenological RG analysis of \cite{zhang2016many}. We note that the physical reason for the occurrence of a stretched exponential decay in the MBL phase is in our case is the same as in \cite{zhang2016many}: thermal inclusions exist in the MBL phase and the probability of observing a thermal inclusion of length $k$ decays subexponentially with $k$ due to their fractale structure.

\section{Numerics} \label{sec:numerics}

\begin{figure}
\centerline{\includegraphics[width=13cm]{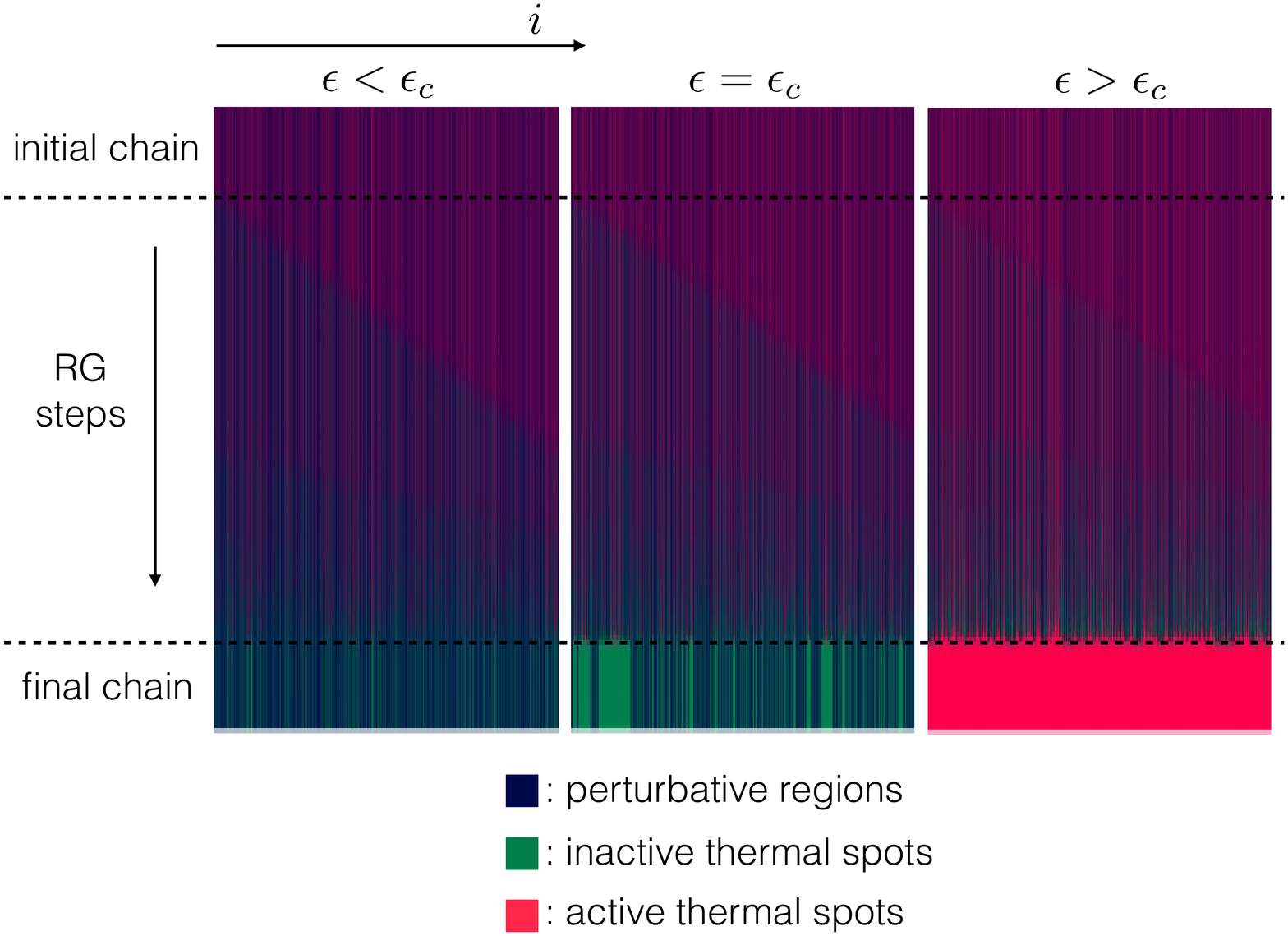} } 
\caption{Cartoon of the RG process for chains of length $L=16000$ and $\epsilon = 0.15$ (left), $\epsilon=0.169$ (middle) and $\epsilon = 0.2$. At the beginning the chain contains perturbative (blue) and active thermal regions (red). The non-perturbative diagonalization of an active thermal spot either leads to a propagation of the spot or renders the spot inactive (green). For $\epsilon = 0.15$ the final chain contains both perturbative and (inactive) thermal regions. At the critical point $\epsilon \simeq 0.169$ the final chain look strongly inhomogeneous with large (inactive) thermal spots included in perturbative regions. Above the critical $\epsilon = 0.2$ the chain is fully thermal. The diagonal structures that can be distinguished are due to the fact that the algorithm first diagonalize resonant regions of size $1$ in order of appearance along the chain, before starting to diagonalize resonant regions of size $2$ and so on.}
\label{Fig:cartoonnumerics}
\end{figure}

We performed numerical simulations of our scheme following the algorithm of Sec.~\ref{subsec:MainAlgo}. We consider chains of length $L$ with periodic boundary conditions and initial two-body couplings are declared resonant with probability $\epsilon$. We take for the bare inverse localization length $\alpha_1(\epsilon) = - \log(\epsilon)$, which simply amounts to a choice of the constant $C,K$ of Sec.~\ref{sec:AnalysisGen} as $\frac{K}{C}d^{1+b}=1$. The decay of the norm of perturbative operators through treated resonant region is taken as $\alpha_t=-\alpha_c/2$, as was evaluated using microscopically motivated rules in Sec.~\ref{sec:AnalysisGen}. We perform simulations for system of size $L \in \{250,500, 1000,2000,4000,8000,16000\}$, and $\epsilon$ in the range $\epsilon \in [0.06,0.26]$. Averages over disorder are performed using $5 \times 10^5$ samples for $\epsilon \in [0.168,0.172]$ and $2\times10^5$ samples otherwise. A cartoon illustrating the algorithm is given in Fig.~\ref{Fig:cartoonnumerics}. We compare the numerical results to the prediction of the mean-field analysis (\cite{ThieryHuveneersMullerDeRoeck2017} and Sec.~\ref{sec:MF}) focusing on several observables: the probability to observe the chain in the thermal phase (Sec.~\ref{subsec:Num:pth}), the typical localization length (Sec.~\ref{subsec:Num:LL}), the distribution of inactive thermal spots in localized samples (Sec.~\ref{subsec:Num:pofS}) and the distribution of effective bare spots (Sec.~\ref{subsec:Num:pofk}).

\subsection{Thermal probability} \label{subsec:Num:pth}

\begin{figure}
\centerline{\includegraphics[width=8cm]{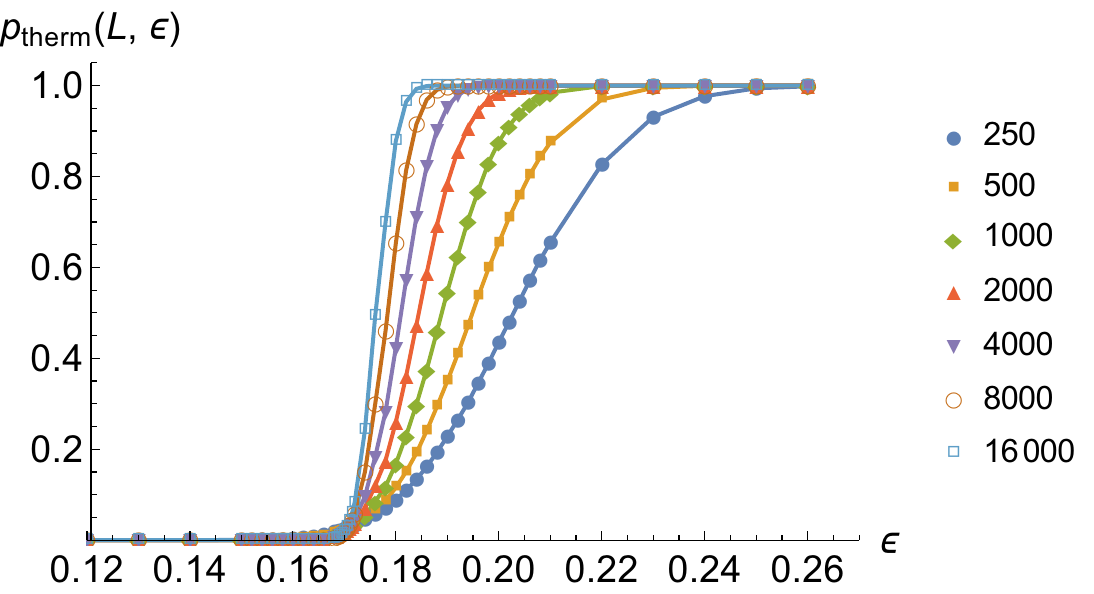} \includegraphics[width=8cm]{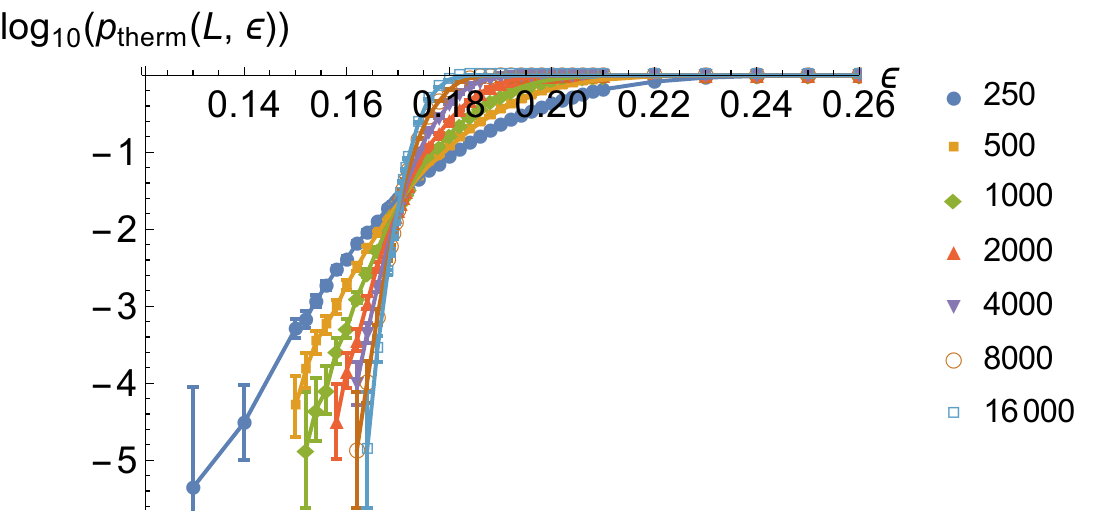}} 
\caption{Left: Linear plot of $p_{{\rm therm}}(L,\epsilon)$, the probability of observing the system in the thermal phase. Right: Logarithmic plot of $p_{{\rm therm}}(L,\epsilon)$. Error-bars are $3-$sigma Gaussian estimates.}
\label{Fig:pthermallin}
\end{figure}

\begin{figure}
\centerline{\includegraphics[width=8cm]{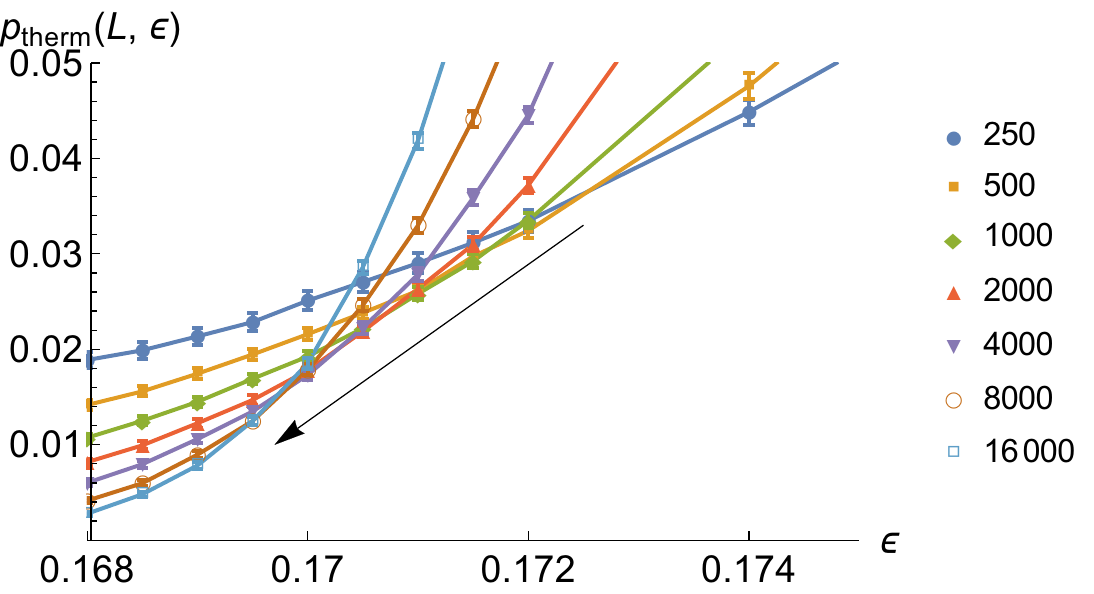}  \includegraphics[width=8cm]{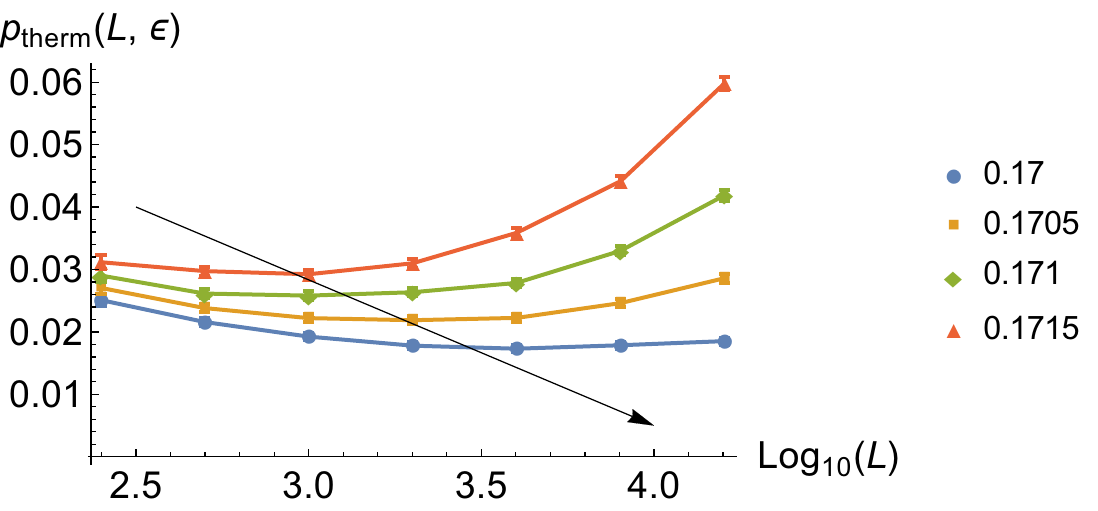}} 
\caption{Left: Plot of $p_{{\rm therm}}(L,\epsilon)$ for $\epsilon \in [0.168,0.174]$: the crossing point shifts towards the bottom left of the picture, in agreement with our prediction that the critical point is localized. Right: Non-monotonic behavior of $p_{{\rm therm}}(L,\epsilon)$ as a function of $L$ for $\epsilon$ slightly larger than the critical point: $p_{{\rm therm}}(L,\epsilon)$ first decreases up to a scale $L_+(\epsilon)$ that diverges as $\epsilon \to \epsilon_c^+$. Error-bars are $3-$sigma Gaussian estimates.}
\label{Fig:pthermallogzoom}
\end{figure}

The scheme is runned for each sample and we determine whether or not the system is in the MBL or in the thermal phase. That allows the determination of $p_{{\rm therm}}(L,\epsilon)$, the probability of observing the system in the thermal phase for a given $(L,\epsilon)$. Its linear and logarithmic plots are shown in Fig.~\ref{Fig:pthermallin}. While it would naively appear that the curves exhibit a crossing point around $\epsilon =0.17$, zooming in on the curve reveals that the crossing point shifts (see Fig.~\ref{Fig:pthermallogzoom}): $p_{{\rm therm}}(L,\epsilon)$ and $p_{{\rm therm}}(2 L,\epsilon)$ are equal for an L-dependent value $\epsilon(L)$ with $\epsilon(L) > 0.172$ for $L =250$ and $\epsilon(L) < 0.170$ for $L=8000$. The value of the probability at the crossing point shifts towards $0$ with $p_{{\rm therm}}(L,\epsilon(L)) \simeq (3.3\pm0.1)\times10^{-2}$ for $L=250$ and $p_{{\rm therm}}(L,\epsilon(L)) \simeq (1.2\pm0.1)\times10^{-2}$ for $L=8000$. That is coherent with the existence of a critical point $\epsilon_c$ smaller than $0.17$ such that 
\bea
&& \lim_{L \to \infty} p_{{\rm therm}}(L,\epsilon ) = 0 \quad \text{ for } \epsilon \leq \epsilon_c \nn 
&& \lim_{L \to \infty} p_{{\rm therm}}(L,\epsilon ) = 0 \quad \text{ for } \epsilon > \epsilon_c \nonumber \, ,
\eea
and the transition appears in this sense {\it continuous from the MBL side but discontinuous from the thermal one}. Getting a precise estimate of the position $\epsilon_c$ of the critical point from the datas for $p_{{\rm therm}}(L,\epsilon )$ is difficult and we evaluate it from the study of the localization length as $\epsilon_c \simeq 0.169 \pm 0.001$ (see next section). Indeed, important finite size effects exist on the thermal side: while on the MBL side $\epsilon \leq \epsilon_c$, $p_{{\rm therm}}(L,\epsilon )$ monotonously decreases with $L$, on the thermal side $\epsilon > \epsilon_c$, $p_{{\rm therm}}(L,\epsilon )$ first decreases with $L$ up to a scale $L_+(\epsilon)$, before it starts increasing (see Fig.~\ref{Fig:pthermallogzoom}). The scale $L_+(\epsilon)$ diverges as $\epsilon \to \epsilon_c^+$. This scale plays a role similar to the scale $\epsilon^{-k_*(\epsilon)}$ in the mean-field analysis \cite{ThieryHuveneersMullerDeRoeck2017}: for small $L$ and on the thermal side, the system first think that he is in the MBL phase, before thermal spots large enough to delocalize the system appear. Such important finite size effects on the thermal side were already discussed in the MBL context in \cite{devakul2015early}, but also reminds us of the case of Anderson localization on random regular graphs \cite{GarciaMataGiraudGeorgeotMartinDubertrandLemarie2017,TikhonovMirlinSkvortsov2016}. We report a discussion of the finite size scaling analysis of these datas to Sec.~\ref{subsec:Num:pthFSC}.

\subsection{Localization length} \label{subsec:Num:LL}

\begin{figure}
\centerline{\includegraphics[width=8cm]{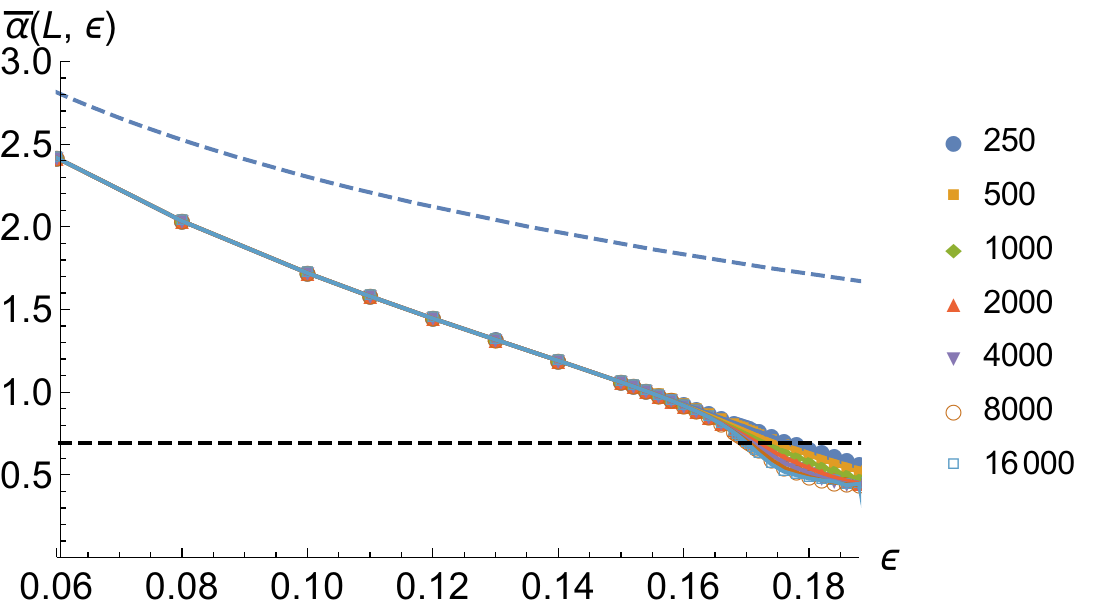} \includegraphics[width=8cm]{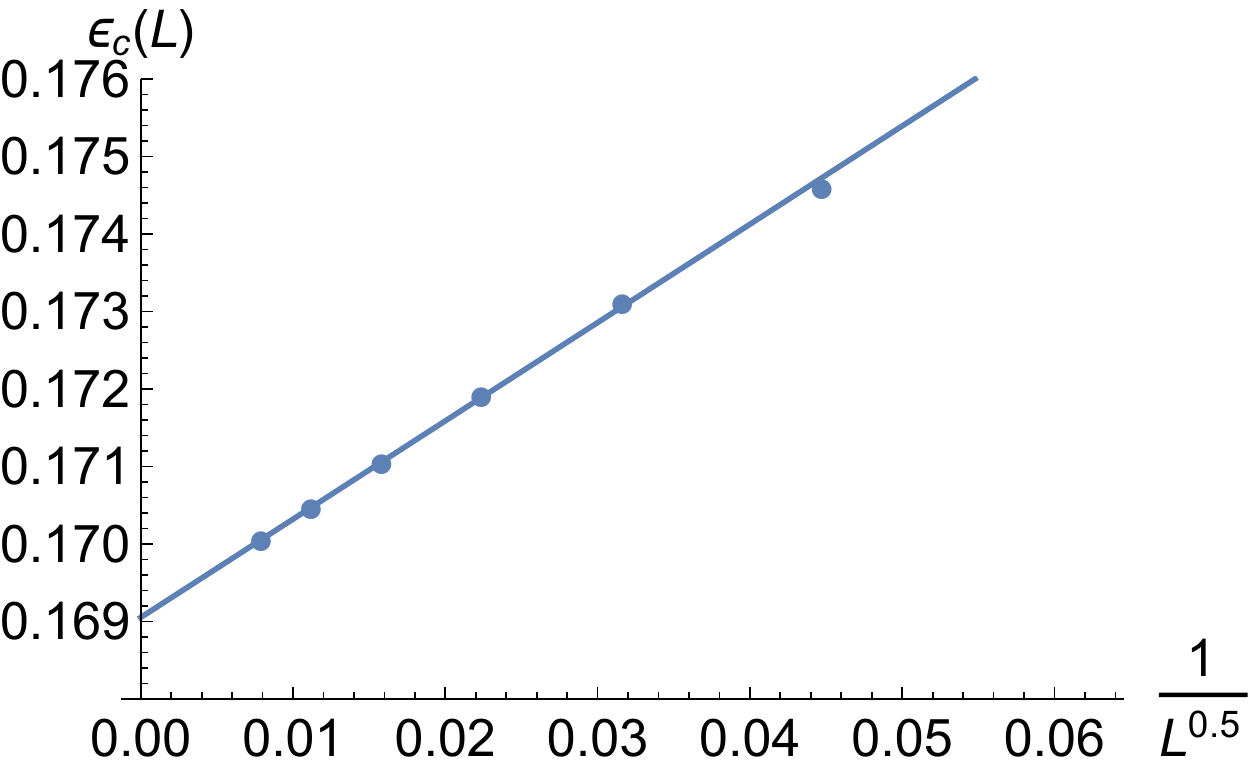}} 
\caption{Left: Plain lines and markers: measured average localization length for MBL sample.  The black-dashed line corresponds to $\alpha_c$, the blue-dashed one to the bare inverse localization length $\alpha_1(\epsilon)$. Right: Study of $\epsilon_c(L)$, defined as $\overline{\alpha}(L,\epsilon_c(L)) = \alpha_c$. The extrapolated limit $\lim_{L \to \infty} \epsilon_c(L)$ gives our measurement of the critical point. Error-bars smaller than marker size.}
\label{Fig:alpha}
\end{figure}

For sample in the MBL phase, we evaluate the inverse localization length $\alpha$ at the end of the procedure as $\alpha = (1-\rho) \alpha_1(\epsilon) + \rho \alpha_t$, with $\rho \in ]0,1[$ the fraction of sites in inactive spots. The average $\overline{\alpha}(L,\epsilon)$ (note that this is here an average conditioned on being in the MBL phase) is plotted on Fig.~\ref{Fig:alpha}. Comparing this with Fig.~\ref{Fig:pthermallin}, this clearly confirms our scenario of having a transition driven by the avalanche instability with at the critical point $\overline{\alpha}(+\infty,\epsilon_c) = \alpha_c \simeq 0.69$ and $\epsilon_c$ around $0.170$. The fact that the critical point must be localized with $\overline{\alpha}(+\infty,\epsilon_c) = \alpha_c$  was argued based on the mean-field analysis (see \cite{ThieryHuveneersMullerDeRoeck2017} and Sec.~\ref{sec:MF}) but can easily be argued to hold more generally within our scheme. Based on this observation that is now confirmed by numerical simulations, we measure the position of the critical point by studying the function $\epsilon_c(L)$ defined by $\overline{\alpha}(L,\epsilon_c(L)) = \alpha_c$, which satisfy by construction $\lim_{L \to \infty} \epsilon_c(L) = \epsilon_c$. Our datas appear coherent with $\epsilon_c(L) \simeq \epsilon_c + C/\sqrt{L}$ with $\epsilon_c \simeq 0.169 \pm 0.001$ and $C \simeq 1.17$ (se Fig.~\ref{Fig:alpha}). Taking this value as the definition of the critical point, we measure the convergence of the localization length at the critical point to its critical value. Our datas are coherent with the scaling form $\overline{\alpha(\epsilon_c,L)} \simeq 0.68+1.5/L^{1/5}$, in satisfying agreeement with $\alpha_c=\log(2) \simeq 0.69$. Note that this also implies that the density of inactive spots at the critical point is $\rho_c = \frac{\alpha_1(\epsilon_c)-\alpha_c}{\alpha_1(\epsilon_c)-\alpha_t} \simeq 0.51$. Let us also mention here that the mean-field analysis of Sec.~\ref{sec:MF} predicts a critical point at $\epsilon_c^{{\rm MF}} \sim 0.177$, not too far from the true value.

\subsection{Inactive spot distribution}  \label{subsec:Num:pofS}

\begin{figure}
\centerline{\includegraphics[width=8cm]{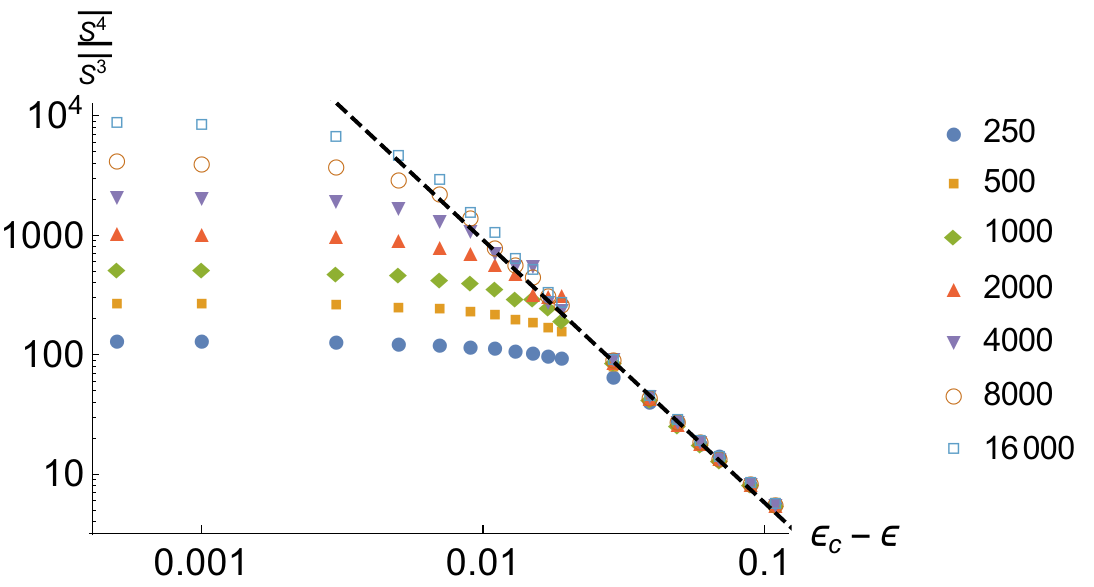}} 
\caption{Measurement of $\int S^4 p_{L,\epsilon}(S) dS / \int S^3 p_{L,\epsilon}(S) dS$ below criticality. For large $L \gg S_-^*(\epsilon)$ that gives a measurement of $ S_-^*(\epsilon)$. For smaller $L$ that  exhibits the cutoff due to the finite system size in \eqref{Eq:pofSfinitesize}. The dashed-line is a power-law behavior $1/(\epsilon_c-\epsilon)^{2.2}$}
\label{Fig:cutoffpofS}
\end{figure}

We study $p_{L,\epsilon}(S)$, the distribution of ergodic (inactive) spots in the end of the diagonalization procedure for systems of size $L$ and bare spot probability $\epsilon$, again conditioned on the system being in the MBL phase. For $\epsilon \leq \epsilon_c$ our results suggest (for $S \gg 1$, we do not study the small scales) the rather unusual scaling behavior
\bea \label{Eq:pofSfinitesize}
p_{L,\epsilon}(S) \sim \frac{1}{S^{\tau(\epsilon)}} {\cal F}_1(\epsilon, S/S_-^*(\epsilon)) {\cal F}_2(S/L) \, ,
\eea
a power-law distributions that is cutoff at large scales by two mechanism (never seen simulateneously). The power-law exponent appears to vary continuously in the MBL phase with for example $\tau(0.164) \simeq 2.6 \pm 0.1$ and $\tau(\epsilon_c) \simeq 2.25 \pm 0.1$. The cutoff ${\cal F}_2(S/L)$ is due to the finite system size and ${\cal F}_2(y)$ is a scaling function with ${\cal F}_2(0) =1$ and ${\cal F}_2(y) $ quickly decaying to $0$ for $y$ larger than $1$. The second cutoff scale $S_-^*(\epsilon)$ only depends on $\epsilon$ and diverges as $\epsilon \to \epsilon_c^-$. The cutoff function itself appears to vary continuously with $\epsilon$ and exhibit a stretched exponential decay ${\cal F}_1(\epsilon,y) \sim e^{-y^{\hat{\gamma}(\epsilon)}}$ with $0<\hat{\gamma}(\epsilon) <1$ and $\hat{\gamma}(\epsilon)$ deaying with $\epsilon$. Assuming\footnote{That is not exactly correct since the $\epsilon$ dependence of the cutoff function ${\cal F}_1(\epsilon, y)$ can in principle also contributes to the divergence of high order moments of $p_{\epsilon,L=+\infty}(S)$.} that $S_-^*(\epsilon)$ can be measured using $S_-^*(\epsilon) \sim \int S^4 p_{L,\epsilon}(S) dS / \int S^3 p_{L,\epsilon}(S) dS$ for $S_-^*(\epsilon) \ll L$, our datas appear coherent with $S_-^*(\epsilon) \sim (\epsilon_c-\epsilon)^{- 2.2\pm 0.2}$ (see Fig.~\ref{Fig:cutoffpofS}). Comparing with the mean-field results of \cite{ThieryHuveneersMullerDeRoeck2017} (recalled in Sec.~\ref{sec:MF}) that gives a measurement of the exponent $\nu_- \simeq - 2.2\pm 0.2$, now in agreement with the Chayes-Harris bound \cite{chandran2015finite}.

\medskip

\begin{figure}
\centerline{\includegraphics[width=8cm]{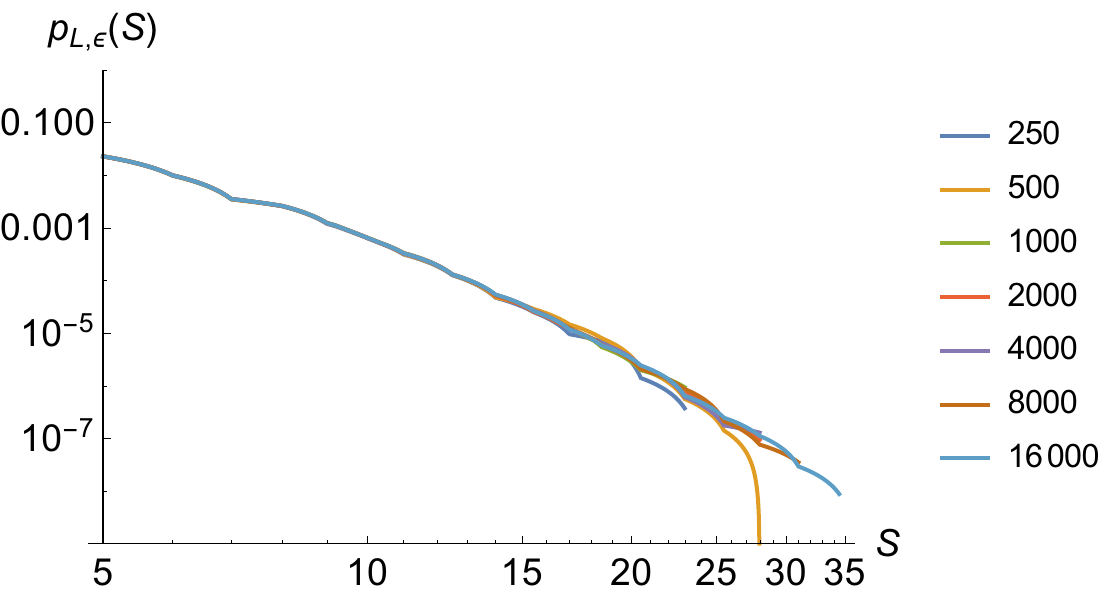} \includegraphics[width=8cm]{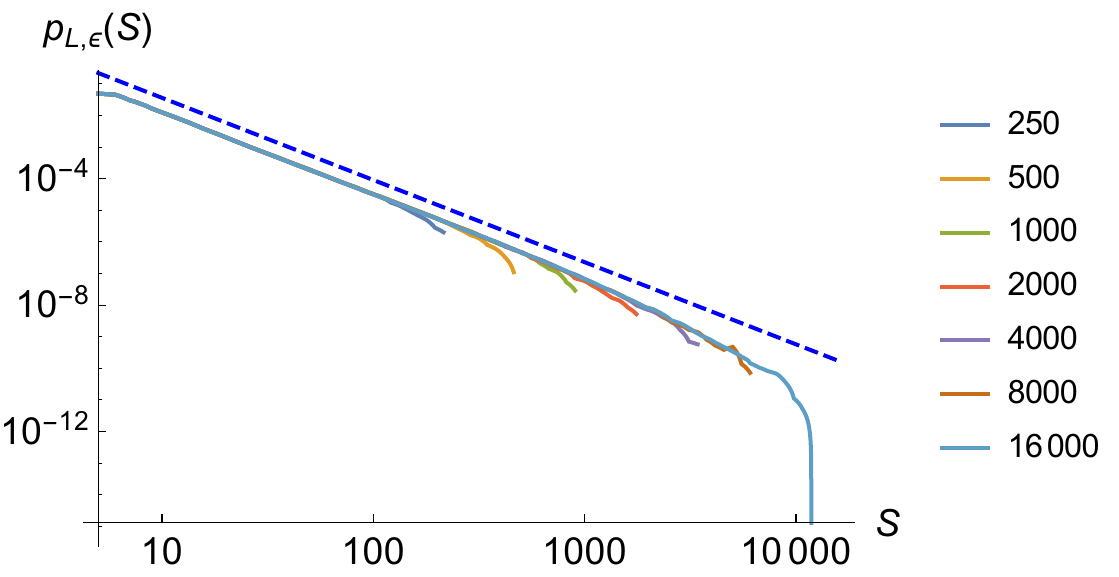}} 
\caption{Left: Distribution of inactive spots far below criticality, for $\epsilon = 0.06$. Right: Distribution of inactive spots close but below criticality, for $\epsilon = 0.164$. The blue-dashed line corresponds to a power-law $p(S) \sim S^{-2.6}$.}
\label{Fig:pofS006} 
\end{figure}

\begin{figure}
\centerline{\includegraphics[width=8cm]{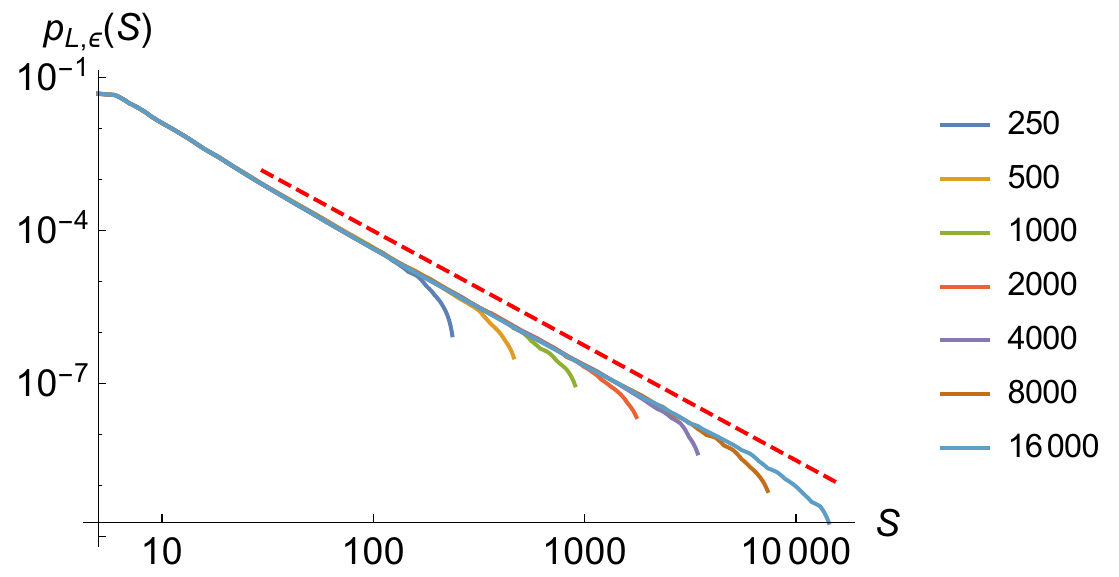}} 
\caption{Distribution of inactive spots at criticality, for $\epsilon = 0.169$. The red-dashed line corresponds to a power-law $p(S) \sim S^{-2.25}$.}
\label{Fig:pofS0169}
\end{figure}

\begin{figure}
\centerline{\includegraphics[width=8cm]{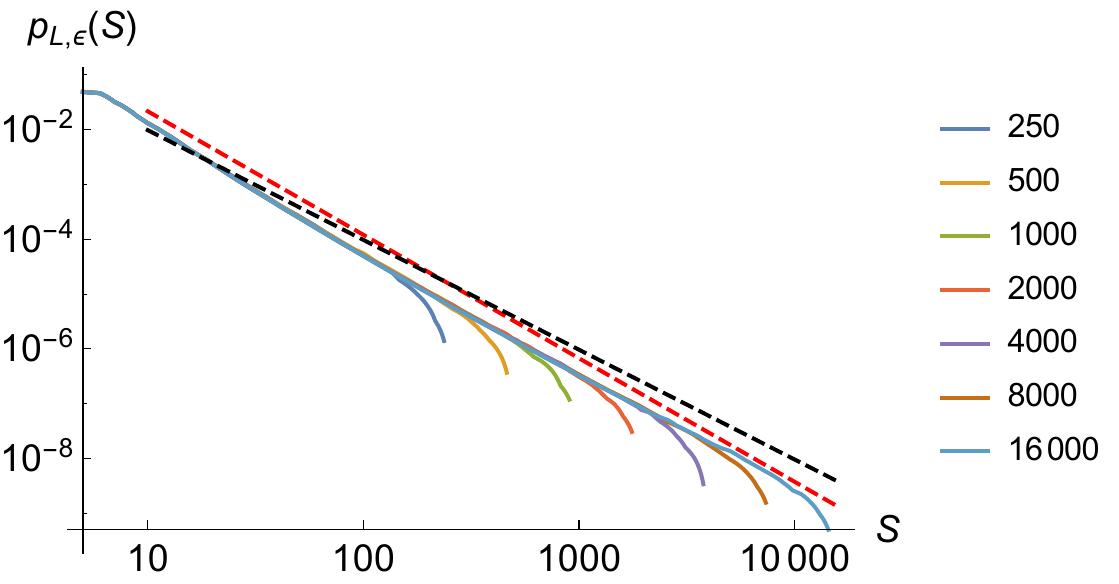} \includegraphics[width=8cm]{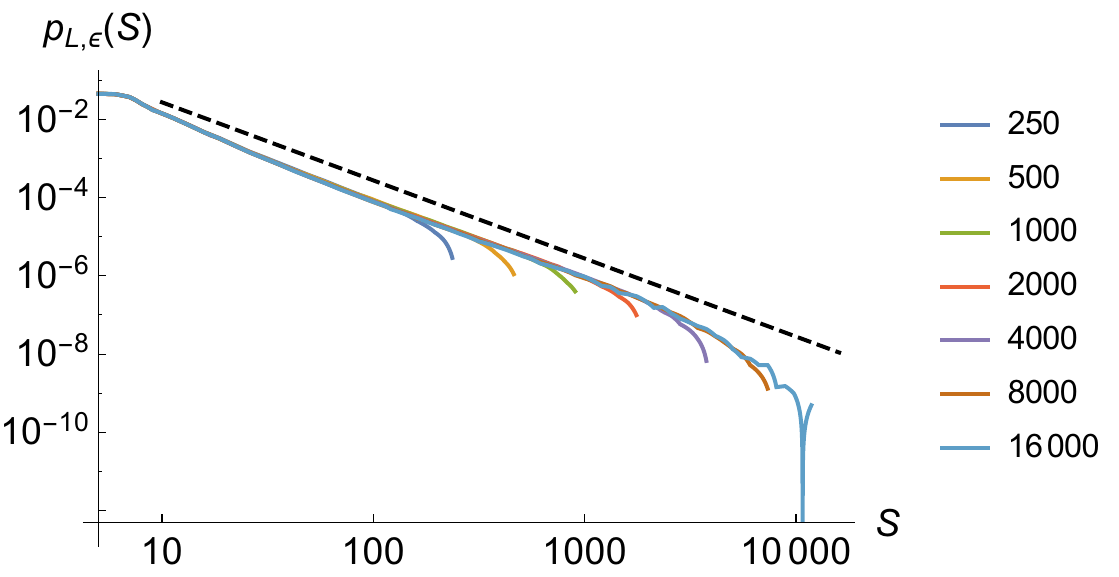}} 
\caption{Left: Distribution of inactive spots slightly above criticality, for $\epsilon = 0.171$. The red-dashed line corresponds to a power-law $p(S) \sim S^{-2.25}$, the black-dashed line to a power-law $p(S) \sim S^{-2}$. Right: Distribution of inactive spots above criticality, for $\epsilon = 0.182$. The black-dashed line corresponds to a power-law $p(S) \sim S^{-2}$.}
\label{Fig:pofS0171}
\end{figure}

Exactly at criticality and for an infinite system we obtain a pure power-law decay
\bea
\lim_{L \to \infty} p_{L,\epsilon_c}(S) \sim 1/S^{\tau(\epsilon_c)}  \, ,
\eea
with $\tau(\epsilon_c) \simeq 2.25 \pm 0.1$. That should be compared with the MF result (see \cite{ThieryHuveneersMullerDeRoeck2017} and Sec.~\ref{sec:MF})  where we obtained $\tau(\epsilon_c)=3$.
Above the critical point, we observe the emergence of a new power law at intermediate scales. Our results suggest the following: there is another diverging length scale $S_+^*(\epsilon)$ with $\lim_{\epsilon \to \epsilon_c^+} S_+^*(\epsilon) = + \infty $ such that:
For $S \ll S_+^*(\epsilon)$ we still observe a power-law that can be consistent with the one measured at the critical point, i.e. $ p_{L,\epsilon_c}(S) \sim_{S \ll S_+^*(\epsilon)} 1/S^{\tau(\epsilon_c)}$. For $S \gg S_+^*(\epsilon)$ there appears a new power-law $ p_{L,\epsilon_c}(S) \sim 1/S^{\tau_2}$ with a power-law exponent that is measured as $\tau_2 = 2\pm0.1$. Finally for $S \sim L$ we again observe a cutoff similar to the subcritical regime. We plot $p_{L,\epsilon}(S)$ for $\epsilon = 0.06$ deep in the MBL phase, close to the critical point on the MBL side for $\epsilon =0.164 $ (Fig.~\ref{Fig:pofS006}), at the critical point $\epsilon = 0.169$ (Fig.~\ref{Fig:pofS0169}), and above the critical point at $\epsilon = 0.171$ and $\epsilon = 0.182$ (Fig.~\ref{Fig:pofS0171}).

\medskip

Let us now discuss these results. Below criticality, the presence of the first cutoff ${\cal F}_1(\epsilon, S/S_-^*(\epsilon))$ with a continuously varying streched exponential exponent should be expected from our discussion of Sec.~\ref{subsec:renormalizedbarespot2}. The divergence of $S_-^*(\epsilon)$ should be related to the convergence of $\overline{\alpha}(\infty,\epsilon)$ to $\alpha_c$ for $\epsilon \to \epsilon_c^-$ but is not studied here. The second cutoff ${\cal F}_2(S/L)$ (that is only seen sufficiently close to the critical point when $S_-^*(\epsilon) \geq L$) is due to the conditioning on being in the MBL phase: sample with an inactive spot of a size comparable to the system size are very likely to end up in the thermal phase. At first sight the apparent existence of a continuously varying power-law exponent is somewhat surprising. However, we discussed in Sec.~\ref{Sec:FirstPassageTimeInterp} the relation between our problem and a first passage time problem for a discrete time random walk. Approaching criticality, the distribution of the $\alpha_i$ (jump of the random walker) exhibits power-law correlations (since $\alpha_i = \alpha_t$ for all $i$ inside an inactive spot). In \cite{LeDoussalWiese2009} it was shown that the power-law exponent of the distribution of first passage times for a drifted random walker with Levy distributed jumps varies continuously with the drift. Making the parallel with our case, our finding is thus not so surprising. For that reason we also expect $\tau(\epsilon)$ to be non-universal, even at the critical point. The fact that it is bounded as $\tau(\epsilon_c) >2$ is however expected. In the infinite system at criticality $p(\epsilon) \sim 1/S^{\tau(S)}$. Taking any finite portion of size $L$ of the infinite system, one expects that the probability $\tilde{p}(L)$ that the portion is included in an inactive spot scales as $\tilde{p}(L) \sim \int_{S>L} (S-L) p_{L=+\infty,\epsilon_c}(S) dS \sim L^{2-\tau(\epsilon_c)}{\cal F}_1(\epsilon,L/S_-^*(\epsilon))$. Since the critical point is localized, one expects this probability to decay to $0$, hence the bound $\tau(\epsilon_c) >2$. Above criticality, the appearance of a new exponent $\tau_2 \leq 2$ is for the same reason not surprising, otherwise $\tilde{p}(L) \sim L^{2-\tau_2}$ would converge to $0$, in contradiction with the assumption of being in the thermal phase. Our datas suggest exactly $\tau_2 = 2$ (marginal case) but we do not have a clear explanation for this fact. Finally the fact that this new power-law is only observed for large enough $S \geq S_+^*(\epsilon)$ with a diverging scale $S_+^*(\epsilon)$ should be thought of in analogy with the non-monotonic behavior of $p_{{\rm therm}}(L,\epsilon)$ as a function of $L$ for $\epsilon > \epsilon_c$.
 
\subsection{Bare spots distribution} \label{subsec:Num:pofk}

\begin{figure}
\centerline{\includegraphics[width=8cm]{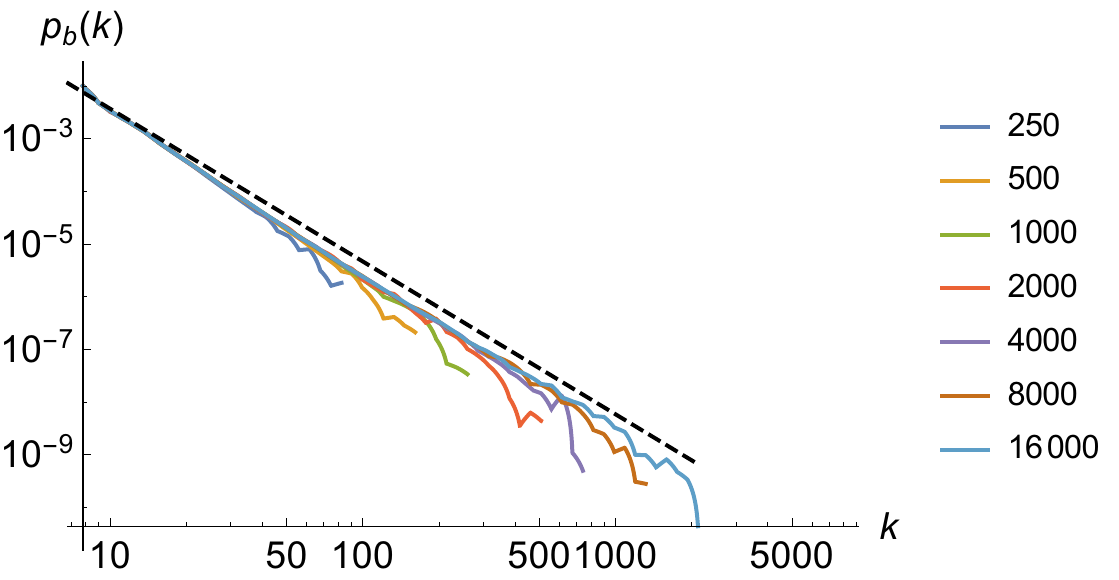}} 
\caption{Effective bare spots distribution at criticality. The dashed-line corresponds to a power-law $\sim 1/k^{2.9}$.}
\label{Fig:pofk0169}
\end{figure}

For each spot that becomes inactive during the RG, we also store the size of the effective bare spot it originates from, and investigate the effective bare spots distribution $p_b^{\epsilon,L}(k)$ (here not conditioned on being in the MBL phase). We find a behavior similar to the inactive spot distribution, and we only give here the main difference. Below criticality one observes a power-law with an exponent $\tau_b(\epsilon)$ that is continuously varying with $\epsilon$ and cutoff at an epsilon-dependent cutoff and at another $L-$dependent cutoff. This $L$ dependent cutoff $k^*(L)$ does not diverge linearly with the system size but rather as $L^{0.8 \pm 0.05}$. At criticality, the power-law exponent is measured as $\tau_b(\epsilon_c) \simeq 2.9 \pm 0.1$ (see Fig.~\ref{Fig:pofk0169}). Going above criticality, we do not observe important change in this exponent. Well above criticality, one observes a new cutoff smaller than $k^*(L)$ and that diminishes as $\epsilon$ grows. This is likely due to the fact that too big effective bare spots are then never completely diagonalized. 

\subsection{Finite size scaling of the thermal probability} \label{subsec:Num:pthFSC}

\begin{figure}
\centerline{\includegraphics[width=8cm]{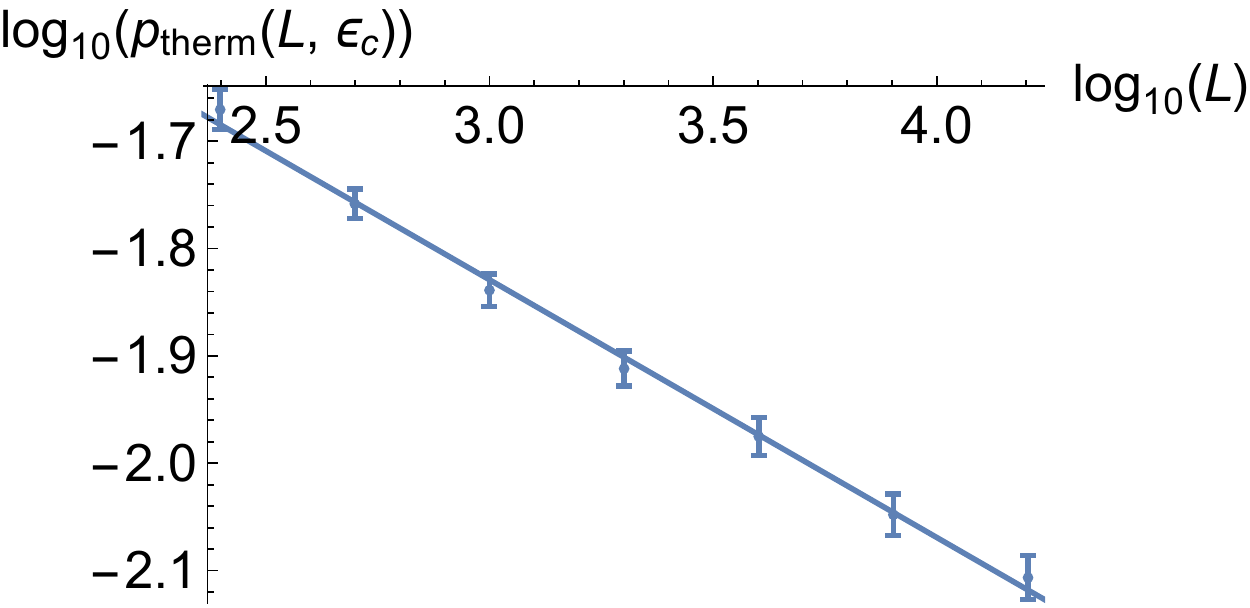} \includegraphics[width=8cm]{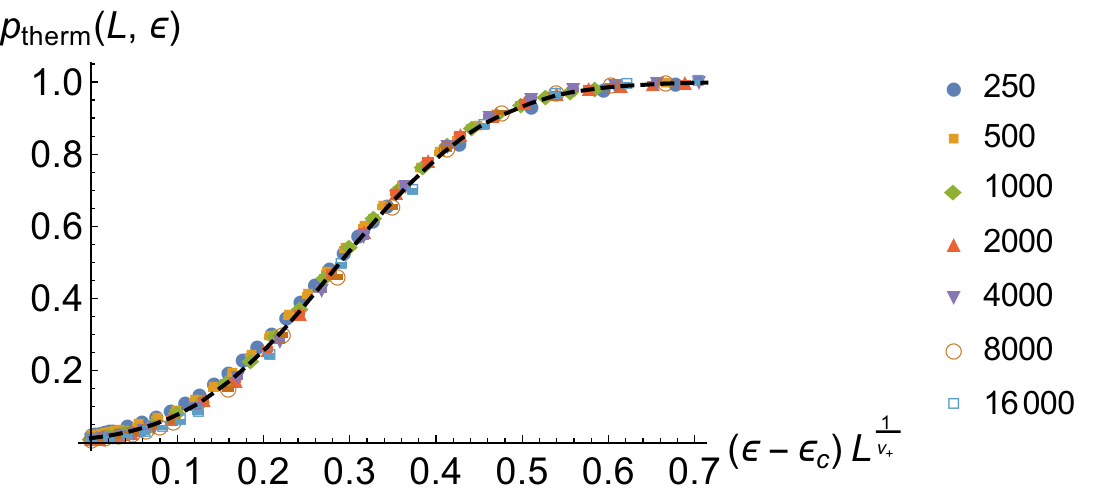}} 
\caption{Left: Logarithmic plot of $p_{{\rm therm}}(L,\epsilon_c)$. The dashed-line corresponds to the power-law $1/L^{0.25}$, in agreement with our measurement of $\tau(\epsilon_c)$ and the scaling conjecture \eqref{Eq:num:scalingMBL}-\eqref{Eq:aofeps}. Error-bars are $3-$sigma Gaussian estimates. Right: Scaling collapse of $p_{{\rm therm}}(L,\epsilon )$ above the critical point according to \eqref{Eq:num:scalingThermal} using here $\nu_+ = 2.6$. Markers correspond  to our datas, the dashed-line is a guide to the eye. Error-bars smaller than marker size.}
\label{Fig:pthermalatcriticality}
\end{figure}

Below criticality we can identify $p_{{\rm therm}}(L,\epsilon )$ with the probability to observe in the infinite system a finite portion of size $L$ inside an ergodic spot. That leads to $p_{{\rm therm}}(L,\epsilon ) \sim \int_{S>L} (S-L) \, p_{L=+\infty,\epsilon}(S) dS$ and
\bea \label{Eq:num:scalingMBL}
p_{{\rm therm}}(L,\epsilon ) \sim \frac{1}{L^{a(\epsilon)}} {\cal F}_1(\epsilon,L/S^*(\epsilon)) \, ,
\eea
with
\bea \label{Eq:aofeps}
a(\epsilon) = \tau(\epsilon)-2 \, .
\eea
Hence it appears meaningless to try fitting $p_{{\rm therm}}(L,\epsilon )$ with a single-parameter scaling form in the MBL phase. Exactly at criticality this predicts $p_{{\rm therm}}(L,\epsilon_c ) \sim \frac{1}{L^{a(\epsilon_c)}}$ with $a(\epsilon_c) \simeq 0.25 \pm 0.1$ (from our previous measurement of $\tau(\epsilon_c)$). This is confirmed in Fig.~\ref{Fig:pthermalatcriticality}.

Above criticality, one cannot either obtain a scaling form for $p_{{\rm therm}}(L,\epsilon )$ working $\forall \epsilon \geq \epsilon_c$ and $\forall L \gg 1$, as was already the case in the mean-field analysis. Still, our datas are consistent with the scaling form 
\bea \label{Eq:num:scalingThermal}
p_{{\rm therm}}(L,\epsilon ) = {\cal G}_+((\epsilon-\epsilon_c) L^{1/\nu_+}) \, ,
\eea
valid for $\epsilon>\epsilon_c$ and $(\epsilon-\epsilon_c) L^{1/\nu_+}$ of order $1$. The scaling function ${\cal G}_+(y)$ satisfies ${\cal G}_+(y) \in [0,1]$, ${\cal G}_+(0)=0$ and ${\cal G}_+(+\infty)=1$. The fit of our datas according to \eqref{Eq:num:scalingThermal} is shown on Fig.~\ref{Fig:pthermalatcriticality} and we obtain $\nu_+ \simeq 2.6 \pm 0.1$. Note that, contrarily to the mean-field predictions, this value is now in agreement with the Chayes-Harris bound \cite{chandran2015finite}. For $ (\epsilon-\epsilon_c)^{-\nu_+} \gg  L \gg 1 $ one expects the scaling $p_{{\rm therm}}(L,\epsilon ) \sim 1/L^{a(\epsilon_c)}$ as for MBL system at criticality.

\section{Conclusion}

Taking the eigenstate thermalization hypothesis has an effective description of ergodic inclusions in MBL systems has remarkable consequences for localization physics, the main one being the instability of localization with respect to the inclusion of ergodic grains \cite{de_roeck_stability_2016,luitz2017}. In this work and in \cite{ThieryHuveneersMullerDeRoeck2017} we have shown that this instability drives the MBL/ETH transition. The main consequence for the critical behavior is that the critical point is localized, on the verge of stability, with infinite response properties with respect to the inclusion of ergodic spots. We showed that this implies a breakdown of single parameter scaling around the transition, and through numerical analysis and the link between our work and first passage time problems, suggested that the critical exponents characterizing the transition are non-universal. While our scenario implies a bound on the typical localization length in the MBL phase that is saturated at the transition, we exhibited (mostly discussed in \cite{ThieryHuveneersMullerDeRoeck2017}) several diverging length or time scales: the average localization length and eigenstate correlation length in the MBL phase , also linked to the cutoff of the distribution of ergodic inclusions; the length scale at which avalanches are seen in the thermal phase, related to the average eigenstate end-to-end correlation length; the thermalization time on the thermal side. 

Interesting directions for future research are new checks of our working hypothesis or the study of the MBL/ETH transition in quasi-periodic systems where the nature of the transition could be different \cite{PhysRevLett.119.075702}.

\section*{Acknowledgements}
We are grateful to A. Scardicchio,  D. Huse and V. Khemani for helpful insights and discussions.  WDR acknowledges the support of the Flemish Research Fund FWO under grant G076216N. TT and WDR have been supported by the InterUniversity Attraction Pole phase VII/18 dynamics, geometry and statistical physics of the Belgian Science Policy. TT is a postdoctoral fellow of the Research Foundation, Flanders (FWO).

\bibliography{loclibrary}

\begin{thebibliography}{10}
\providecommand{\url}[1]{\texttt{#1}}
\providecommand{\urlprefix}{URL }
\expandafter\ifx\csname urlstyle\endcsname\relax
  \providecommand{\doi}[1]{doi:\discretionary{}{}{}#1}\else
  \providecommand{\doi}{doi:\discretionary{}{}{}\begingroup
  \urlstyle{rm}\Url}\fi
\providecommand{\eprint}[2][]{\url{#2}}

\bibitem{anderson_absence_1958}
P.~W. Anderson,
\newblock \emph{Absence of {Diffusion} in {Certain} {Random} {Lattices}},
\newblock Phys. Rev. \textbf{109}(5), 1492 (1958),
\newblock \doi{10.1103/PhysRev.109.1492}.

\bibitem{fleishman_interactions_1980}
L.~Fleishman and P.~W. Anderson,
\newblock \emph{Interactions and the {Anderson} transition},
\newblock Phys. Rev. B \textbf{21}(6), 2366 (1980),
\newblock \doi{10.1103/PhysRevB.21.2366}.

\bibitem{basko2006metal}
D.~Basko, I.~Aleiner and B.~Altshuler,
\newblock \emph{Metal--insulator transition in a weakly interacting
  many-electron system with localized single-particle states},
\newblock Annals of physics \textbf{321}(5), 1126 (2006).

\bibitem{gornyi2005interacting}
I.~Gornyi, A.~Mirlin and D.~Polyakov,
\newblock \emph{Interacting electrons in disordered wires: Anderson
  localization and low-t transport},
\newblock Physical review letters \textbf{95}(20), 206603 (2005).

\bibitem{znidaric_many-body_2008}
M.~{\v Z}nidari{\v c}, T.~Prosen and P.~Prelov{\v s}ek,
\newblock \emph{Many-body localization in the {Heisenberg} ${XXZ}$ magnet in a
  random field},
\newblock Phys. Rev. B \textbf{77}(6), 064426 (2008),
\newblock \doi{10.1103/PhysRevB.77.064426}.

\bibitem{oganesyan_localization_2007}
V.~Oganesyan and D.~A. Huse,
\newblock \emph{Localization of interacting fermions at high temperature},
\newblock Phys. Rev. B \textbf{75}(15), 155111 (2007),
\newblock \doi{10.1103/PhysRevB.75.155111}.

\bibitem{pal_many-body_2010}
A.~Pal and D.~A. Huse,
\newblock \emph{Many-body localization phase transition},
\newblock Phys. Rev. B \textbf{82}(17), 174411 (2010),
\newblock \doi{10.1103/PhysRevB.82.174411}.

\bibitem{ros2015integrals}
V.~Ros, M.~M{\"u}ller and A.~Scardicchio,
\newblock \emph{Integrals of motion in the many-body localized phase},
\newblock Nuclear Physics B \textbf{891}, 420 (2015).

\bibitem{kjall2014many}
J.~A. Kj{\"a}ll, J.~H. Bardarson and F.~Pollmann,
\newblock \emph{Many-body localization in a disordered quantum ising chain},
\newblock Physical review letters \textbf{113}(10), 107204 (2014).

\bibitem{luitz_many-body_2015}
D.~J. Luitz, N.~Laflorencie and F.~Alet,
\newblock \emph{Many-body localization edge in the random-field {Heisenberg}
  chain},
\newblock Phys. Rev. B \textbf{91}(8), 081103 (2015),
\newblock \doi{10.1103/PhysRevB.91.081103}.

\bibitem{nandkishore_many-body_2015}
R.~Nandkishore and D.~A. Huse,
\newblock \emph{Many-{Body} {Localization} and {Thermalization} in {Quantum}
  {Statistical} {Mechanics}},
\newblock Annual Review of Condensed Matter Physics \textbf{6}(1), 15 (2015),
\newblock \doi{10.1146/annurev-conmatphys-031214-014726}.

\bibitem{altman_universal_2015}
E.~Altman and R.~Vosk,
\newblock \emph{Universal {Dynamics} and {Renormalization} in
  {Many}-{Body}-{Localized} {Systems}},
\newblock Annu. Rev. Condens. Matter Phys. \textbf{6}(1), 383 (2015),
\newblock \doi{10.1146/annurev-conmatphys-031214-014701}.

\bibitem{abanin_recent_2017}
D.~A. Abanin and Z.~Papi{\'c},
\newblock \emph{Recent progress in many-body localization},
\newblock arXiv:1705.09103  (2017),
\newblock ArXiv: 1705.09103.

\bibitem{luitz_ergodic_2017}
D.~J. Luitz and Y.~Bar~Lev,
\newblock \emph{The ergodic side of the many-body localization transition},
\newblock Ann. Phys. (Berlin)  (2017),
\newblock \doi{10.1002/andp.201600350}.

\bibitem{agarwal_rare-region_2017}
K.~Agarwal, E.~Altman, E.~Demler, S.~Gopalakrishnan, D.~A. Huse and M.~Knap,
\newblock \emph{Rare-region effects and dynamics near the many-body
  localization transition},
\newblock Ann. Phys. (Berlin)  (2017),
\newblock \doi{10.1002/andp.201600326}.

\bibitem{imbrie_review:_2016}
J.~Z. Imbrie, V.~Ros and A.~Scardicchio,
\newblock \emph{Review: {Local} {Integrals} of {Motion} in {Many}-{Body}
  {Localized} systems},
\newblock arXiv:1609.08076  (2016),
\newblock ArXiv: 1609.08076.

\bibitem{huse2014phenomenology}
D.~A. Huse, R.~Nandkishore and V.~Oganesyan,
\newblock \emph{Phenomenology of fully many-body-localized systems},
\newblock Physical Review B \textbf{90}(17), 174202 (2014).

\bibitem{serbyn_local_2013}
M.~Serbyn, Z.~Papi{\'c} and D.~A. Abanin,
\newblock \emph{Local {Conservation} {Laws} and the {Structure} of the
  {Many}-{Body} {Localized} {States}},
\newblock Phys. Rev. Lett. \textbf{111}(12), 127201 (2013),
\newblock \doi{10.1103/PhysRevLett.111.127201}.

\bibitem{deutsch1991quantum}
J.~Deutsch,
\newblock \emph{Quantum statistical mechanics in a closed system},
\newblock Physical Review A \textbf{43}(4), 2046 (1991).

\bibitem{srednicki1994chaos}
M.~Srednicki,
\newblock \emph{Chaos and quantum thermalization},
\newblock Physical Review E \textbf{50}(2), 888 (1994).

\bibitem{rigol2008thermalization}
M.~Rigol, V.~Dunjko and M.~Olshanii,
\newblock \emph{Thermalization and its mechanism for generic isolated quantum
  systems},
\newblock Nature \textbf{452}(7189), 854 (2008).

\bibitem{d2015quantum}
L.~D'Alessio, Y.~Kafri, A.~Polkovnikov and M.~Rigol,
\newblock \emph{From quantum chaos and eigenstate thermalization to statistical
  mechanics and thermodynamics},
\newblock arXiv preprint arXiv:1509.06411  (2015).

\bibitem{imbrie2016many}
J.~Z. Imbrie,
\newblock \emph{On many-body localization for quantum spin chains},
\newblock Journal of Statistical Physics \textbf{163}(5), 998 (2016).

\bibitem{imbrie2016review}
J.~Z. Imbrie, V.~Ros and A.~Scardicchio,
\newblock \emph{Review: Local integrals of motion in many-body localized
  systems},
\newblock arXiv preprint arXiv:1609.08076  (2016).

\bibitem{grover2014certain}
T.~Grover,
\newblock \emph{Certain general constraints on the many-body localization
  transition},
\newblock arXiv preprint arXiv:1405.1471  (2014).

\bibitem{serbyn2015criterion}
M.~Serbyn, Z.~Papi{\'c} and D.~A. Abanin,
\newblock \emph{Criterion for many-body localization-delocalization phase
  transition},
\newblock Physical Review X \textbf{5}(4), 041047 (2015).

\bibitem{potter2015universal}
A.~C. Potter, R.~Vasseur and S.~Parameswaran,
\newblock \emph{Universal properties of many-body delocalization transitions},
\newblock Physical Review X \textbf{5}(3), 031033 (2015).

\bibitem{vosk2015theory}
R.~Vosk, D.~A. Huse and E.~Altman,
\newblock \emph{Theory of the many-body localization transition in
  one-dimensional systems},
\newblock Physical Review X \textbf{5}(3), 031032 (2015).

\bibitem{khemani2017critical}
V.~Khemani, S.-P. Lim, D.~N. Sheng and D.~A. Huse,
\newblock \emph{Critical properties of the many-body localization transition},
\newblock Phys. Rev. X \textbf{7}(2), 021013 (2017).

\bibitem{PhysRevLett.119.075702}
V.~Khemani, D.~N. Sheng and D.~A. Huse,
\newblock \emph{Two universality classes for the many-body localization
  transition},
\newblock Phys. Rev. Lett. \textbf{119}, 075702 (2017),
\newblock \doi{10.1103/PhysRevLett.119.075702}.

\bibitem{PhysRevB.96.104205}
F.~Setiawan, D.-L. Deng and J.~H. Pixley,
\newblock \emph{Transport properties across the many-body localization
  transition in quasiperiodic and random systems},
\newblock Phys. Rev. B \textbf{96}, 104205 (2017),
\newblock \doi{10.1103/PhysRevB.96.104205}.

\bibitem{kulshreshtha2017behaviour}
A.~K. Kulshreshtha, A.~Pal, T.~B. Wahl and S.~H. Simon,
\newblock \emph{Behaviour of l-bits near the many-body localization
  transition},
\newblock arXiv preprint arXiv:1707.05362  (2017).

\bibitem{parameswaran_eigenstate_2017}
S.~A. Parameswaran, A.~C. Potter and R.~Vasseur,
\newblock \emph{Eigenstate phase transitions and the emergence of universal
  dynamics in highly excited states},
\newblock Ann. Phys. (Berlin)  (2017),
\newblock \doi{10.1002/andp.201600302}.

\bibitem{zhang2016many}
L.~Zhang, B.~Zhao, T.~Devakul and D.~A. Huse,
\newblock \emph{Many-body localization phase transition: A simplified
  strong-randomness approximate renormalization group},
\newblock Physical Review B \textbf{93}(22), 224201 (2016).

\bibitem{dumitrescu2017scaling}
P.~T. Dumitrescu, R.~Vasseur and A.~C. Potter,
\newblock \emph{Scaling theory of entanglement at the many-body localization
  transition},
\newblock Phys. Rev. Lett. \textbf{119}, 110604 (2017),
\newblock \doi{10.1103/PhysRevLett.119.110604}.

\bibitem{devakul2015early}
T.~Devakul and R.~R. Singh,
\newblock \emph{Early breakdown of area-law entanglement at the many-body
  delocalization transition},
\newblock Physical review letters \textbf{115}(18), 187201 (2015).

\bibitem{PhysRevB.94.045111}
S.~P. Lim and D.~N. Sheng,
\newblock \emph{Many-body localization and transition by density matrix
  renormalization group and exact diagonalization studies},
\newblock Phys. Rev. B \textbf{94}, 045111 (2016),
\newblock \doi{10.1103/PhysRevB.94.045111}.

\bibitem{wegner1994flow}
F.~Wegner,
\newblock \emph{Flow-equations for hamiltonians},
\newblock Annalen der physik \textbf{506}(2), 77 (1994).

\bibitem{PhysRevLett.116.010404}
L.~Rademaker and M.~Ortu\~no,
\newblock \emph{Explicit local integrals of motion for the many-body localized
  state},
\newblock Phys. Rev. Lett. \textbf{116}, 010404 (2016),
\newblock \doi{10.1103/PhysRevLett.116.010404}.

\bibitem{PhysRevLett.119.075701}
D.~Pekker, B.~K. Clark, V.~Oganesyan and G.~Refael,
\newblock \emph{Fixed points of wegner-wilson flows and many-body
  localization},
\newblock Phys. Rev. Lett. \textbf{119}, 075701 (2017),
\newblock \doi{10.1103/PhysRevLett.119.075701}.

\bibitem{monthus2016flow}
C.~Monthus,
\newblock \emph{Flow towards diagonalization for many-body-localization models:
  adaptation of the toda matrix differential flow to random quantum spin
  chains},
\newblock J. Phys. A Math. Theor \textbf{49}, 305002 (2016).

\bibitem{rademaker2017many}
L.~Rademaker, M.~Ortu{\~n}o and A.~M. Somoza,
\newblock \emph{Many-body localization from the perspective of integrals of
  motion},
\newblock Annalen der Physik p. 1600322 (2017).

\bibitem{de2017stability}
W.~De~Roeck and F.~Huveneers,
\newblock \emph{Stability and instability towards delocalization in many-body
  localization systems},
\newblock Physical Review B \textbf{95}(15), 155129 (2017).

\bibitem{chandran2016many}
A.~Chandran, A.~Pal, C.~Laumann and A.~Scardicchio,
\newblock \emph{Many-body localization beyond eigenstates in all dimensions},
\newblock arXiv preprint arXiv:1605.00655  (2016).

\bibitem{ponte2017thermal}
P.~Ponte, C.~Laumann, D.~A. Huse and A.~Chandran,
\newblock \emph{Thermal inclusions: how one spin can destroy a many-body
  localized phase},
\newblock arXiv preprint arXiv:1707.00004  (2017).

\bibitem{luitz2017}
D.~Luitz, W.~De~Roeck and F.~Huveneers,
\newblock \emph{How a small quantum batch can thermalize long localized
  chains},
\newblock https://arxiv.org/abs/1705.10807  (2017).

\bibitem{ThieryHuveneersMullerDeRoeck2017}
T.~{Thiery}, F.~{Huveneers}, M.~{M{\"u}ller} and W.~{De Roeck},
\newblock \emph{{Many-body delocalization as a quantum avalanche}},
\newblock ArXiv e-prints  (2017),
\newblock \eprint{1706.09338}.

\bibitem{de2016absence}
W.~De~Roeck, F.~Huveneers, M.~M{\"u}ller and M.~Schiulaz,
\newblock \emph{Absence of many-body mobility edges},
\newblock Physical Review B \textbf{93}(1), 014203 (2016).

\bibitem{chandran2015finite}
A.~Chandran, C.~R. Laumann and V.~Oganesyan,
\newblock \emph{Finite size scaling bounds on many-body localized phase
  transitions},
\newblock arXiv preprint arXiv:1509.04285  (2015).

\bibitem{MajumdarLectureNotes2010}
S.~N. Majumdar,
\newblock \emph{Universal first-passage properties of discrete-time random
  walks and lévy flights on a line: Statistics of the global maximum and
  records},
\newblock Physica A: Statistical Mechanics and its Applications
  \textbf{389}(20), 4299  (2010),
\newblock \doi{https://doi.org/10.1016/j.physa.2010.01.021},
\newblock Proceedings of the 12th International Summer School on Fundamental
  Problems in Statistical Physics.

\bibitem{LeDoussalWiese2009}
P.~Le~Doussal and K.~J. Wiese,
\newblock \emph{Driven particle in a random landscape: Disorder correlator,
  avalanche distribution, and extreme value statistics of records},
\newblock Physical Review E \textbf{79}(5), 051105 (2009).

\bibitem{GarciaMataGiraudGeorgeotMartinDubertrandLemarie2017}
I.~Garcia-Mata, O.~Giraud, B.~Georgeot, J.~Martin, R.~Dubertrand and
  G.~Lemari\'e,
\newblock \emph{Scaling theory of the anderson transition in random graphs:
  Ergodicity and universality},
\newblock Phys. Rev. Lett. \textbf{118}, 166801 (2017),
\newblock \doi{10.1103/PhysRevLett.118.166801}.

\bibitem{TikhonovMirlinSkvortsov2016}
K.~S. Tikhonov, A.~D. Mirlin and M.~A. Skvortsov,
\newblock \emph{Anderson localization and ergodicity on random regular graphs},
\newblock Phys. Rev. B \textbf{94}, 220203 (2016),
\newblock \doi{10.1103/PhysRevB.94.220203}.

\bibitem{de_roeck_stability_2016}
W.~De~Roeck and F.~Huveneers,
\newblock \emph{Stability and instability towards delocalization in many-body
  localization systems},
\newblock Phys. Rev. B \textbf{95}(15), 155129 (2017).

\end{thebibliography}
\bibliographystyle{SciPost_bibstyle}

\nolinenumbers

\end{document}